\documentclass[twocolumn, twocolappendix,times]{aastex631}
\DeclareUnicodeCharacter{02BC}{'}
\usepackage{amsmath}	
\usepackage{amssymb}	
\usepackage{soul}
\usepackage[normalem]{ulem}

\newcommand{\pc}{\ensuremath{\,{\rm pc}\ }}
\newcommand{\kpc}{\ensuremath{\,{\rm kpc}\ }}

\newcommand{\kms}{\,{\rm km}\,{\rm s}$^{-1}$}

\def\apjl{Astrophysical Journal Letters}
\def\aap{Astronomy and Astrophysics}

\def\apjs{The Astrophysical Journal Supplement Series}

\begin{document}

\title{Dynamical Models of the Milky Way in Action Space with LAMOST DR8 and GAIA EDR3}

\author[0000-0002-1398-5588]{Guang-Chen Sun}
\affiliation{Key Laboratory of Cosmology and Astrophysics (Liaoning)\\
\& College of Sciences, Northeastern University, Shenyang 110819, China\\}
\affiliation{National Astronomical Observatories, Chinese Academy of Sciences, Beijing 100101, China\\}
\affiliation{School of Astronomy and Space Science, University of Chinese Academy of Sciences, Beijing 100049, China\\}

\author[0000-0003-2153-7758]{Qiao Wang}
\affiliation{National Astronomical Observatories, Chinese Academy of Sciences, Beijing 100101, China\\}
\affiliation{School of Astronomy and Space Science, University of Chinese Academy of Sciences, Beijing 100049, China\\}

\author{Shude Mao}
\affiliation{Department of Astronomy, Tsinghua University, Beijing 100084, China\\}

\author[0000-0003-1962-2013]{Yichao Li}
\affiliation{Key Laboratory of Cosmology and Astrophysics (Liaoning)\\
\& College of Sciences, Northeastern University, Shenyang 110819, China\\}

\author[0000-0002-8559-0067]{Richard J. Long}
\affiliation{Department of Astronomy, Tsinghua University, Beijing 100084, China\\}

\author[0000-0003-3907-697X]{Ping-Jie Ding}
\affiliation{Purple Mountain Observatory, Chinese Academy of Sciences, Nanjing 210023, Peopleʼs Republic of China\\}

\correspondingauthor{Yougang Wang}
\author[0000-0003-0631-568X]{Yougang Wang}
\email{wangyg@bao.ac.cn}
\affiliation{National Astronomical Observatories, Chinese Academy of Sciences, Beijing 100101, China\\}
\affiliation{Key Laboratory of Radio Astronomy and Technology, Chinese Academy of Sciences, A20 Datun Road, Chaoyang District, Beijing 100101, China\\}
\affiliation{School of Astronomy and Space Science, University of Chinese Academy of Sciences, Beijing 100049, China\\}
\affiliation{Key Laboratory of Cosmology and Astrophysics (Liaoning)\\
\& College of Sciences, Northeastern University, Shenyang 110819, China\\}

\correspondingauthor{Xin Zhang}
\author[0000-0002-6029-1933]{Xin Zhang}
\email{zhangxin@mail.neu.edu.cn}
\affiliation{Key Laboratory of Cosmology and Astrophysics (Liaoning)\\
\& College of Sciences, Northeastern University, Shenyang 110819, China\\}
\affiliation{National Frontiers Science Center for Industrial Inteligence and Systems Optimization, Northeastern University, shenyang 110819, China}
\affiliation{Key Laboratory of Data Analytics and Optimization for Smart Industry (Ministry of Education), Northeastern University, Shenyang 110819, China}

\correspondingauthor{Xuelei Chen}
\author[0000-0001-6475-8863]{Xuelei Chen}
\email{xuelei@cosmology.bao.ac.cn}
\affiliation{National Astronomical Observatories, Chinese Academy of Sciences, Beijing 100101, China\\}
\affiliation{Key Laboratory of Radio Astronomy and Technology, Chinese Academy of Sciences, A20 Datun Road, Chaoyang District, Beijing 100101, China\\}
\affiliation{School of Astronomy and Space Science, University of Chinese Academy of Sciences, Beijing 100049, China\\}
\affiliation{Key Laboratory of Cosmology and Astrophysics (Liaoning)\\
\& College of Sciences, Northeastern University, Shenyang 110819, China\\}



\begin{abstract}
This work explores dynamical models of the Milky Way (MW) by analyzing a sample of 86,109 K giant stars selected through cross-matching the LAMOST DR8 and Gaia EDR3 surveys. 
Our earlier torus models in \citet{Wang2017} did not include Gaia data, making them incompatible with the new sample’s proper motion distributions. Here, we refine the construction of action-based, self-consistent models to constrain the three-dimensional velocity distribution of K giants over a larger parameter space, drawing on a series of existing MW models. This approach produces several new MW models.
Our best-fit model for the local kinematics near the Sun indicates a MW virial mass of 1.35 $\times 10^{12} M_\odot$, a local stellar density of 0.0696 $\rm M_\odot pc^{-3}$, and a local dark matter density of 0.0115 $\rm M_\odot pc^{-3}$.
Our main conclusion supports a thicker and more extended thick disk, alongside a cooler thin disk, compared to the best-fitting model in \citet{Wang2017}. Near the Sun, our model aligns well with observations, but is less satisfactory at distances far from the Galactic center, perhaps implying unidentified structures. 
Further high-precision observations will be critical for understanding the dynamics in these outer Galactic regions, and will require a more realistic model.
\end{abstract}

\keywords{Galaxy dynamics (591) --- Milky Way Galaxy (1054) --- Milky Way mass (1058)---- Galaxy kinematics (602) --- Milky Way Galaxy physics (1056) --- Galaxy physics (612)}


\section{Introduction}
\label{sec:introduction}

To unravel the mysteries of the Milky Way (MW), astrophysicists have long sought to develop dynamical models that accurately reflect its structure and formation. The advent of high-precision observing campaigns, especially the Gaia mission, has precipitated an unprecedented influx of data, characterizing over a billion stars with previously unmatched astrometric and photometric precision \citep{2016A&A...595A...1G, gaia2018a, GaiaEDR32020}. Combining observations from large spectroscopic surveys, such as the Large Area Multi-Object Spectroscopic Telescope (LAMOST) Experiment for Galactic Understanding and Exploration ~\citep{Deng2012}, a large number of stars with six-dimensional phase space information has become available. It provides a unique opportunity to refine our understanding of the MW's fundamental properties, including its mass distribution, kinematic behaviors, and the elusive dark matter halo surrounding it.

Many techniques have been proposed to model the dynamics of the MW. The first technique is the Jeans method, based on moments of the Jeans equation and the adopted density and kinematics~\citep{Jeans1915}. Most mass measurements of the MW are obtained from the Jeans equation~\citep[e.g.][]{Xue2008, Kafle2014, Bird2022, Sun2023}. The second method is Schwarzschild's orbit-superposition technique~\citep{1979ApJ...232..236S,1993ApJ...409..563S}, which builds a stead-state model by calculating orbits in a fixed gravitational potential and determining the orbit weights required to fit the observational constraints. Using this method, a three-dimensional steady-state stellar dynamical model of the Galactic bar has been constructed ~\citep{Zhao1996}, and two relatively new Galactic bar models were constructed by ~\cite{Wang2012, Wang2013}. The third method is the made-to-measure (M2M) method, first proposed by~\cite{1996MNRAS.282..223S}. This method is close to Schwarzschild's orbit-superposition technique, the only difference being how the Schwarzschild orbit and M2M particle weights are obtained. M2M techniques have also been used to construct the Galactic bar ~\citep{Long2013, Hunt2013, Zhu2014,2017MNRAS.470.1233P, Webb2023}.     

\cite{1990MNRAS.244..634M} proposed a numerical method to construct action-angle tori in general gravitational potentials, and a series of following studies~\citep{1993MNRAS.261..584B,1994MNRAS.268.1033K,1999ASPC..182..178V,2008MNRAS.390..429M} described tori construction and their application to galactic dynamics.  An orbital torus is associated with specific values of the actions $\boldsymbol{J}$.  Once a torus has been so specified, the star's position and velocity is determined, and the contribution to the local density from any value of the star's angle variables $\theta$ can be obtained~\citep{2016MNRAS.456.1982B}.  This torus and action-based distribution function method is useful and powerful for modeling the dynamics of galaxies ~\citep{2016MNRAS.457.2107S}, and has distinct advantages~\citep{2016MNRAS.456.1982B}.  
Recently, new self-consistent models of the MW have been introduced by the application of action-based distribution functions (DFs)~\citep{binney2023, binney2024}.  Meanwhile, \cite{Robin2022} used the Stäckel approximation to build a self-consistent dynamical model of the MW disk, \textbf{defining a DF in terms of three integrals of motion, energy $E$, angular momentum $L_z$, and a third integral $I_3$ which is assumed to be close to the truly conserved integral in the the Stäckel potential.}  
These studies underscore the efficacy of action-based DF methods for Galactic modeling. By leveraging integrals of motion and iteratively refining the gravitational potential, researchers can build models that remain dynamically self-consistent while aligning closely with observational constraints.

We have constructed torus models of the MW in a large volume by using the K giant stars selected from the LAMOST DR 2 catalogue~\citep{Wang2017}. We found that the outer disk is much thicker than previously thought, or alternatively that the outer structure is not a conventional disk at all. 
Here we return to this topic for three main reasons. First, the sample of K giants has significantly increased; second, proper motion constraints were not used in \cite{Wang2017}; and, third, dynamical modeling in action space has also been improved which enables us to construct self-consistent models in action space.

The structure of this paper is as follows. Section~\ref{sec:Observations} describes the data samples used. 
In Section~\ref{sec:methods}, we detail our models and methods. In Section~\ref{sec:results}, we present the dynamical models of the Milky Way. Section~\ref{sec:conclusions} summarizes our main results.

\section{Data Sample}
\label{sec:Observations}

In this study, we focus on K-type giant stars, which are highly luminous and have long lifespans, making them excellent tracers for Galactic observations. Our primary dataset comes from the catalog by \cite{Ding2021}, where K-type giants are identified following the criteria of \cite{Liu2014}, which is based on the effective temperature $(T_{\rm eff})$ and the surface gravity $(\log g)$, $4000\ {\rm K}<T_{\rm eff}< 4600\ {\rm K}$ with $\log g<3.5$ and $4600\ {\rm K}<T_{\rm eff}< 5600\ {\rm K}$ with $\log g<4$.
The spectroscopic data from the LAMOST eighth data release (LAMOST DR8). This identification is complemented by Gaia EDR3 proper motions and parallaxes, ensuring high accuracy. The line-of-sight velocity ($v_{\rm LOS}$) measurements are taken directly from LAMOST DR8.

Neither LAMOST nor Gaia datasets are immune to systematic errors, which must be carefully addressed in the analysis \citep{anguiano2018, lingdian2021}. We correct the LAMOST $v_{\rm LOS}$ measurements by 5.7 \kms, following \cite{Tian2015} to account for observed systematic offsets. Recently, \cite{lingdian2021} introduced a method to correct the Gaia EDR3 parallax zero-point, considering source magnitude, color, and sky position. This approach is crucial for improving distance estimates and enhancing spatial analyses. Using the corrected parallax $\varpi$, we derive the inverse parallax distance $D_{\varpi} = 1/\varpi$. Additionally, we follow \cite{Ding2021}, which employed a method developed by \cite{carlin2015} to estimate photometric distances ($D_{\rm Carlin}$) using the LAMOST spectra.
This dual approach to distance estimation provides a robust assessment of the spatial distribution of K giants. Section~\ref{sec:suitable distance} further discusses the optimization of these distances for specific regions of interest.

To ensure high-quality data, we adhere to the following standards: for LAMOST DR8 data, we apply $S/N > 5$; for Gaia EDR3 data, we apply \texttt{ASTROMETRIC\_GOF\_AL} $\leq 3.0$, \texttt{ASTROMETRIC\_EXCESS\_NOISE\_SIG} $\leq 2.0$, and a renormalized unit weight error $(RUWE) \leq 1.4$, selecting stars with reliable spectra and astrometry. Here, $S/N$ represents the signal-to-noise ratio in the G-band. We compile a comprehensive six-dimensional phase space dataset of 607,833 K giants, which forms the basis for all subsequent analyses unless otherwise stated.

\begin{figure}
\includegraphics[width= \columnwidth]{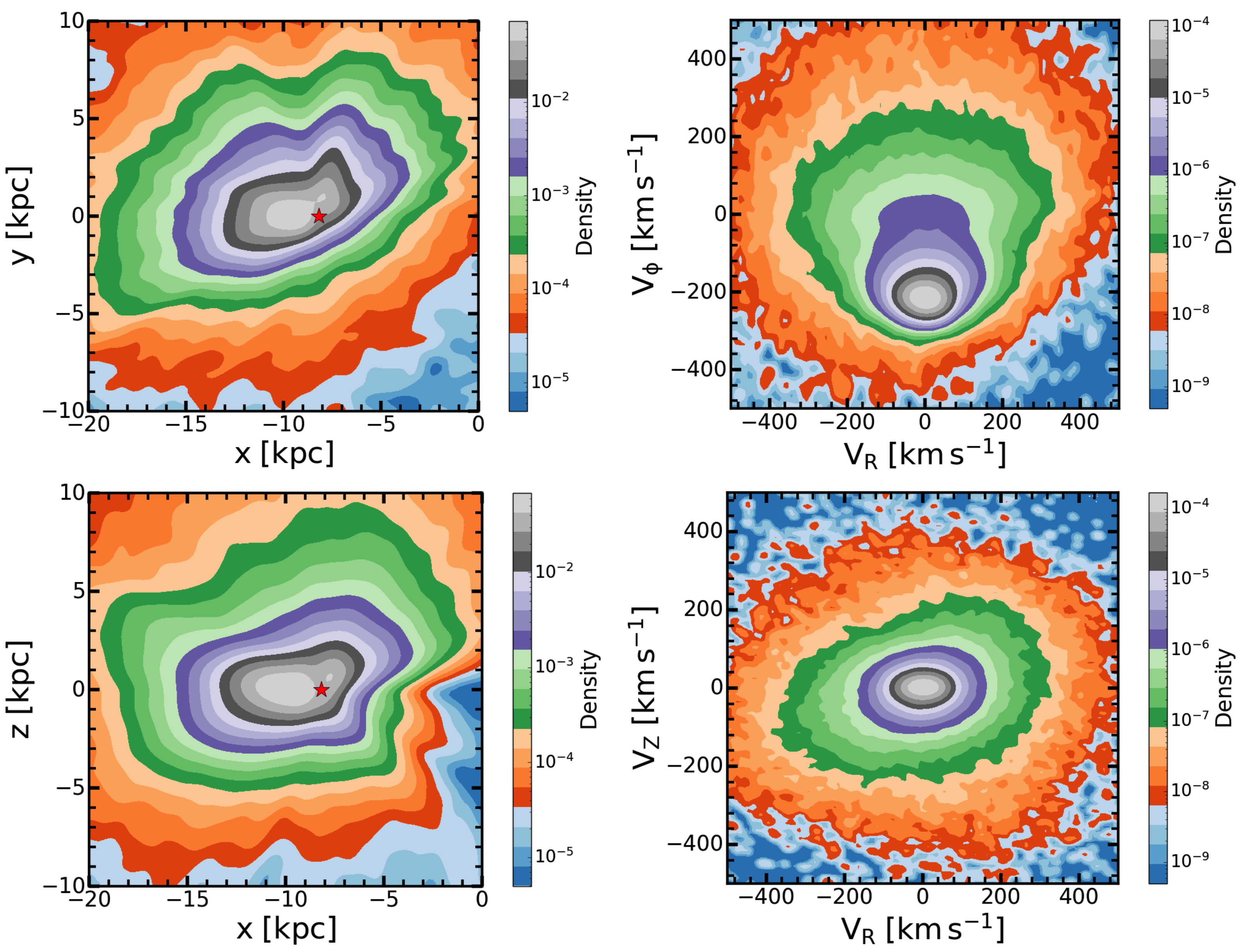}
    \caption{The projected distribution in the x-y plane (upper left), x-z plane (lower left), $V_R$-$V_{\phi}$ plane (upper right), and $V_R$-$V_Z$ plane (lower right) for all 607,833 K giants in our sample. The red stars in the left panels indicate the position of the Sun. The color mapping represents the observed density of K giants, calculated using the kernel-density estimation method with Gaussian kernels. It is important to note that this method may smooth out some fine details.}
    \label{fig:local}
\end{figure}

We use the \texttt{Astropy} library \citep{astropy2013, astropy2018} to transform ($\alpha$, $\delta$, $D_{\rm Carlin}$, $v_{\rm LOS}$, $\mu_\alpha$, $\mu_\delta$) into a Cartesian coordinate system $(x, y, z, v_x, v_y, v_z)$ centered on the Galactic center. In this system, $\alpha$ and $\delta$ are right ascension (R.A.) and declination, respectively, $x$ points from the Sun towards the Galactic center, $y$ aligned with Galactic rotation, and $z$ directs towards the Northern Galactic Pole. Velocities $v_x$, $v_y$, and $v_z$ are measured in the Galactic rest frame. $\mu_\alpha$ and $\mu_\delta$ correspond to $\alpha$ and $\delta$. Figure~\ref{fig:local} shows the spatial distribution of all 607,833 K giants.

The parameters required for coordinate transformation include the distance from the Sun to the Galactic center, which we set to $R_{\sun} = 8.178 \pm 0.035$ \kpc \citep{GravityCollaboration2019}, and the Sun's height above the Galactic plane is set to $z_{\sun} = 20.8 \pm 0.3$ \pc \citep{zsun2019}. The motion of the Sun in the Galactic frame is represented as $v_{\sun} = (U_{\sun}, V_{\sun}, W_{\sun}) = (11.10 \pm 1.75, 12.24 \pm 2.47, 7.25 \pm 0.87)$ \kms \citep{schonrich2010}, where $U_{\sun}$, $V_{\sun}$, and $W_{\sun}$ represent the components toward the Galactic center, in the direction of Galactic rotation, and towards the Galactic north pole, respectively. 
The initial potential model, hereafter Mc17 \citep{McMillan2017}, represents the MW as an axisymmetric system comprising a bulge, thin and thick stellar disks, two gas disks, and a spherical dark matter halo. In Mc17, the circular velocity at the Sun’s position is $v_0 = 232.8 \pm 3.0$ \kms.

\subsection{Suitable distance}
\label{sec:suitable distance}

Distances for K-type giant stars can be estimated using two methods: the inverse parallax distance from Gaia EDR3 ($D_{\varpi}$), and the photometric distance ($D_{\rm Carlin}$) obtained from absolute magnitudes and stellar parameters based on LAMOST spectra \citep{carlin2015}. While $D_{\varpi}$ benefits from Gaia’s high-precision astrometric measurements (particularly for stars located nearer to the Sun), its reliability decreases for more distant objects or those with lower astrometric accuracy \citep{2016ApJ...832..137A, Luri2018}. In contrast, $D_{\rm Carlin}$ is less affected by these observational constraints but is more sensitive to uncertainties arising from extinction and stellar atmosphere modeling.

\begin{figure}
\includegraphics[width= \columnwidth]{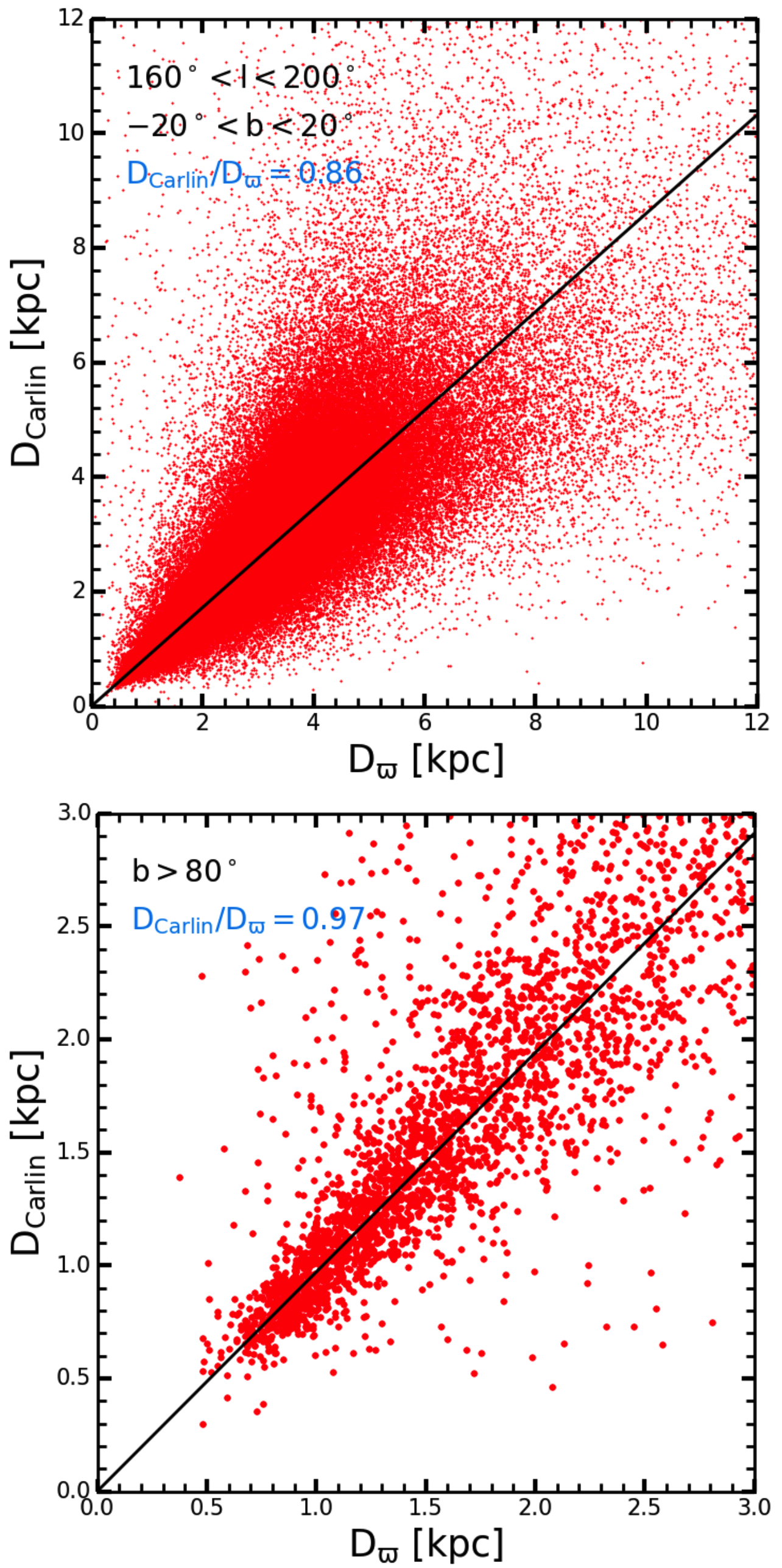}
    \caption{The comparison between photometric distance ($D_{\rm Carlin}$) and parallax distance ($D_{\varpi}$) within the specified range reveals interesting trends. In the direction of the Galactic anticenter (upper panel), we observe $D_{\rm Carlin}/D_{\varpi}$=0.86, incorporating approximately 160,000 high-precision parallax measurements for K giants. Conversely, toward the north Galactic pole, this ratio is $D_{\rm Carlin}/D_{\varpi}$=0.97, based on observations of approximately 2,000 K giants.}
    \label{fig:distant}
\end{figure}

As noted by \cite{Ding2021}, there is a close agreement between $D_{\varpi}$ and $D_{\rm Carlin}$, although $D_{\varpi}$ tends to yield slightly smaller values. Moreover, there is a constant scale factor between these two distances. In Figure~\ref{fig:distant}, we focus on stars with small relative parallax uncertainties ($\sigma_\varpi/\varpi < 0.1$) and $D_{\rm Carlin}$ uncertainties  ($\sigma_{D_{\rm Carlin}}/D_{\rm Carlin} < 0.5$). The agreement between these two distance measures is better towards the north Galactic Pole than the Galactic anticenter, possibly due to the significant influence of interstellar dust in the galactic plane on photometric estimates. 

This observation motivates us to apply region-specific $D_{\rm Carlin}/D_{\varpi}$ ratios to reconcile the differences. Following the method of \cite{Ding2021}, we use $D = D_{\rm Carlin}/0.86$ to denote the corrected distance in the direction of the Galactic disk region and $D = D_{\rm Carlin}/0.97$ in the polar direction, ensuring that our distance estimates are consistent and accurately reflect data uncertainties. It is noted that the scale factor 0.86 in the Galactic disk region in this paper is slightly larger $(2\%)$  than the scale factor 0.842 in \cite{Ding2021}, which is used for all samples.

\subsection{Sample selection}
\label{sec:sample selection}

\begin{figure}
\includegraphics[width=\columnwidth]{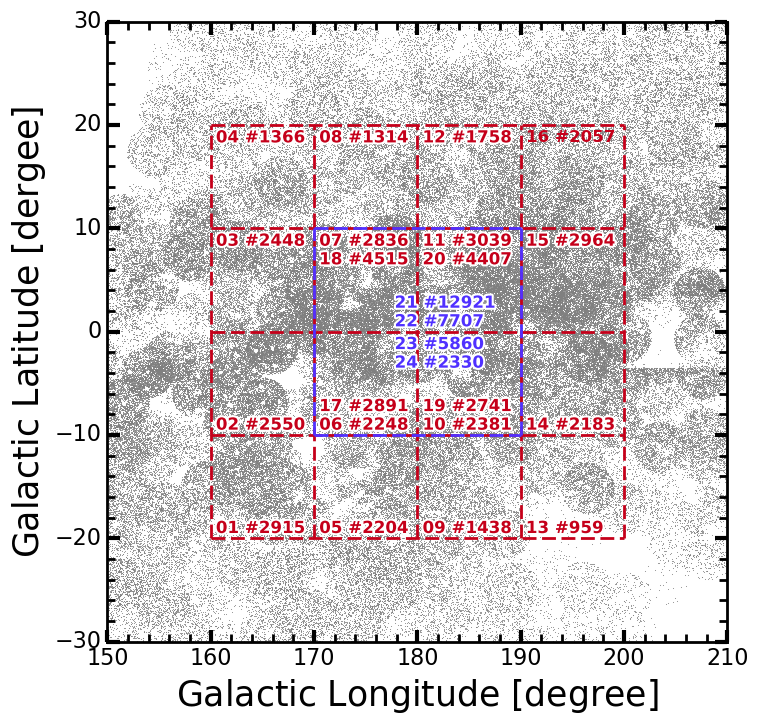}
\caption{The selected sky regions. The figure is similar to Figure 1 in \protect\cite{Wang2017}. The labels denote the region identifier. Following the symbol `\#' is the number of K giants in the region. Regions 01–16 encompass giants within 2 \kpc from the Sun. Regions 17, 18, 19, and 20 cover the same directions as 06, 07, 10, and 11, respectively, but extend to a depth of 2–3 \kpc. Regions 21, 22, 23, and 24 cover a $20^\circ \times,20^\circ$ sky area of the Galactic anticenter from 3–12 \kpc, with the blue box indicating the area covered by regions 21, 22, 23, and 24. Additionally, the grey points represent all K giants in the survey.}
\label{fig:lb}
\end{figure}

Considering the observational limitations of LAMOST, we focus our analysis on K-type giants in the Galactic anticenter to characterize the parameters of the Galactic disk. Figure~\ref{fig:lb} extends Figure 1 from \cite{Wang2017}, with the new observational sample roughly doubling the previous one overall and expanding by a factor of $4–6$ in more distant regions. This enlarged dataset enhances the comparative framework provided by the reference study.
Our study benefits from an expanded dataset of giants, thus enriching the comparative basis established in the reference study.

In characterizing our sample of K giants as disc-specific members, we adopt velocity-dispersion parameters from \cite{Anguiano2020, Ding2021, vieira2022}. A relaxed criterion of $3\sigma$ is adopted to reduce the inclusion of stars with excessively high velocities, thereby setting the constraints as follows: $|V_R| < 200$ \kms, $-350 < V_{\phi} < -100$ \kms, $|V_{\rm z}| < 150$ \kms, $|V_{\rm tot}| < 600$ \kms, and [Fe/H] $> -1$. Here,  $V_{\rm tot} = \sqrt{V_R^2 + V_{\phi}^2 + V_{\rm z}^2}$ represents the total velocity relative to the Galactic center. 

The regions of the area observed are identified in Table~\ref{tab:sky area}. Regions 25 and 26 are specifically allocated to encompass the North Galactic Pole at a distance of up to 3 \kpc to facilitate an extensive examination of the vertical kinematics.  Regions 27 to 30 are strategically allocated to evaluate the validity of the potential and distribution model derived from the fitting process for the first 26 regions. These regions span four arbitrarily chosen regions close to the Galactic plane, covering distances ranging from 0 to 2 \kpc.

\begin{table}
	\centering
	\renewcommand{\arraystretch}{1}
    \setlength\tabcolsep{5pt}
	\caption{The sky area is divided into regions. }
	\label{tab:sky area}
	\begin{tabular}[c]{lllll}
		\hline
		& Longitude range & Latitude range & Distance & Counts\\
		& [deg] & [deg] & [kpc] & \\
        \hline
        01 & (160, 170) & (-20, -10) & [0, 2] & 2550\\
        02 & (160, 170) & (-10, 0)   & [0, 2] & 2915\\
        03 & (160, 170)	& (0, 10)    & [0, 2] & 2448\\
        04 & (160, 170)	& (10, 20)   & [0, 2] & 1366\\
        05 & (170, 180)	& (-20, -10) & [0, 2] & 2204\\
        06 & (170, 180)	& (-10, 0)   & [0, 2] & 2248\\
        07 & (170, 180)	& (0, 10)    & [0, 2] & 2836\\
        08 & (170, 180)	& (10, 20)   & [0, 2] & 1314\\
        09 & (180, 190)	& (-20, -10) & [0, 2] & 1438\\
        10 & (180, 190)	& (-10, 0)   & [0, 2] & 2381\\
        11 & (180, 190)	& (0, 10)    & [0, 2] & 3039\\
        12 & (180, 190)	& (10, 20)   & [0, 2] & 1758\\
        13 & (190, 200)	& (-20, -10) & [0, 2] & 959\\
        14 & (190, 200)	& (-10, 0)   & [0, 2] & 2183\\
        15 & (190, 200)	& (0, 10)    & [0, 2] & 2964\\
        16 & (190, 200)	& (10, 20)   & [0, 2] & 2057\\
        17 & (170, 180)	& (-10, 0)   & [2, 3] & 2891 \\
        18 & (170, 180)	& (0, 10)    & [2, 3] & 4515\\
        19 & (180, 190)	& (-10, 0)   & [2, 3] & 2741\\
        20 & (180, 190)	& (0, 10)    & [2, 3] & 4407\\
        21 & (170, 190)	& (-10, 10)  & [3, 4] & 12921\\
        22 & (170, 190)	& (-10, 10)  & [4, 5] & 7707\\
        23 & (170, 190)	& (-10, 10)  & [5, 7] & 5860\\
        24 & (170, 190)	& (-10, 10)  & [7, 12] & 2330\\
        25 & (0, 360)	& (80, 90)   & [0, 1.5] & 1272\\
        26 & (0, 360)	& (80, 90)   & [1.5, 3] & 1256\\
        27 & (210, 220)	& (10, 20)   & [0, 2] & 1144\\
        28 & (140, 150)	& (-10, 0)   & [0, 2] & 1839\\
        29 & (130, 140)	& (-20, -10) & [0, 2] & 1206\\
        30 & (200, 210)	& (0, 10)    & [0, 2] & 1360\\
		\hline
	\end{tabular}
\end{table}

\section{Methods}
\label{sec:methods}

In exploring galactic dynamics,  \texttt{AGAMA} (Action-based Galaxy Modeling Architecture)  offers a sophisticated platform for developing self-consistent models of galaxies~\citep{agama2019}. By utilizing action-angle variables, \texttt{AGAMA} provides a detailed and accurate method for modeling the intricate gravitational interactions that govern stellar motions and dark matter distributions, which are vital for understanding galaxy formation and evolution. For the avoidance of doubt, \textit{self-consistent} in this context is defined as in \cite{binney2011galactic} and is the definition used in \cite{agama2019}. In this paper, we use \texttt{AGAMA} for model construction.

\subsection{Action-based self-consistent methods}
Accurately modeling galaxies presents a significant challenge due to the complex interactions among billions of stars influenced by their collective gravitational fields and dark matter.  For an axisymmetric potential, there are three integrals of motion $I_1, I_2, I_3$,  and, by the strong Jeans Theorem, an equilibrium model can be taken as a function $f(I_1, I_2, I_3)$ of these integrals.   Action integrals have several advantages over the integrals of the motion~\citep{Binney2014,agama2019} so a distribution function in \texttt{AGAMA} is a function of the action integrals $f(\boldsymbol{J})$. Density, mean velocity, the second moment of velocity, and velocity dispersion are then defined as 
\begin{equation}
    \rho(\boldsymbol{x}) = \iiint \mathrm {d}^3 v\; f\big(\boldsymbol{J}[\boldsymbol{x},\boldsymbol{v}]\big),
	\label{eq:den}
\end{equation}
\begin{equation}
    \overline{\boldsymbol{v}} =\frac{1}{\rho} \iiint \mathrm{d}^3 v\; \boldsymbol{v}\; f\big(\boldsymbol{J}\big),
	\label{eq:vel}
\end{equation}
\begin{equation}
    \overline{v^2_{ij}} =\frac{1}{\rho} \iiint \mathrm{d}^3v\; v_i v_j\; f\big(\boldsymbol{J}\big)
	\label{eq:sec_vel}
\end{equation}
\begin{equation}
    \sigma^2_{ij} =\overline{v^2_{ij}} - \overline{v_{i}}\,\overline{v_{j}}.
	\label{eq:sig_vel}
\end{equation}
Generally, a galaxy model can be defined by specifying the DFs of each major component of stars and dark matter ~\citep{Binney2018}. However, a single DF can yield different density distributions depending on the potential ~\citep{agama2018}. \texttt{AGAMA} provides a self-consistent modelling approach,  through which self-consistent density-potential pairs can be constructed. The procedure for constructing a self-consistent model has the following steps.

(i) Create a reasonable initial guess for the total density (or potential) of the whole system. 

(ii) Adopt an action finder for computing the actions in the given potential.

(iii) Compute the density distribution from the DFs via Eq.~\ref{eq:den}. In practice, this step is carried out by sampling random particles in $(\boldsymbol{x}, \boldsymbol{v})$ space. Each $(\boldsymbol{x}, \boldsymbol{v})$ point is transformed into $(\boldsymbol{J}, \theta)$ space using the action finder.

(iv) Recalculate the potential by solving the Poisson equation with the updated density distribution.

(v) Repeat from step (ii), until the changes in potential are negligible, and the potential has converged as described in Section 5.2 of \cite{agama2019}. Typically, five iterations are enough with a good initial guess, and 10 iterations are sufficient in practice even with a poor guess.

For computational details, the reader is referred to \cite{agama2018}. The operational steps in this work are as follows.

(1) Initial preparations: Gather observational data, calibrate distances, and divide the sky into specific regions (see Section~\ref{sec:Observations}).

(2) Create velocity distributions: For each sky region, bin the three velocity components ($v_{\rm LOS}$, $\mu_\alpha$, $\mu_\delta$) of the observed K giants in International Celestial Reference System (hereafter ICRS) coordinates. The resulting histograms visualize the velocity distributions in each direction. 

(3) The parameters of the relevant distribution functions are initially set using values of density and distribution function parameters from existing references as initial values. Utilize the \texttt{AGAMA} code to construct a self-consistent model, and use this model to produce 300 million simulated data points, each representing a K giant, to create the simulated (model) distribution histograms. 
\textbf{In order to increase the calculation speed, the number of iterations is set to three.}

(4) Perform least $\chi^2$ fits. Evaluate how well the model fits the data using the Nelder–Mead optimization method to search for $\chi^2$ minima. The DF parameters are set as free, and we use step (3) to construct the self-consistent model. The $\chi^2$ is defined as
\begin{equation}
    \chi^2=\sum_{n} \frac{(p_n^{\rm data}-p_n^{\rm mod})^2}{\sigma_n^2},
	\label{eq:chi2}
\end{equation}
where $p_n^{\rm data}$ denotes the observed velocity dispersion, and $\sigma_n$ the corresponding errors. $p_n^{\rm mod}$ represents the model prediction. The overall $\chi^2$ used for model comparison is computed as the average of the reduced $\chi^2$ values across all regions.

\begin{figure*}
\includegraphics[width= \textwidth]{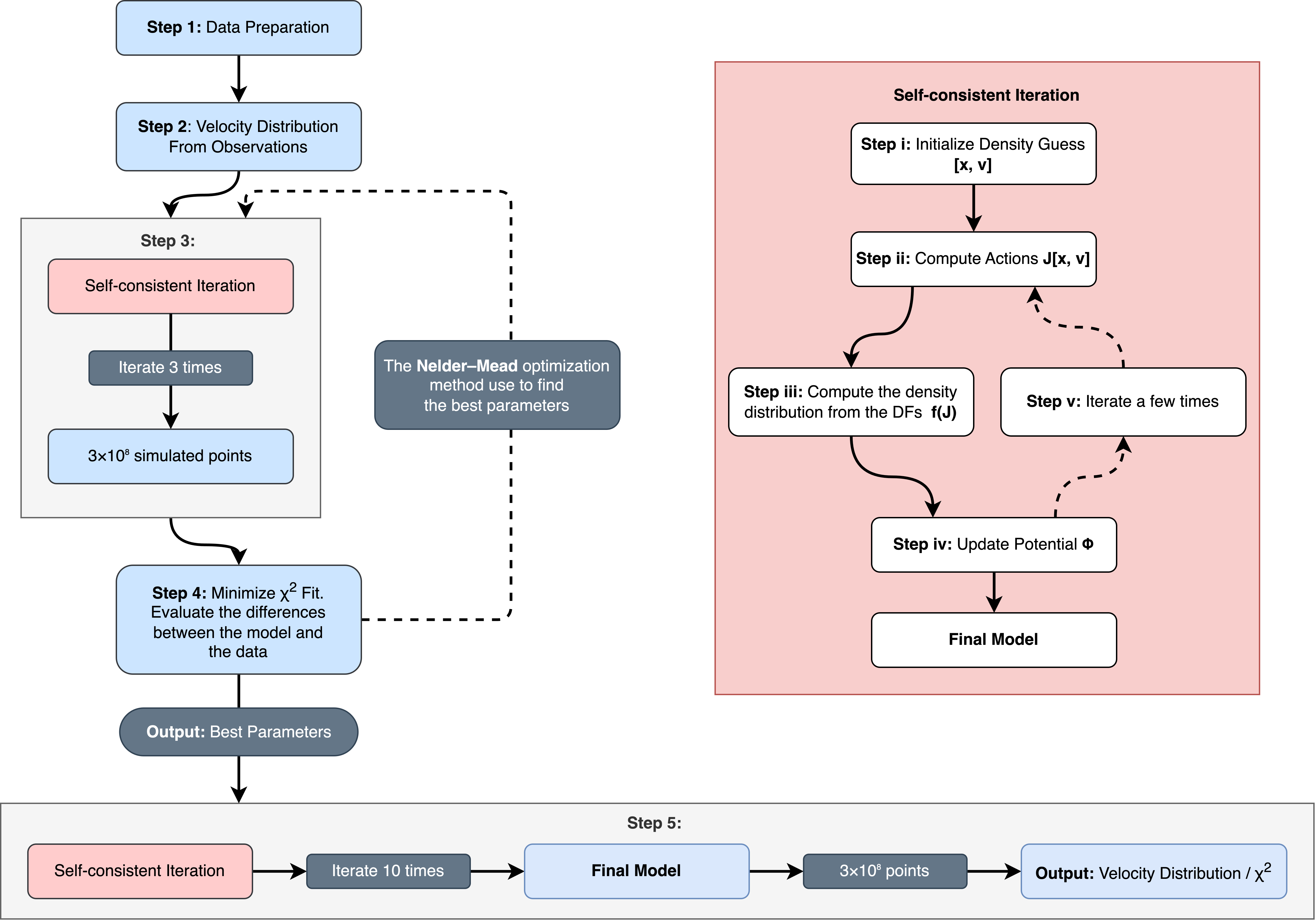}
    \caption{\textbf{Flowchart for constructing the best-fitting dynamical model. The left and bottom parts are the scheme for our calculating process, while the red top-right part represents the process of building a self-consistent model using AGAMA, which is an extension of the red box on the left. The flowchart of this study consists of five stages: data preparation (Step 1), velocity distribution modeling (Step 2), self-consistent iteration (Step 3), $\chi^2$ optimization (Step 4), and final model output (Step 5).}}
    \label{fig:flow}
\end{figure*}

(5) After step (4), we have a series of DFs from the preliminary self-consistent models. \textbf{We now use these DFs to construct the self-consistent models again.} We iterate this process 10 times and take the final self-consistent model for each DF distribution into our analyses. All results shown in this paper are based on 10 iterations for the construction of the self-consistent model of the MW, unless stated otherwise. \textbf{The $\chi^2$ values are obtained by generating 300 million points from  the final self-consistent model and fitting the kinematics predicted by these points.
The scheme of the operational steps is summarized in Figure ~\ref{fig:flow}.}

As shown in Figure~\ref{fig:distant}, the stellar distances are not very precise, and cause some stars to move into adjacent bins, thus distorting the line-of-sight velocity and PM distributions. Based on the position and velocity of each star, we assume a normal distribution for using the observed velocity values as the means and the observational errors as the standard deviations. We then randomly generate 1,000 values for each velocity, resulting in a new sample pool that is 1,000 times larger than the original. Subsequently, we perform repeated sampling from this pool, drawing a dataset equivalent in size to the original sample each time to compute the distribution. This sampling process is repeated 1,000 times, after which the median and error of these 1,000 samples are calculated, yielding the final $p_n^{\rm data}$  and  $\sigma_n$ values.

\subsection{Density and potential}
\label{sec:potential}

The density $\rho$ and potential $\Phi$ profiles of a galaxy model are related to each other by the Poisson's equation
\begin{equation}
    \nabla^2\Phi(\mathbf{x})=4\pi G \rho(x).
	\label{eq:dendisk}
\end{equation}
The MW model adopted in this paper is assumed to be axisymmetrical and comprises a
spheroidal bulge and a spheroidal dark matter halo, two stellar disks, an HI gas disk, and a molecular gas disk.

For disk components, the density profile is modeled as \citep{agama2019}
\begin{equation}
\begin{split}
    &\rho=\Sigma_0\exp{\left(-\frac{R}{R_d}-\frac{R_{\rm cut}}{R}\right)}\\
    &\times \left\{ \begin{array}{lll}
    \delta(z) & \mbox{if} & h=0,\\
    \frac{1}{2h}\exp{\left(-\left|\frac{z}{h}\right|\right)} & \mbox{if} & h>0,\\
    \frac{1}{4|h|}\mbox{sech}^2\left(\left|\frac{z}{2h}\right|\right) & \mbox{if} & h<0,
    \end{array}\right.
	\label{eq:dendisk}
\end{split}
\end{equation}
where $R=\sqrt{x^2+y^2}$ represents the cylindrical radius, $R_d$ is the scale radius, $\Sigma_0$ is the central surface density (value at $R=0$), and $h$ is the scale height. The \texttt{AGAMA} framework designates $h>0$ corresponds to an exponential disk, typically applied to stellar disks, while $h<0$ corresponds to an isothermal disk, typically applied to gas disks, with $R_{\rm cut}$ indicating the inner cutoff radius, adjusted according to the gas disks.

The spheroidal components, including the bulge and dark matter halo, have a density profile expressed as
\begin{equation}
    \rho = \rho_0\left(\frac{\tilde{r}}{r_0}\right)^{-\gamma}\left(1+\frac{\tilde{r}}{r_0}\right)^{\gamma-\beta}\times\exp{\left[-\left(\frac{\tilde{r}}{r_{\rm cut}}\right)^2\right]},
	\label{eq:densph}
\end{equation}
where
\begin{equation}
    \tilde{r} = \sqrt{x^2+y^2+\left(z/q\right)^2}
	\label{eq:tilder}
\end{equation}
defines the ellipsoidal radius, $\rho_0$ is the volume density at scale radius $r_0$, and $q$ is the axial ratio $z/x$ of the isodensity surfaces. The parameters $\beta$ and $\gamma$ represent the power-law indices of the density profile in the outer and inner regions, respectively, with $r_{\rm cut}$ marking the outer cutoff radius.

\subsection{Distribution function}
\label{sec:distribution function}

\begin{figure}
\includegraphics[width= \columnwidth]{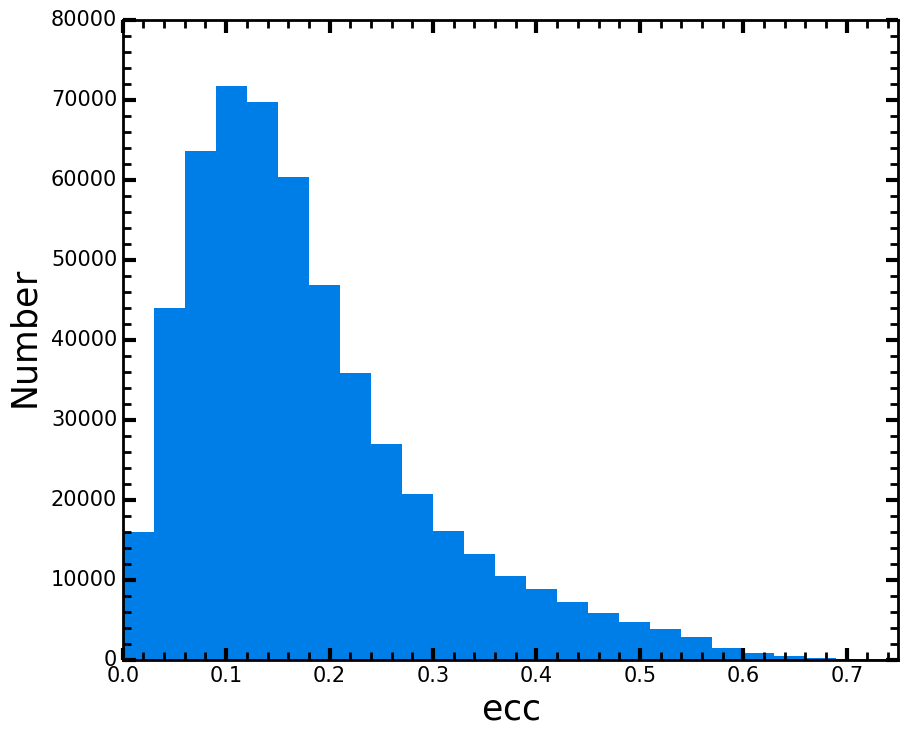}
    \caption{The distribution of recognized orbital eccentricities for 531,521 disk K giants. The horizontal axis represents the eccentricity values, while the vertical axis indicates the corresponding count of giants.}
    \label{fig:ecc}
\end{figure}

In the analysis of MW dynamics, the distribution function (DF) of disk components is modeled using the quasi-isothermal framework as described in several studies \citep{binney2010, binney2011, binney2012, piffl2014}. This model assumes that the orbits are predominantly circular, thus orbits with high eccentricity and orbits that are significantly tilted to the disk plane are physically improbable \citep{binney2023}. To prove that this model is suitable for our study, we evaluate the eccentricity of disk giants in our dataset, as shown in Figure~\ref{fig:ecc}. Most of these orbits exhibit low eccentricities, confirming that the quasi-isothermal model is suitable for our analysis.

The quasi-isothermal DF is mathematically expressed as
\begin{equation}
\begin{split}
    &f(\boldsymbol{J}) = \frac{\widetilde{\Sigma}\Omega}{2\pi^2\kappa^2}\times\frac{\kappa}{\widetilde{\sigma}^2_r}\exp\left(-\frac{\kappa J_r}{\widetilde{\sigma}^2_r}\right)\times\frac{\nu}{\widetilde{\sigma}^2_z}\exp\left(-\frac{\mu J_z}{\widetilde{\sigma}^2_z}\right)\\
    &\times \left\{ \begin{array}{ll}
    1 & \mbox{if}\ J_\phi >0,\\
    \exp\left(\frac{2\Omega J_\phi}{\widetilde{\sigma}^2_r}\right) & \mbox{if}\ J_\phi <0,\\
    \end{array}\right.\\
    &\widetilde{\Sigma}(R_{\rm c})\equiv\Sigma_0\exp(R_{\rm c}/R_{\rm d}),\\
    &\widetilde{\sigma}^2_r(R_{\rm c})\equiv\sigma^2_{r0}\exp(-2R_{\rm c}/R_{\sigma_r})+\sigma^2_{\rm min},\\
    &\widetilde{\sigma}^2_z(R_{\rm c})\equiv\sigma^2_{z0}\exp(-2R_{\rm c}/R_{\sigma_z})+\sigma^2_{\rm min},
	\label{eq:dfdisk}
\end{split}
\end{equation}
where $J_R$, $J_z$, and $J_\phi$ correspond to the radial, vertical, and azimuthal components, respectively, within a cylindrical coordinate system. The azimuthal action $J_\phi$ is calculated as $J_\phi \equiv L_z = x \times v_y - y \times v_x$. $L_z$ is the $z$-component of the angular momentum $L$ in Cartesian coordinates. In addition, $J_R$ and $J_z$ denote radial and vertical actions, respectively. $\Sigma$, $\kappa$, and $\nu$ are azimuthal, radial, and vertical epicyclic frequencies, respectively. $\Sigma_0$ is the overall normalization of the surface density profile, consistent with the $\Sigma_0$ provided in Section~\ref{sec:potential}. $R_c(J_\phi \equiv L_z)$, the radius for circular orbits, is a function of $J_\phi$ and is derived from the galactic potential.

Several parameters are specific to the DF and need to be provided in a dynamic model. $\sigma_{r0}$, $R_{\sigma_r}$, $\sigma_{z0}$, and $R_{\sigma_z}$ control the radial and vertical velocity dispersion profile. $\sigma_{r0}$ and $\sigma_{z0}$ follow an exponential pattern with central values, while $R_{\sigma_r}$ and $R_{\sigma_z}$ represent radial scale length. 
We note that \cite{binney2011} used a fixed relation $R_{\sigma_r} = R_{\sigma_z} = R_{\rm d} / 0.45$. However, subsequent research by \cite{Wang2017} indicated that the values of $R_{\sigma_r}$ and $R_{\sigma_z}$ do not maintain a fixed relationship with $R_{\rm d}$, and are actually smaller than those estimated based on $R_{\rm d}$.

Consequently, it was determined that the disk stars exhibited a greater degree of diffusion in velocity. In this work, we persist in adopting $R_{\sigma_r} = R_{\sigma_z} = R_\sigma$ for the following reasons: 

(1) physically, the two parameters should not deviate significantly; 

(2) our tests with these two parameters as independent free variables demonstrate that the differences in the fit results are small; 

(3) this choice aims to reduce the number of free parameters and thus optimize computational efficiency.

The minimum velocity dispersion, $\sigma_{\rm min}$, is set as $\sigma_{\rm min}=1\% \times (\sigma_{r0}+\sigma_{z0})/2 $ \kms to prevent the DF from reaching unphysical values at extreme $J_\phi$ without radial or vertical motion, thus ensuring the physical plausibility of the model.

\subsection{The contraction of dark matter halo}
\label{sec:The contraction of dark matter halo}
In addressing the modeling of the dark matter (DM) halo, this section focuses on incorporating the effect of baryonic matter contraction, which may be important for an accurate representation of the galaxy's mass distribution within the $\Lambda$CDM cosmological framework.

Traditionally, MW mass estimates have often relied on the Navarro-Frenk-White (NFW) profile \cite{navarro1997}, aimed at delineating the DM distribution. This approach, however, has led to inferred halo concentrations exceeding those predicted by cosmological simulations \citep{hellwing2016, klypin2016, McMillan2017, monai2018, lin2019}, suggesting potential inaccuracies in the halo's depicted structure. The discrepancy primarily lies in the NFW profile's limitations in effectively capturing the dynamic influence of baryonic matter, such as stars and gas, on the surrounding DM, especially within the inner regions of a galaxy \citep{schaller2015, dutton2016, lovell2018}.

To mitigate this limitation, the contracted halo model proposed by \cite{Cautun2020} underscores the significant gravitational impact of baryonic matter on the DM distribution. The gravitational pull from the condensation of baryonic matter into stars and other structures draws in nearby DM, leading to a denser and more concentrated halo profile. This adiabatic contraction mechanism surpasses the simplistic assumptions of the NFW profile, providing a model that may accurately reflect the observed structure of the MW.

This research meticulously fits physically motivated models to the Gaia DR2 Galactic rotation curve \citep{eilers2019} and additional data, leveraging hydrodynamical simulations to illustrate how baryons, specifically within a radius of roughly 20\kpc, induce a notable contraction of the DM distribution. The analytical expression derived by \cite{Cautun2020} elucidates the relationship between the baryonic distribution and the consequential alteration in the DM halo profile. Notably, for the MW, this contraction significantly amplifies the enclosed DM halo mass by approximate factors of 1.3, 2, and 4 at radial distances of 20, 8, and 1 \kpc, respectively, compared to an uncontracted halo, emphasizing the importance of model adjustments to rectify systematic biases in halo mass and concentration estimates due to overlooking baryonic effects.

The corrected mass is provided by the following equation \citep{Cautun2020}
\begin{equation}
    M_{\rm DM}(<r)=M_{\rm DM}^{\rm DMO}(<r)[0.45+0.38(\eta_{\rm bar}+1.16)^{0.53}],
	\label{eq:dmhalo}
\end{equation}
where $\eta_{\rm bar}=M_{\rm bar}(<r)/M_{\rm bar}^{\rm DMO}(<r)$ is defined as the ratio of the baryonic masses enclosed within a specific region, as observed in hydrodynamic simulations, to those in dark matter only (DMO) simulations, and $M_{\rm bar}^{\rm DMO}=f_{\rm bar}M_{\rm tot}^{\rm DMO}$. A value $f_{\rm bar}=15.7 \%$ represents the cosmic mean baryon fraction according to \cite{planck2014} cosmology. In this equation, the subscript `DM' refers to the mass of dark matter, `bar' to the mass of baryonic matter, and `tot' to the combined mass of baryonic and dark matter. The superscript `DMO' denotes the mass in the context of a dark matter numerical simulation, specifically under the NFW profile. The absence of a superscript indicates the mass as described by the new model.

\subsection{Circumgalactic medium}
\label{sec:Circumgalactic medium}

We also consider the Circumgalactic Medium (CGM) component in our study. As shown in \cite{Cautun2020}, the CGM constitutes an extensive gas halo that surrounds a galaxy, extending hundreds of kiloparsecs beyond the visible stars and interstellar matter. 

The radial density profile of the CGM has a power-law dependence on distance ~\citep{Cautun2020}, which can be written as: 
\begin{equation}
    \rho_{\text{CGM}} = 200\rho_{\text{crit}} A_{\text{CGM}} f_{\text{bar}} \left(\frac{r}{R_{200}}\right)^{\beta_{\rm CGM}},
	\label{eq:cgm}
\end{equation}
where $\rho_{\rm crit} = 3 H_0^2/8\pi G$ denotes the Universe's critical density, with Hubble constant $H_0=67.66$\kms {\rm Mpc}$^{-1}$ \citep{planck2020}. $A_{\text{CGM}}=0.190$ is a normalization factor, and $\beta_{\rm CGM}=-1.46$ is the exponent of the power-law. In addition, $R_{200}$ is defined as the radius within which the mean density is 200 times $\rho_{\rm crit}$. The mass of the CGM enclosed within a radius $r$ is calculated as:
\begin{equation}
    M_{\rm CGM}(<r)=\frac{3A_{\rm CGM}}{\beta_{\rm CGM}+3}f_{\rm bar}M_{200}^{\rm tot}\left(\frac{r}{R_{200}}\right)^{\beta_{\rm CGM}+3},
	\label{eq:mcgm}
\end{equation}
where $M_{200}^{\rm tot}$ signifies the total mass enclosed within the halo radius $R_{200}$.

\section{Dynamics of the Milky Way}
\label{sec:results}

\subsection{Initialization parameters}
\label{sec:initialization parameters}
We aim to find the best fitting parameters for the phase space distribution function and the density distribution so that the corresponding velocity distribution fits the observations well. Although we have made efforts to reduce unnecessary parameters, there are still as many as 15 free parameters to be determined.
Given computational limitations, we decided to try using existing models and their parameters as a foundational starting point. This work involved an in-depth analysis of several established models, including those proposed by \cite{piffl2014}, \citet[MWPotential2014, hereafter Bo15]{Bovy2015}, \cite{binney2017}, \citet[hereafter Mc17]{McMillan2017}, \cite{Price-Whelan2017}\footnote{MilkyWayPotential from Gala \citep{Price-Whelan2017}, incorporating the disk model from `Bo15'.}, and \citet[hereafter Ca20]{Cautun2020}, with the initial distribution function parameters sourced from \cite{Wang2017}. Unfortunately, most of these parameters failed to produce satisfactory results.

As a consequence, our approach focuses primarily on adjusting parameters relevant to the thin disk, thick disk, and dark matter halo, as they are the primary mass contributors to the MW. 
Notably, for the $q$ parameter of the dark matter halo, after reviewing various studies \citep{posti2019, hattori2021, das2023, Chen2023} and our analyses, we found that the influence on our results was small. Thus, we set $q=1$ in all our models. Since there is a significant degree of degeneracy between the distribution function parameters $R_{\sigma_r}$ and $R_{\sigma_z}$, we merged these two parameters into a single parameter, $R_\sigma$.

We also took into account a detailed assessment of model reliability, including an evaluation of circular velocity near the position of the Sun and the coherence of inferred gravitational mass with established astrophysical knowledge. The culmination of these efforts was the identification and documentation of optimal model parameters, as detailed in Table~\ref{tab:parameter}, accompanied by the $\chi^2$ values derived from these models, presented in Table~\ref{tab:chi2}. 

Although the iterative DF-based technique is used to construct the self-consistent model, we do not use the DFs to name the model, but rather the initial potentials. There are three reasons. Firstly, some components, such as the two gas disks and the CGM, are fixed during the iteration. Secondly, we have arrived at the best-fitting DF models by varying the parameters in a large parameter space. Thirdly, although the final potential deviates from the initial one, the differences are small for our selected parameters.

In the case of `Wang17', the results were recalculated based on its new model parameters. While some modeling details may introduce minor differences, these do not significantly impact the overall model outcomes. 
Additionally, it should be noted that the definition of $\sigma_{r0}$ and $\sigma_{z0}$ in this paper differs from that in \cite{Wang2017}: their values correspond to the location of the Sun, whereas our values pertain to the Galactic Center.

\subsection{Model iterations}
\label{sec:Model iteration}

The journey of our modeling begins with the parameters of the density and distribution functions, as shown in Table~\ref{tab:parameter}. These initial parameters were subjected to a thorough investigation and selection process, marking the origin of our model's evolutionary path. Following the structured approach described by \cite{agama2018}, our model went through a series of ten iterative refinements. This iterative refinement process was designed to rapidly bring the model into self-consistency, typically achieving substantial convergence within the first three iterations.

\begin{figure}
\includegraphics[width=\columnwidth]{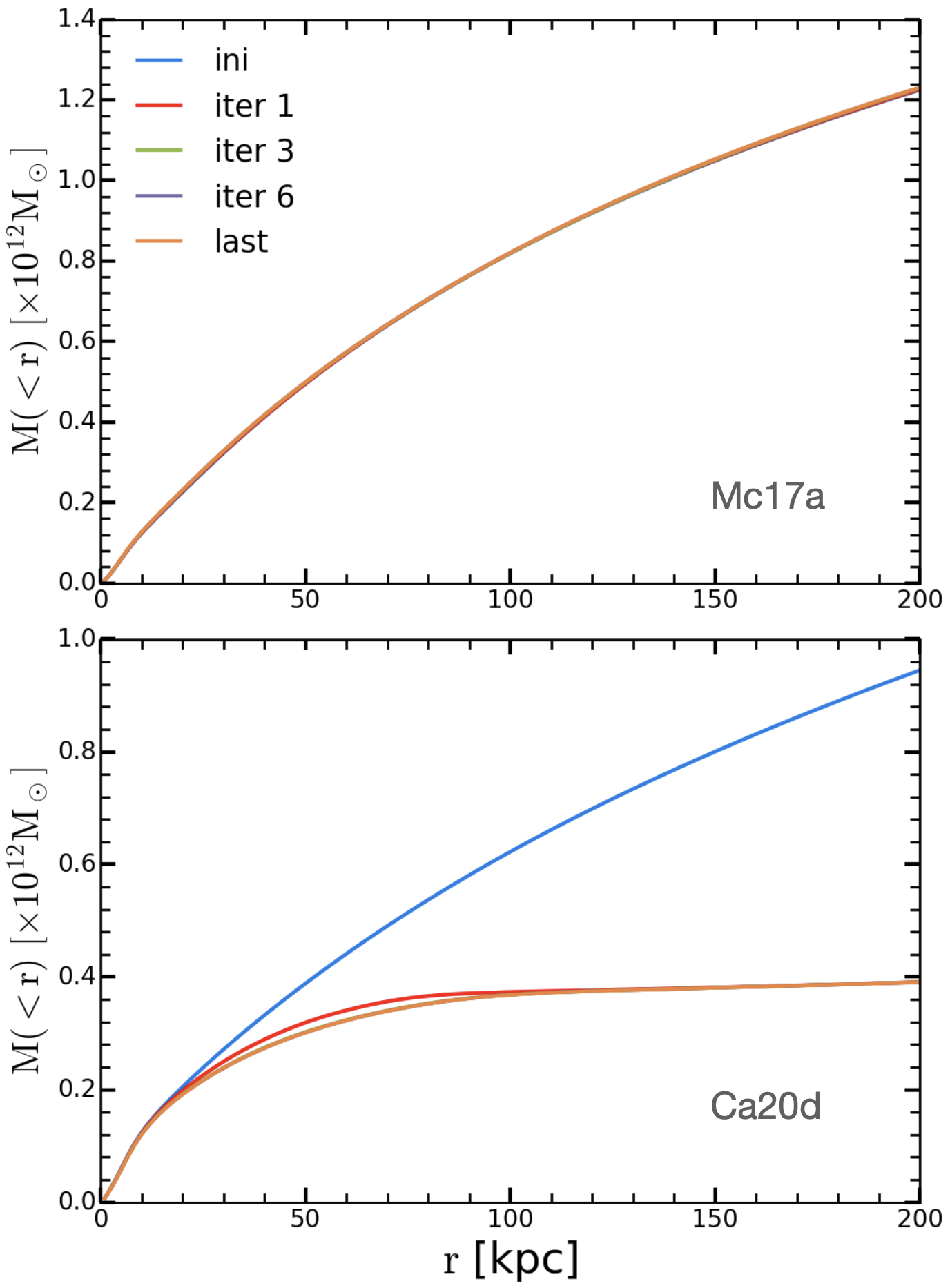}
\caption{The mass distribution at different stages of the iteration process for the model is depicted.  Various colors denote different stages of iteration, showcasing the mass distribution from the initial to the final iteration. `ini' represents the model constructed with the initial parameters, and `last' indicates our final model result. To avoid clutter in the visuals, we only provide the results after the 1st, 3rd, 6th, and 10th iterations. The upper panel corresponds to the `Mc17a' model, where the results of multiple iterations almost overlap. The lower panel shows the results for the `Ca20d' model, where three iterations are sufficient for convergence.}
\label{fig:mass}
\end{figure}

In Figure~\ref{fig:mass}, we present the results for this iterative process, tracing the evolution of the model mass distribution from its initial state to the final iteration. The upper and lower panels show the results for the `Mc17a' and `Ca20d' models, respectively. We note that most models undergo negligible adjustments during the iterations, with minimal changes in the upper panel. In contrast, the `Ca20d' model, as depicted in the lower panel, shows a marked reduction in mass following just one iteration. This exception highlights the efficiency of our iterative process in achieving rapid convergence toward a model that reflects a self-consistent state, with significant modifications occurring after just a single iteration cycle.

\begin{figure}
\includegraphics[width=\columnwidth]{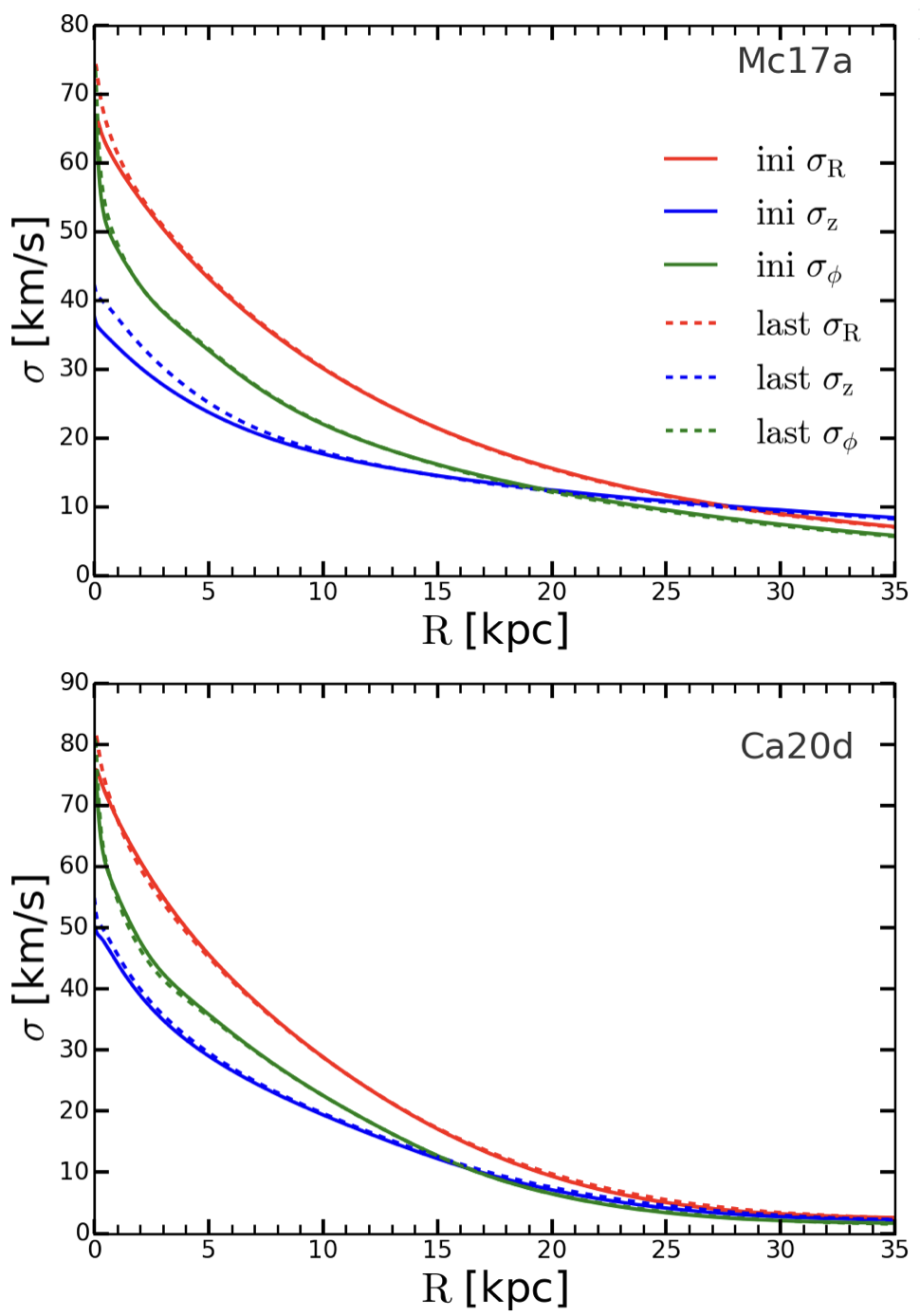}
\caption{The variation of velocity dispersion with radius before and after iteration. The solid lines represent the initial parameter distribution, while the dashed lines indicate the distribution after iteration. Different colors represent dispersions in different directions: red for $\sigma_R$, blue for $\sigma_z$, and green for $\sigma_\phi$. The upper panel corresponds to the `Mc17a' model, whlie the lower panel shows the results for the `Ca20d' model.}
\label{fig:sigma}
\end{figure}

Figure~\ref{fig:sigma} illustrates the variation of velocity dispersion with radius before and after iteration. It can be seen that the iterations generally have a minimal impact on the velocity dispersion distribution. Therefore, the prior values in Table~\ref{tab:parameter} can also serve as approximate references for the self-consistent model.

Finally, we employed Markov Chain Monte Carlo (MCMC) sampling to estimate the uncertainties in the fitted parameters. The results indicate that the model’s uncertainties are small, generally less than 10\% of the parameter values. However, due to the extremely time-intensive nature of MCMC computations, specific MCMC results are not included in this paper.

\begin{splitdeluxetable*}{lccccccccBcccccccccc}
\tabletypesize{\scriptsize}
\tablewidth{0pt} 
    \tablecaption{The density and distribution function parameters of the best-fitting models are presented. The first column lists the names of the initial model parameters, where `Wang17' refers to the best model described in \protect\cite{Wang2017}, `Mc17' denotes the results based on \protect\cite{McMillan2017}, with `Mc17a (fid)' indicating our considered best-fitting model, and `Ca20' represents the results based on \protect\cite{Cautun2020}. Other models with poorer fits are not shown here. Columns (2) to (9) list the parameters for the density models, columns (11) represents the distance from the Sun to the Galactic Center, columns (12) to (17) detail the parameters related to the distribution function, column (18) shows the circular velocity calculated at point $d_\odot$, and column (19) provides the virial mass given by the models, with a Hubble constant $H_0=67.66$\kms {\rm Mpc}$^{-1}$ \citep{planck2020}. The calculated values of the $\chi^2$ parameter for each model are provided in Table~\protect\ref{tab:chi2}. Columns $(2)-(9)$ and $(11)-(17)$ represent the free parameters of our model, while the last two columns show the results calculated based on the previous parameters.}
    \label{tab:parameter}
    \tablehead{
    \colhead{Model} & \colhead{$\Sigma_{\rm 0, thin}$} & \colhead{$R_{\rm d, thin}$} & \colhead{h$_{\rm thin}$} & \colhead{$\Sigma_{\rm 0, thick}$} & \colhead{$R_{\rm d, thick}$} & \colhead{h$_{\rm thick}$} & \colhead{$\rho_{\rm 0, halo}$} & \colhead{$r_{\rm 0, halo}$} & \colhead{Model} & \colhead{d$_\odot$} & \colhead{$\sigma_{\rm r0, thin}$} & \colhead{$\sigma_{\rm z0, thin}$} & \colhead{$R_{\rm \sigma, thin}$} & \colhead{$\sigma_{\rm r0, thick}$} & \colhead{$\sigma_{\rm z0, thick}$} & \colhead{$R_{\rm \sigma, thick}$}& \colhead{$v_{\rm cric}$} & \colhead{$M_{200}$} \\
        \colhead{} & \colhead{($M_\odot$ \kpc$^{-2}$)} & \colhead{(\kpc)}       &\colhead{(\kpc)}      &\colhead{($M_\odot$ \kpc$^{-2}$)} & \colhead{(\kpc)}          &\colhead{(\kpc)}         & \colhead{($M_\odot$ \kpc$^{-3}$)} & \colhead{(\kpc)}  &       & \colhead{(\kpc)} & \colhead{(\kms)}           & \colhead{(\kms)}             & \colhead{(\kpc)}           & \colhead{(\kms)}            &  \colhead{(\kms)} & \colhead{(\kpc)} & \colhead{(\kms)} & \colhead{($\times 10^{12} M_\odot$)}\\    
    }
    \colnumbers
    \startdata
    Wang17&8.17e8& 2.68/2.41$^{*}$ & 0.30 & 2.09e8 & 3.31/4.07$^{*}$ & 0.90 & 8.46e6 & 20.22 & Wang17 & 8.50 & 61.84 & 91.48 & 10.80 & 77.30 & 121.14 & 19.30 & 244.5 & 1.48\\
    Mc17a (fid) & 8.95e8 & 2.48 & 0.30 & 1.87e8 & 3.05 & 0.92 & 8.52e6 & 19.25 & Mc17a (fid) & 8.11 & 63.99 & 30.22 & 12.52 & 82.82 & 132.99 & 18.73 & 231.7 & 1.31 \\
    Mc17b & 9.06e8 & 2.44 & 0.31 & 1.83e8 & 3.06 & 0.90 & 8.58e6 & 19.38 & Mc17b & 8.04 & 70.54 & 30.40 & 11.74 & 75.62 & 132.55 & 20.97 & 231.7 & 1.35\\
    Mc17c & 9.09e8 & 2.49 & 0.30 & 1.87e8 & 3.03 & 0.91 & 8.49e6 & 19.13 & Mc17c & 8.12 & 79.90 & 36.38 & 10.40 & 87.55 & 146.86 & 18.09 & 231.3 & 1.28 \\
    Mc17d &  8.91e8&  2.49&  0.31&  1.87e8&  3.07& 0.91&   8.55e6& 18.80& Mc17d& 8.29&  77.76 & 41.49&  10.90&  73.32&   132.01&  19.18 & 229.6 &1.23 \\
    Mc17e &  8.96e8&  2.48&  0.31&  1.83e8&  3.05& 0.91&  8.51e6&  19.25& Mc17e & 8.20&  83.58 & 50.72&  9.54&  89.95& 135.07&  21.54 & 229.9 & 1.31\\
    Mc17f &  9.07e8&  2.50&  0.31&  1.87e8&  3.04& 0.91&  8.63e6&  18.65& Mc17f & 7.94&  84.66& 39.46&  10.54& 81.36& 134.60&  19.82 & 230.4 & 1.22\\
    Mc17g &  8.85e8&  2.53&  0.31& 1.79e8& 3.02&  0.93&  8.71e6&  18.75& Mc17g & 8.19&  74.68 &57.26&  9.35&  93.28&  143.13& 17.89 & 229.6 & 1.25\\
    Mc17h & 8.86e8&  2.56&  0.30&  1.84e8&  3.03& 0.93&  8.38e6&  18.79& Mc17h & 8.25&  77.23&57.13&  9.18&  90.24&  133.35&  19.56 & 229.2 & 1.20\\
    Mc17i & 8.89e8&  2.57&  0.31&  1.84e8&  3.10& 0.93&  8.35e6&  19.04& Mv17i & 8.08&  80.50&  56.20&  9.97&  90.08&  138.59& 19.14 & 230.4 & 1.25\\
    Mc17j & 9.30e8  & 2.56  & 0.30 & 1.85e8 & 3.00  & 0.93  & 8.40e6  & 18.43 & Mc17i & 7.96 & 82.81  & 50.20  & 10.11  & 91.39  & 141.77  & 18.00 & 230.1 & 1.14\\
    Mc17k$^{**}$ & 9.43e8 & 2.36  & 0.31  & 1.90e8  & 3.11   & 1.12   & 8.09e6 & 12.48 & Mc17k$^{**}$ & 8.40 & 102.94 & 62.66  & 7.64   & 70.79   & 86.20   & 20.96 & 234.3 & 0.41 \\
    Ca20a & 1.07e9 & 2.39  & 0.31  & 1.16e8 & 3.92   & 0.90   & 1.29e7 & 14.00 & Ca20a & 8.21 & 103.33 & 93.33  & 8.85   & 204.23  & 331.38  & 4.14 & 229.1 & 0.88 \\
    Ca20b & 1.08e9 & 2.46  & 0.31  & 1.14e8 & 3.83   & 0.93   & 1.30e7  & 13.74 & Ca20b& 7.90 & 83.19  & 39.88  & 11.01  & 86.48   & 143.28  & 23.04 & 233.2 & 0.85 \\ 
    Ca20c & 1.08e9 & 2.45  & 0.31  & 1.16e8 & 3.88   & 0.92   & 1.27e7 & 13.83 & Ca20c & 8.04 & 86.14  & 35.79  & 9.52   & 255.20  & 367.38  & 5.71 & 230.9 & 0.84\\
    Ca20d$^{**}$ & 7.33e8 & 2.68  & 0.30  & 1.01e8 & 3.86   & 0.90   & 3.97e6 & 23.75 & Ca20d$^{**}$ & 8.19 & 72.35  & 45.80  & 9.82   & 229.87  & 285.41  & 6.05 & 230.3 & 0.39  \\
    Ca20e$^{**}$ & 7.37e8 & 2.63 & 0.31 & 1.03e8 & 3.86 & 0.92 & 4.05e6 & 22.39 & Ca20e$^{**}$& 8.12 & 72.88 & 39.48 & 11.18 & 84.81 & 130.09 & 21.99 & 232.8 & 0.84\\
    \hline
    \multicolumn{9}{l}{$^{*}$ On the left are the parameters used in the density model, and on the right are the parameters used in the distribution function.}\\
    \multicolumn{9}{l}{$^{**}$ In these models, we have employed contracted dark matter halos, see \ref{sec:The contraction of dark matter halo}.}\\
    \hline
    \enddata
\end{splitdeluxetable*}

\begin{table}
	\centering
    \setlength{\tabcolsep}{3pt}
	\caption{The reduced $\chi^2$ values calculated by the model in Table~\ref{tab:parameter}. The estimation of $\chi^2_{\rm in}$ is derived from the first 16 sky regions as defined in Table~\ref{tab:sky area}, $\chi^2_{\rm min}$ originates from sky regions 17-20, $\chi^2_{\rm out}$ is from sky regions 21-24, $\chi^2_{\rm pole}$ is estimated from sky regions 25-16, and $\chi^2_{\rm tot}$ represents the cumulative result from all sky regions. `dof' stands for the degrees of freedom of the model. Compared with Table~\ref{tab:parameter}, two additional rows labeled `BV2023' and `BV2024' have been added to illustrate the differences between the best-fit models from \cite{binney2023} and \cite{binney2024} and the observational data presented in this study.}
	\label{tab:chi2}
	\begin{tabular}{lccccr} 
		\hline
		Model & $\rm \chi^2_{\rm in}/dof$ & $\rm \chi^2_{\rm mid}/dof$ & $\rm \chi^2_{\rm out}/dof$ & $\rm \chi^2_{\rm pole}/dof$ & $\rm \chi^2_{\rm tot}/dof$\\
		\hline
        Wang17&37.21& 83.19& 96.44 & 10.31 & 51.31\\
        BV2023 & 13.58 & 16.34& 35.63 & 60.64 & 21.02\\
        BV2024 & 9.56  & 14.92 & 44.29 & 24.23 & 16.86   \\
        Mc17a (fid) &  5.84 & 5.94 & 27.10 & 21.91 & 10.36\\
        Mc17b & 6.10 & 6.05 & 26.20 & 24.04 & 10.56\\
        Mc17c&5.96 & 7.09 & 26.44 & 25.54 & 10.79\\
        Mc17d& 7.43 & 13.52&  19.50& 20.73& 11.25\\
        Mc17e& 6.95&  10.80& 26.88&  22.39& 11.80\\
        Mc17f &6.90&  11.84& 21.18&  22.89& 11.09\\
        Mc17g & 7.27&  11.16&  34.88& 14.93& 12.71\\
        Mc17h & 7.18&  11.47&  32.05&  16.28& 12.37\\
        Mc17i & 7.35& 13.00& 25.46& 17.90& 11.82\\
        Mc17j &6.40 & 11.67 & 24.44  & 20.67 & 11.08 \\
        Mc17k$^{*}$ & 7.44 & 9.94 & 29.93 & 20.68 & 12.30 \\
        Ca20a   & 12.50 & 27.49 & 61.83 & 8.12 & 22.06 \\
        Ca20b & 7.91 & 12.78    & 18.87 & 36.55 & 12.55 \\
        Ca20c  & 6.95 & 8.88 & 28.59 & 42.62 & 13.32 \\
        Ca20d$^{*}$ & 6.11 & 9.90 & 33.17 & 28.05 & 12.54 \\
        Ca20e$^{*}$ & 6.08 & 7.36 & 19.92 & 26.44 & 9.97\\
		\hline
        \multicolumn{6}{l}{$^{*}$In these models, we have employed contracted dark matter}\\
        \multicolumn{6}{l}{halos, see \ref{sec:The contraction of dark matter halo}.}\\
        \hline
	\end{tabular}
\end{table}

\begin{figure*}
\includegraphics[width=\textwidth]{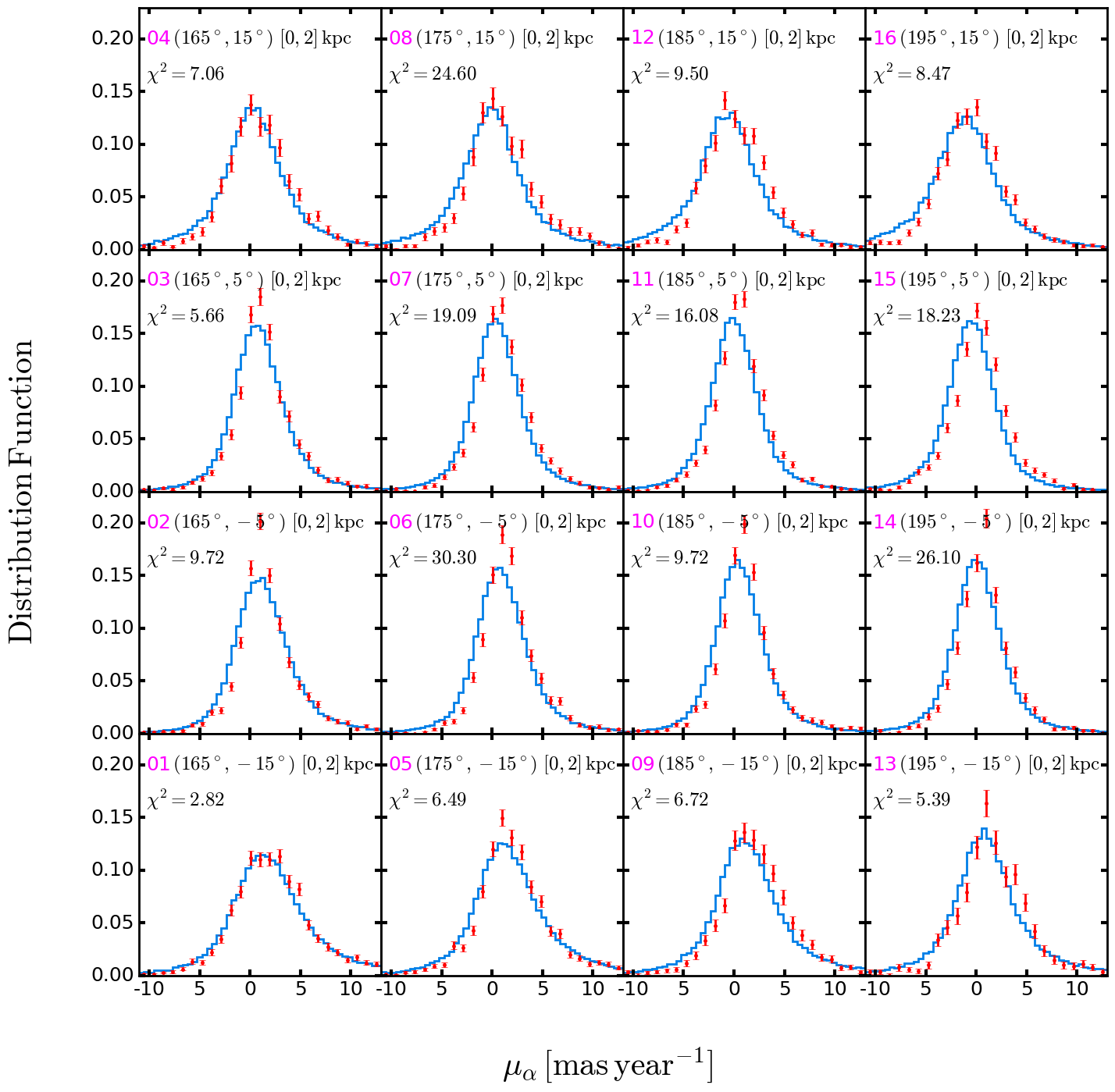}
\caption{The probability distribution function of proper motion along the right ascension ($\mu_\alpha$) for sky regions 01-16, as defined in Table~\ref{tab:sky area}. At the top of each panel, from left to right, are the rank of the sky region, central longitude, and latitude, and the range of distances from the Sun. The blue line represents the predictions by `Mc17a', while the red points with error bars denote the observed data and their errors. Each sky region covers an area of $10^\circ \times 10^\circ$. Notably, due to space constraints, we have displayed only one of the three observational velocity components ($v_{\rm LOS}$, $\mu_\alpha$, $\mu_\delta$) for each set of images. Therefore, please pay careful attention to the labels on the x-axis of each figure. The reduced $\chi^2$ is indicated in each panel.}
\label{fig:ra01-16}
\end{figure*}

\begin{figure}
\includegraphics[width=\columnwidth]{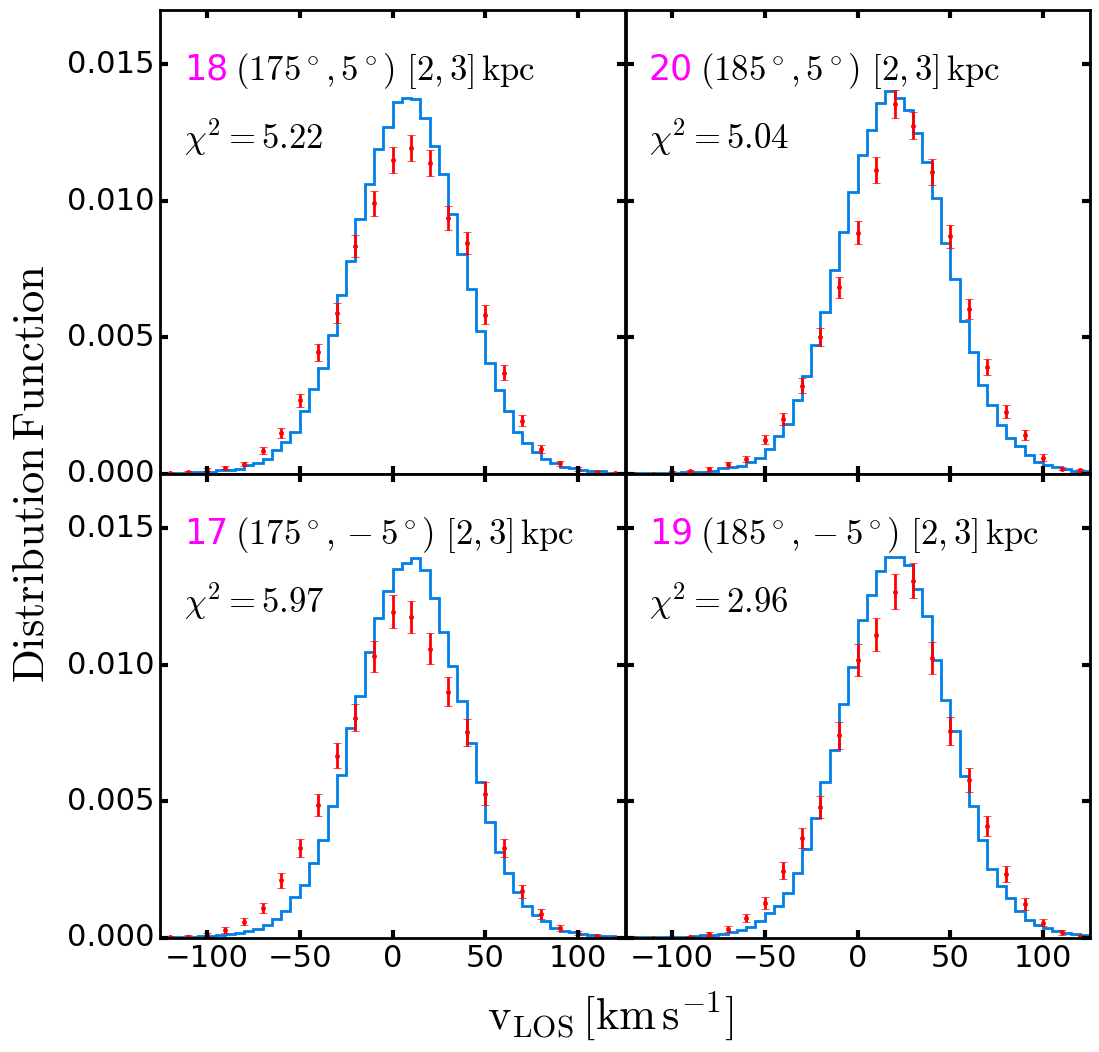}
\caption{The probability distribution function of the line-of-sight velocity ($v_{\rm LOS}$) for sky regions 17-20, as defined in Table~\ref{tab:sky area}. Each sky region covers an area of $10^\circ \times 10^\circ$.}
\label{fig:rv17-20}
\end{figure}

\begin{figure}
\includegraphics[width=\columnwidth]{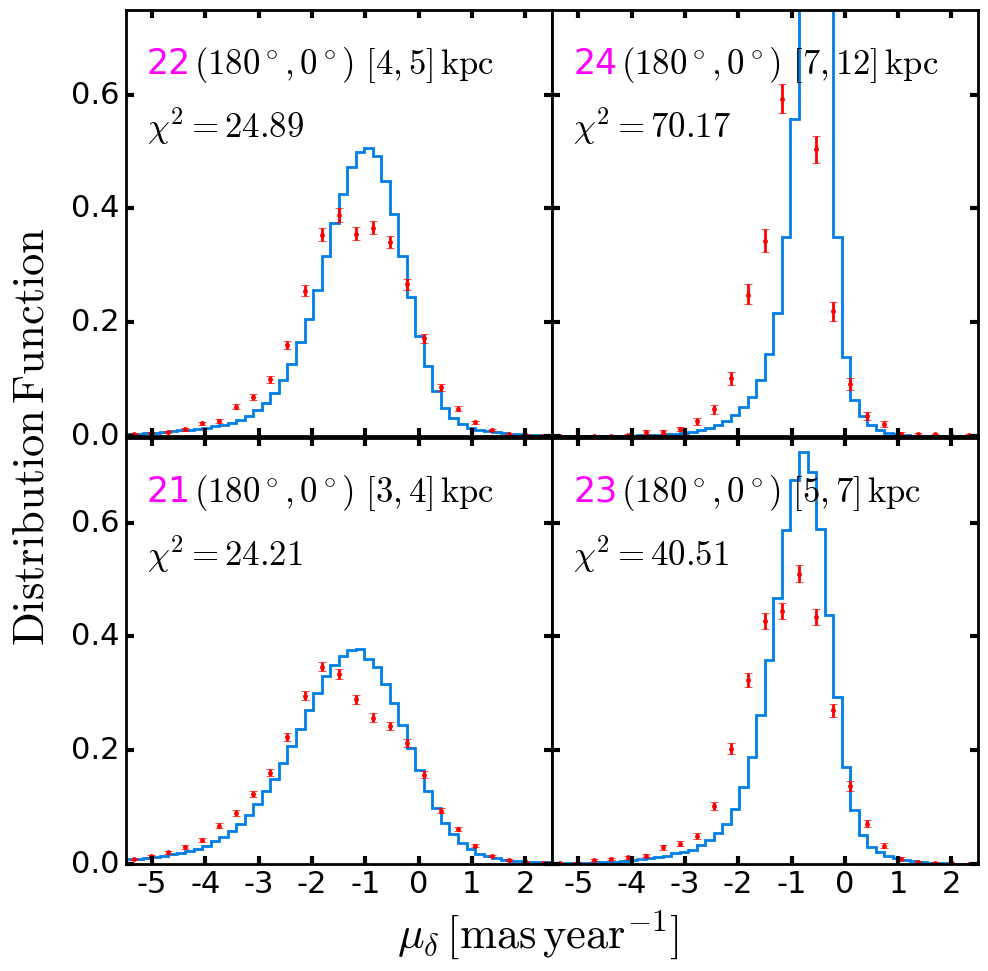}
\caption{The probability distribution function of proper motion in the declination direction ($\mu_\delta$) for sky regions 21-24, as defined in Table~\ref{tab:sky area}. Each sky region covers an area of $20^\circ \times 20^\circ$.}
\label{fig:dec21-24}
\end{figure}

\begin{figure}
\includegraphics[width=\columnwidth]{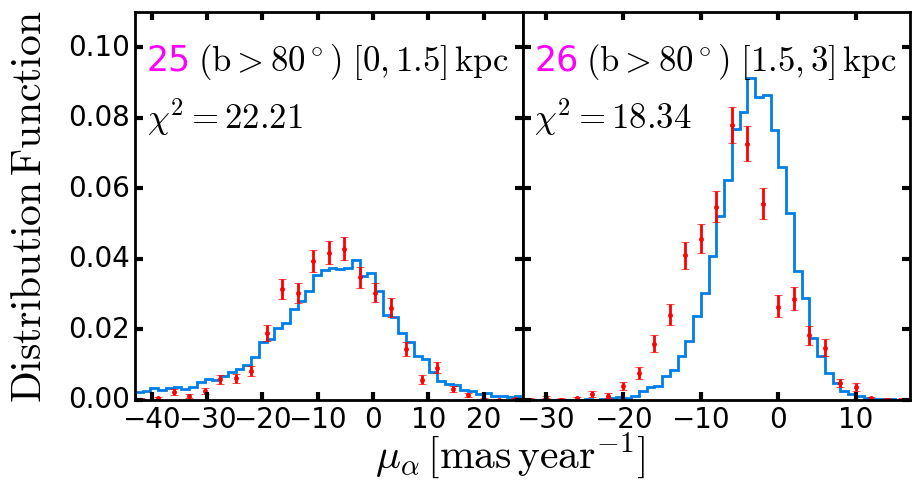}
\caption{The right ascension direction proper motion ($\mu_\alpha$) probability distribution function for sky regions 25 and 26, as defined in Table~\ref{tab:sky area}. Each sky region covers a circle with a diameter of $10^\circ$.}
\label{fig:ra30-32}
\end{figure}

\begin{figure}
\includegraphics[width=\columnwidth]{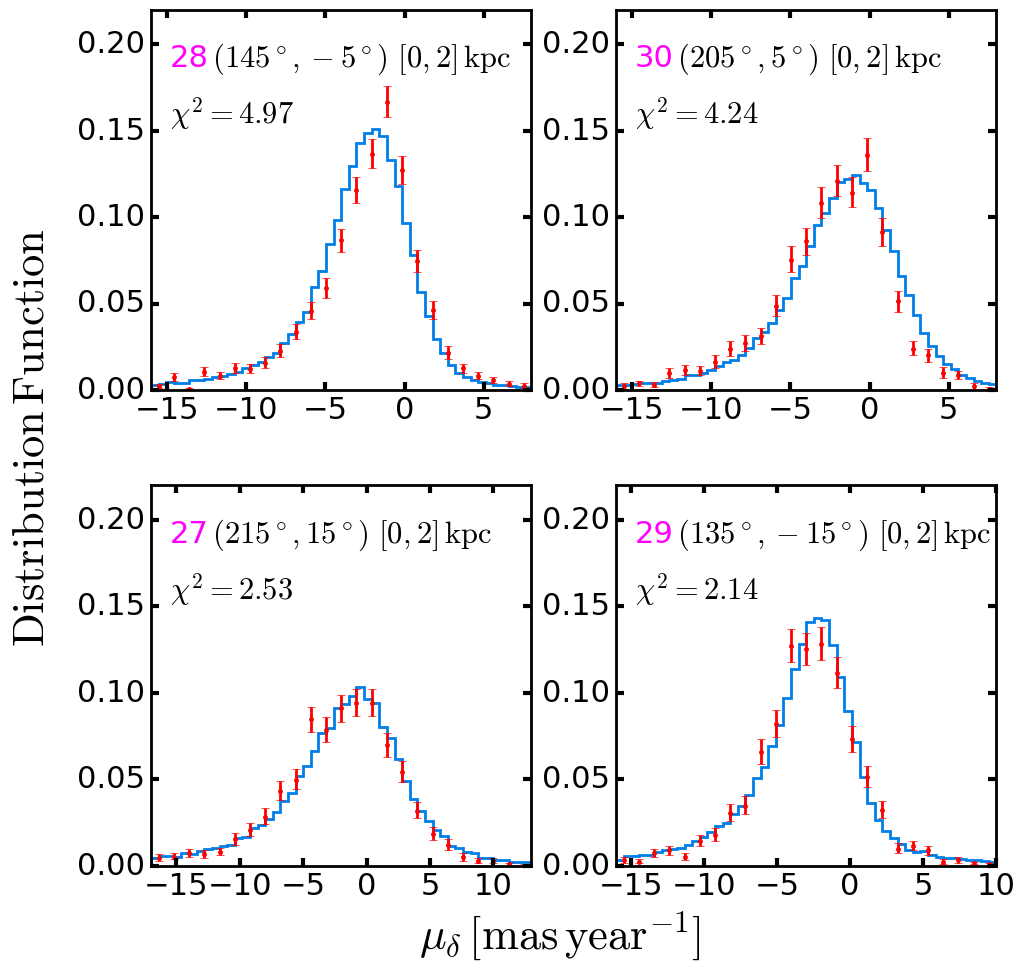}
\caption{The probability distribution function of proper motion in the declination direction ($\mu_\delta$) for sky regions 27-30, as defined in Table~\ref{tab:sky area}. Each sky region covers an area of $10^\circ \times 10^\circ$.}
\label{fig:dec40-43}
\end{figure}

\subsection{Differences between models}
\label{sec:Differences between models}

As shown in Tables~\ref{tab:parameter} and \ref{tab:chi2}, after considering the proper motion distribution, the `Wang17' model is no longer satisfactory. Our new model accommodates the complete velocity distribution, representing an advancement over previous studies. Our main conclusions, however, are not inconsistent with the `Wang17' model. The distribution function parameters consistently suggest a hotter, more extended stellar disk, though the vertical velocity dispersion in the thin disk $\sigma_{\rm z0,\ thin}$ is much smaller. Additionally, our model does not exclude the possibility of smaller disk models, such as `Ca20a', `Ca20c', and `Ca20d'.

For comparison purposes, Table~\ref{tab:chi2} also includes the results from \citeauthor{binney2023} (2023, BV2023) and \citeauthor{binney2024}(2024, BV2024).
Compared with the model in `BV2023', only the bulge in `BV2024' is revised. 
Although \cite{binney2023} used an exponential disk DF model, which differs from the model in our study, their model can still describe our data well, particularly in the solar neighborhood. However, two factors contribute to the higher $\chi^2$ of `BV2023' model. Firstly, in the vertical direction of the Galactic disk far away from the Sun, the predicted kinematics from `BV2023' and `BV2024' deviate from the data significantly.  Secondly, models constructed in their studies are based on giant stars with $\log g<3.5$ and $T_{\rm eff}<5500\ {\rm K}$. We have determined that the kinematics from their giant stars are different from those of the K giant studies in this paper in some regions.

It is also worth mentioning a recent study by \cite{Robin2022}, which introduced an updated `Besançon Galaxy Model' (BGM). 
Instead of relying solely on the action $\boldsymbol{J}$, their DF is a function of two classical integrals plus a third integral from the Stäckel approximation ($E, J_\phi, I_3$). By iteratively updating the gravitational potential, they successfully constructed and refined their model. Additionally, their approach incorporates the age distribution of stars. However, their dataset is limited to more localized regions. Since their method is different from ours, we do not present $\chi^2$ values for their model.

\subsection{Best Model}
\label{sec:best model}

We take model `Mc17a' as our fiducial model because it has the smallest $\chi^2_{\rm in}$ when we fit the models to data in the local region.  
Therefore, `Mc17a' model is used to illustrate our results. Due to the broad range of velocity components ($v_{\rm LOS}$, $\mu_\alpha$, $\mu_\delta$) and sky regions, it is difficult to display all the results. Therefore, as shown in Figs.~\ref{fig:ra01-16}–\ref{fig:dec40-43}, we only present the distribution of a single randomly selected velocity component for each sky region in the main text. The complete velocity distributions from the `Mc17a' model are provided in the appendix~\ref{sec:appendix}.

The three-dimensional velocity distribution indicates that our model agrees well with the observational data from LAMOST DR8 and Gaia EDR3. However, the fit for the $v_{\rm LOS}$ is better than for the proper motion $\mu_\alpha$ and $\mu_\delta$, with the average $\chi^2$ values for the proper motion being 2-5 times higher than those for the $v_{\rm LOS}$. This discrepancy is especially pronounced in regions farther from the Sun. The discrepancy may also indicate that there are still some systematic errors in the $v_{\rm LOS}$ measurements from LAMOST and the proper motion measurements from Gaia.

Figs.~\ref{fig:ra01-16}–\ref{fig:rv17-20} show that the fitting results for the three velocity components in regions 01-20 are fairly accurate. Some observational values are unusually high compared to the model (e.g., region 10 in Figure~\ref{fig:ra01-16}), and systematic offsets are visible in certain areas (e.g., region 15 in Figure~\ref{fig:ra01-16} and Figure~\ref{fig:rv17-20}). This may be due to the influence of anomalous small-scale structures. Queries of relevant catalogs for these regions did not reveal any globular clusters \citep{Sun2023}. However, some bright open clusters and spiral arm structures \citep{Wenger2000, Tarricq2021, Poggio2021, Hao2021, Pang2022, Fu2022} may have caused anomalous peaks in certain sky regions. For example, several nearby regions ($<2$ \kpc) close to the Galactic plane (such as regions 02, 03, and 14) include the Local Spiral arm. More distant regions are also influenced by the Perseus Arm. Additionally, after excluding several K giants associated with open clusters in these regions, this issue was partially alleviated.

In regions farther from the Sun, for example for the anti-Galactic centre direction, as shown in Figure~\ref{fig:dec21-24}. The fitting performance deteriorates, indicating that our model struggles to describe these distant regions accurately. Although the $\chi^2$ values for these regions, as shown in Table~\ref{tab:chi2}, are as poor as those for the North Galactic Pole, some models, such as `Ca20a’, still provide a reasonable fit. None of our attempts could produce an acceptable fit for sky region 24 (Top right panel). 

Figure~\ref{fig:ra30-32} illustrates the distribution of $\mu_\delta$ for giants in sky regions 25 and 26. Given that we have set $R_{\sigma_r}=R_{\sigma_z}$, the fit for results perpendicular to the Galactic plane is acceptable. However, compared to the radial regions, the fit for proper motion data in the vertical direction seems to be better than that for $v_{\rm LOS}$, although the difference is not substantial.

Finally, Figure~\ref{fig:dec40-43} provides a general test of our model by selecting four regions with relatively complete data coverage from different directions, focusing on giants near the Sun. The results indicate that our model performs reliably in regions close to the Sun.

\subsection{Predictions of the models}
\label{sec:Predictions of the models}

In this section, we focus on three selected models for detailed analysis.
In addition to our fiducial model `Mc17a', we also consider the `Ca20d' and `Ca20e' models.   `Ca20d' has the smallest virial mass of among the models shown in Table~\ref{tab:chi2}, and `Ca20e' demonstrates the best overall fit compared to all other models.

\begin{figure}
\includegraphics[width=\columnwidth]{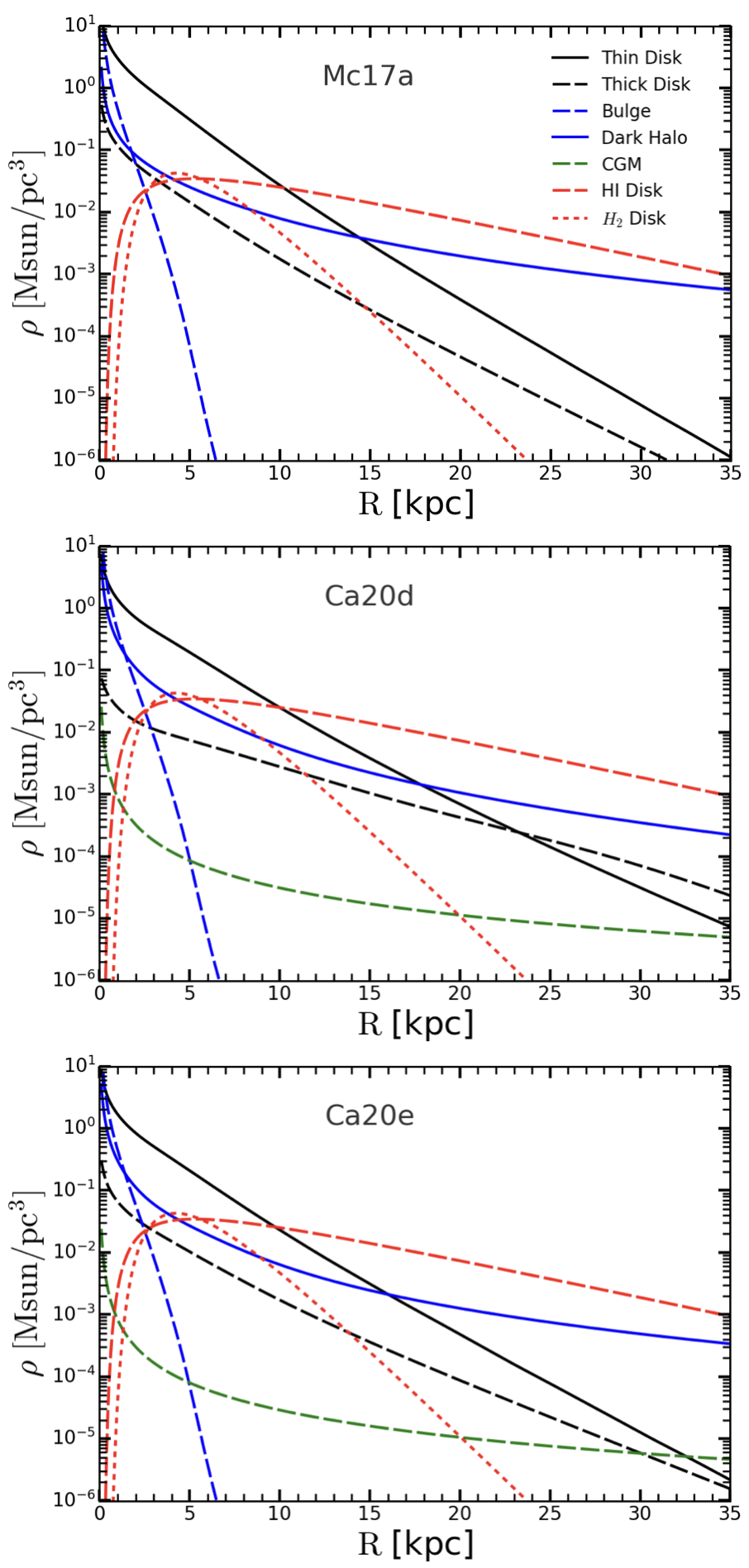}
\caption{Mid-plane density distributions of various components. The black solid and dashed lines denote the thin and thick disks, respectively. The blue solid and dashed lines represent the results for the dark matter halo and bulge, respectively. The red long-dashed line denotes the HI gas disk, the red short-dashed line denotes the $H_2$ gas disk, and the green dashed line denotes the CGM. 
From top to bottom, the results are for `Mc17a', `Ca20d', and `Ca20e' models.}
\label{fig:rou_r}
\end{figure}

Figure~\ref{fig:rou_r} illustrates the mid-plane density distributions of all the components. The black lines represent the two stellar disks; the blue lines denote the two spherical structures, the dark matter halo and the bulge; the red lines correspond to the two gas disks; and the green lines indicate the CGM. Notably, `Mc17a' does not include the CGM structure. It may be seen that the density distributions of the three models do not demonstrate significant differences. Each component's distribution is approximately exponential, and in the vicinity of the Sun, the primary contributions arise from the thin disk and HI gas, with their density contributions being roughly 2-3 times that of the dark halo.

\begin{figure}
\includegraphics[width=\columnwidth]{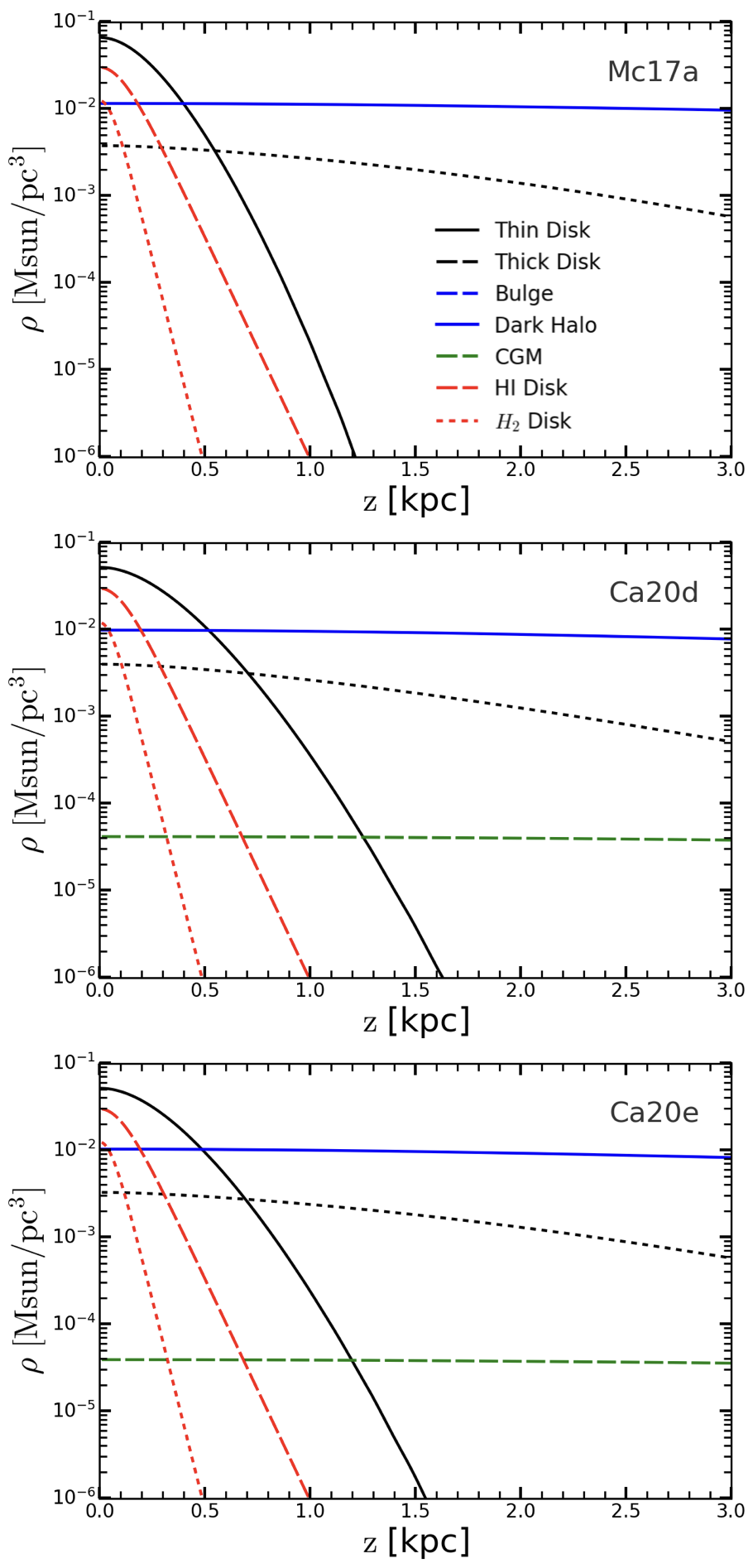}
\caption{Similar to Figure~\ref{fig:rou_r}, but for the density distributions along the height at the Solar Position ($d_{\odot}$).}
\label{fig:rou_z}
\end{figure}

Figure~\ref{fig:rou_z} shows the density distributions as a function of the vertical height at the solar radius. The color scheme is consistent with that of Figure~\ref{fig:rou_r}. The distribution differences between the three models are not substantial. The dark halo and CGM are depicted as nearly horizontal in the figure. Additionally, in the `Mc17a' model, the density of the thin disk decreases more rapidly in the z-direction compared to the other two models.

Table~\ref{tab:rho} presents the densities of the different components of the MW at the position of the Sun. The stellar density (thin disk plus thick disk) in our fiducial model is $0.0696\ {\rm M_{\odot}\ pc^{-3}}$, which is twice the value $0.0317\ {\rm M_{\odot}\ pc^{-3}}$ in `BV2023' obtained using the data from Gaia DR2, while it is close to the results $0.0594\pm0.0008\ {\rm M_{\odot}\ pc^{-3}}$ obtained by using the main-sequence turnoff and subgiant stars from the LAMOST surveys \citep{2018ApJS..237...33X}.  However, the local dark matter density $0.0115\ {\rm M_{\odot}\ pc^{-3}}$ is consistent with the different studies \citep[e.g.][]{McMillan2017, 2020MNRAS.495.4828G, Robin2022, binney2023, 2025JCAP...01..021L}.

\begin{table}
	\centering
    \setlength{\tabcolsep}{3pt}
	\caption{Densities for different models in $M_{\sun}$pc$^{-3}$.}
	\label{tab:rho}
	\begin{tabular}{lccccr} 
		\hline
		Model & Thin Disk & Thick Disk & Dark Halo & HI Disk & $H_2$ Disk\\
		\hline
        Mc17a & 0.0658 & 0.0038 & 0.0115 & 0.0297 & 0.0120\\
        Ca20d & 0.0517 & 0.0040 & 0.0098 & 0.0297 & 0.0120\\
        Ca20e & 0.0515 & 0.0033 & 0.0103 & 0.0299 & 0.0124\\
		\hline
	\end{tabular}
\end{table}

\begin{figure*}
\includegraphics[width=\textwidth]{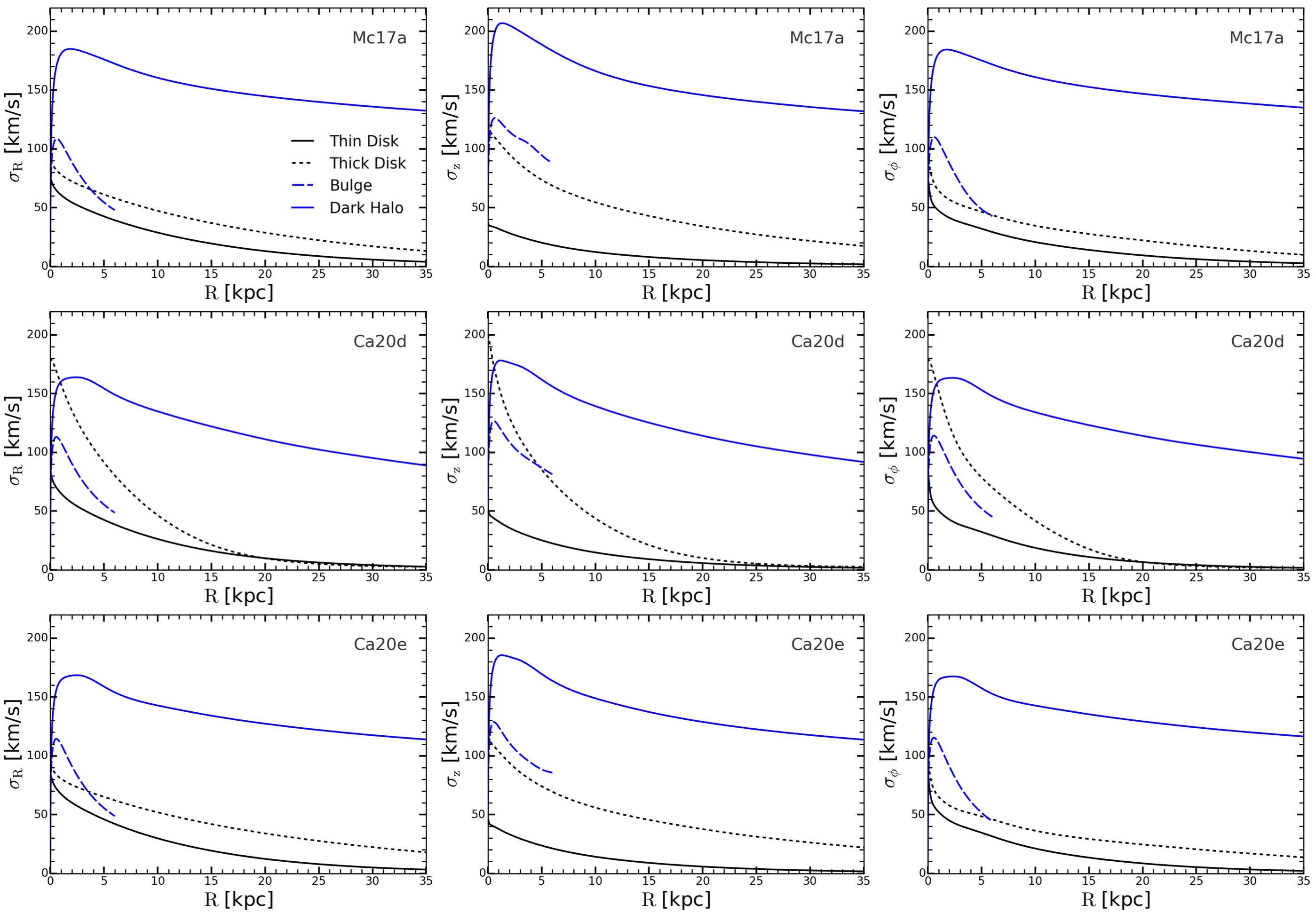}
\caption{From left to right, the radial distributions of the velocity dispersions $\sigma_R$, $\sigma_z$, and $\sigma_{\phi}$ are presented. The color scheme is consistent with that used in Figure~\ref{fig:rou_r}. The black solid line denotes the thin disk, the black dashed line denotes the thick disk, the blue solid line denotes the dark matter halo, the blue dashed line denotes the bulge and reach only to $R=6$\kpc. The upper panel corresponds to the `Mc17a' model, the middle panel shows the results for the `Ca20d' model, whlie the lower panel shows the results for the `Ca20e' model.}
\label{fig:sigma3}
\end{figure*}

In Figure~\ref{fig:sigma3}, we show the velocity dispersions for the three models with only the results of the two stellar disks, the dark matter halo, and the bulge being presented. It is noted that the radial velocity dispersion of the thin disk is smaller than that of the thick disk, except in the `Ca20d' model beyond 20 \kpc. Although the radial velocity dispersion from the thick disk in the `Ca20d' model is highest near the Galactic center, it decreases rapidly with increasing radius. Additionally, the velocity dispersions ($\sigma_R$, $\sigma_z$, $\sigma_\phi$) of the bulge in the central regions (R $<3$ \kpc) are larger than those from the thick disk for the `Mc17a' and `Ca20e'  models, while the opposite is true for the `Ca20d' model.

\begin{figure}
\includegraphics[width=\columnwidth]{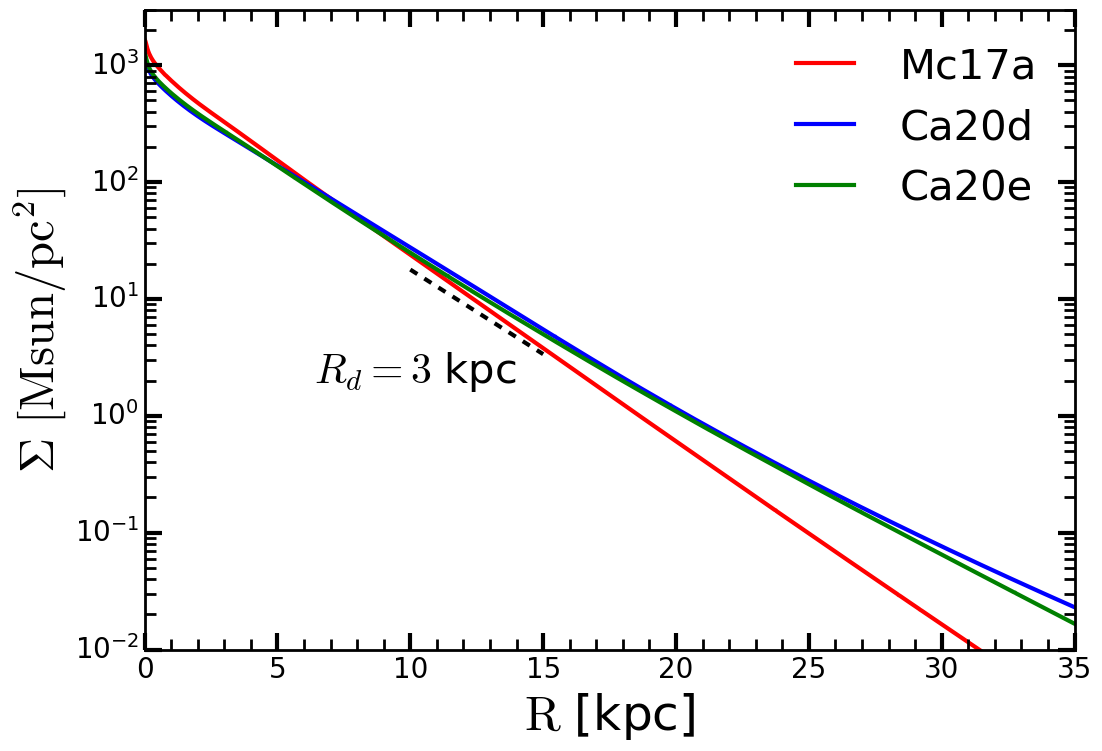}
\caption{Surface density distribution of disk stars for different models. The red, blue and green lines are the results for `Mc17a',`Ca20d' and `Ca20e' models, respectively.}
\label{fig:sigma_all}
\end{figure}

Figure~\ref{fig:sigma_all} illustrates the surface density distributions of disk stars for the three models. In all three, the surface density decreases exponentially with radius. The black dashed line in the figure indicates a scale length of 3 $\rm kpc$, which closely matches the declining trends of the models `Ca20d' and `Ca20e'. This scale length is slightly smaller than the $R_{\rm d}=3.6$\kpc reported by \cite{binney2024}. Conversely, model `Mc17a' exhibits a more rapid decrease in surface density, resulting in a correspondingly larger scale length.

There is considerable variation in the estimated virial mass of the MW across the different models, with the mass ranging from 0.39 to 1.31 $\times 10^{12} M_\odot$. The initial parameters of the `Ca20' model are not significantly different from those of `Mc17'. The primary difference between these models is the inclusion of a CGM component in `Ca20', which is absent in `Mc17'. The fitting results suggest that incorporating the CGM component favors a smaller $r_{\rm 0, halo}$, leading to a reduced mass in the outer regions of the model, and consequently, a smaller estimated virial mass for the MW.

\begin{figure}
\includegraphics[width=\columnwidth]{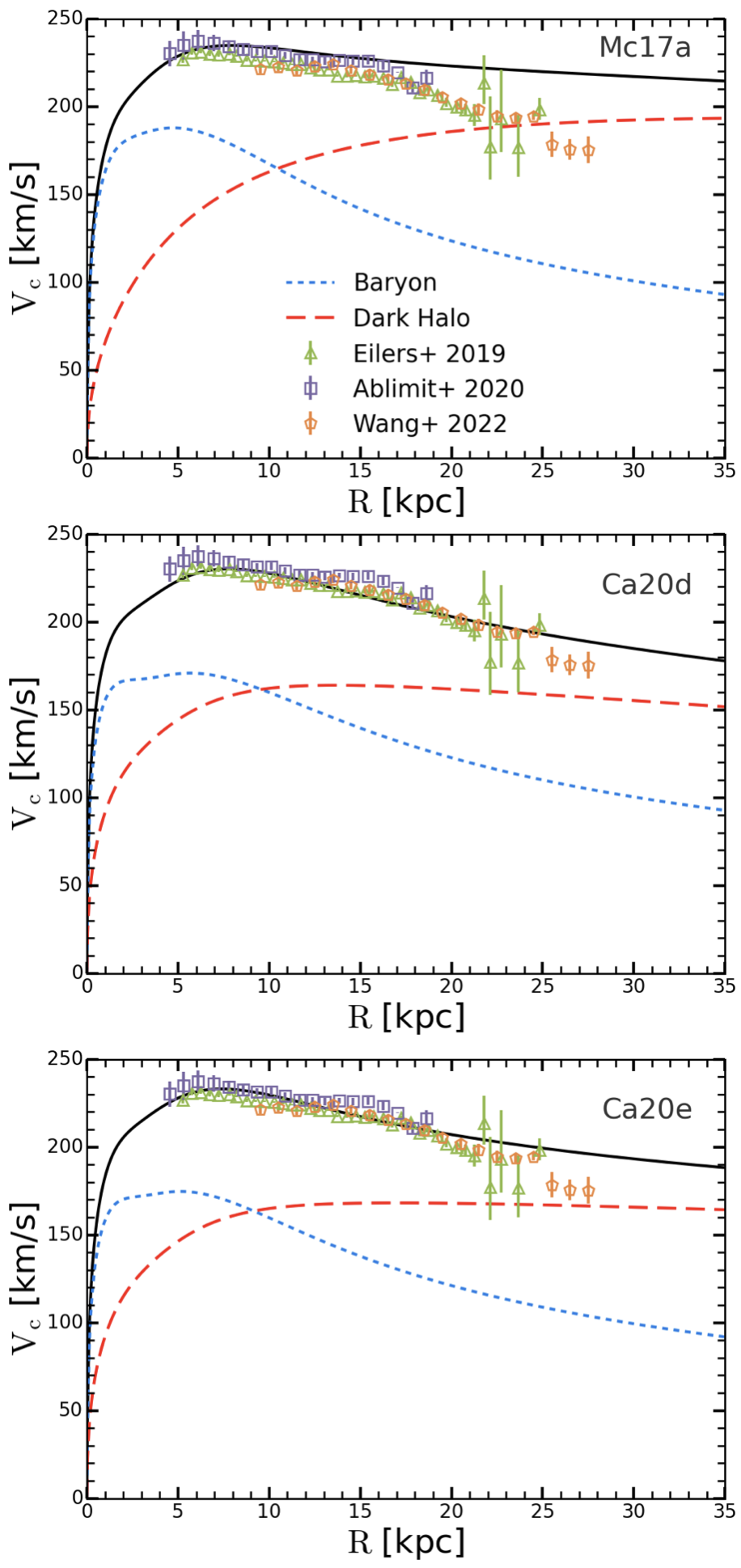}
\caption{Distributions of the circular velocity with radius generated by the  stellar and gas baryonic matter (blue dashed), the dark matter halo (red long-dashed), and the total circular velocity (black solid). The data points represent recent estimates of $V_{\rm c}$ based on three studies using different tracers. From top to bottom, the results are for  `Mc17a', `Ca20d', and `Ca20e' models, respectively.}
\label{fig:vc}
\end{figure}

The black solid lines in Figure~\ref{fig:vc} show the circular velocity curves for the gravitational potentials of the three models, `Mc17a', `Ca20d' and `Ca20e’. All models agree well with the observational data points in the local region, which includes measurements by \cite{eilers2019} using 23,000 red giant stars, \cite{Ablimit2020} using three-dimensional velocity vectors of classical Cepheids, and \cite{Wang2023} calculations of circular velocities based on Gaia DR3.

With similar distributions of baryonic matter, the `Ca20d' and `Ca20e'  models align better with observations of circular velocity in regions far from the Galactic center. However, the masses predicted by the two models `Mc17a' and `Ca20d' in Table~\ref{tab:parameter} differ by more than a factor of three, indicating that the distribution of dark matter is a critical factor influencing the total mass of a galaxy. 
Compared to the classical NFW profile, a contracted dark matter halo \citep{Cautun2020} can significantly reduce the estimated galactic mass while maintaining similar mass in the inner regions.

However, most current studies estimating the mass of the MW rely on extrapolations based on the NFW model \citep{Watkins2010, Vasiliev2019, Wang2020, Sun2023}. Non-model-dependent constraints on the MW’s mass are typically reliable within 60 \kpc \citep{Vasiliev2019, Fritz2020, Sun2023}. From the circular velocity curve of `Ca20d', the MW’s $M_{200}$ could be as low as 0.4 $\times 10^{12} M_\odot$, which is consistent with the result of ~\cite{2022MNRAS.516..731B}.

\begin{figure}
\includegraphics[width=\columnwidth]{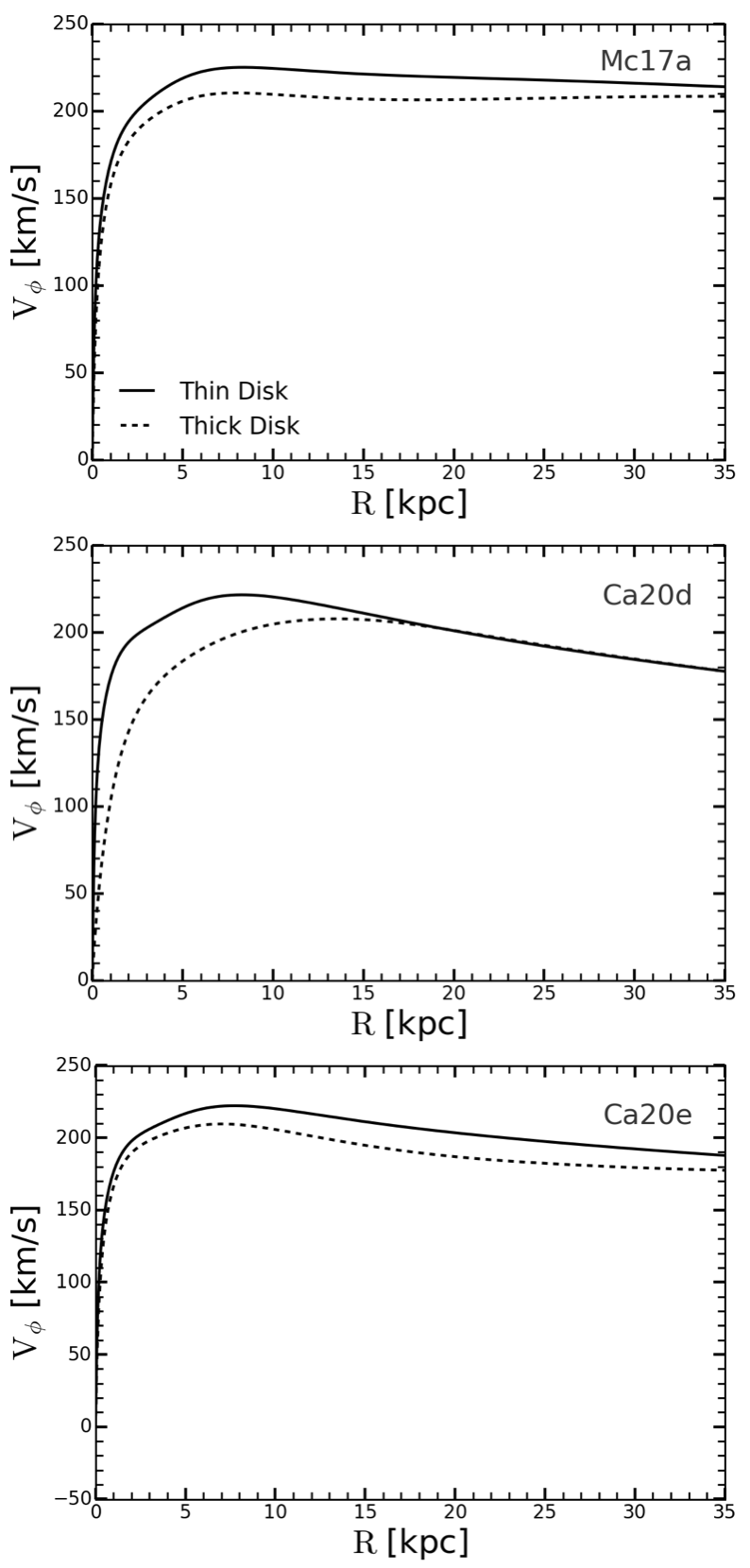}
\caption{Mean steaming velocities of the thick (solid lines) and the thin disk (dashed lines)  in the mid-plane for different Models.}
\label{fig:vphi}
\end{figure}

Figure~\ref{fig:vphi} shows the mean streaming velocities of the thin and thick disks at mid-plane. The thin disk exhibits significantly higher velocities compared to the thick disk. Furthermore, as the distance from the Galactic center increases, the velocity dispersion decreases, and the mean stream velocity slightly surpasses the circular velocity.

\subsection{A hotter outer disk}
\label{sec:A hotter outer disk}

It seems that all the models in Table~\ref{tab:chi2} fail to fit well the outer regions of the MW. The velocities predicted by all models are narrower than those observed at distances far from the Galactic center. This suggests that the MW outer disk, at least within the range of 15–20 \kpc from the Galactic center, is hotter than any of the models we mentioned here. It is known that the outer disk is flared and warped \citep{Mackereth2019, Ablimit2020, Sun2024}. Moreover, the vertical velocity dispersion in the outer disk does not decline with the radius; instead, it slightly increases with the radius beyond 12 \kpc from the Galactic center \citep{2018MNRAS.481.4093S, Mackereth2019}.

Furthermore, we have also attempted to introduce a new disk model. This model, however, does not significantly improve the results, possibly due to the complexity introduced by too many free parameters, making it difficult to achieve a global minimum. Although we did not adopt a three-disk model, we advocate for a new structure with larger velocity dispersions in the outer MW, as existing models cannot provide satisfactory results for the entire velocity distribution. Additionally, the resonant effects of the Galactic bar~\citep{2020MNRAS.495..895B, 2021MNRAS.505.2412C, 2021MNRAS.506.4687M}, along with certain non-equilibrium structures in the MW, such as phase spirals, may influence the observed velocity distribution \citep{Xu2020, Guo2022, McMillan2022, 2024ApJ...960..133G}.

\section{Conclusions}
\label{sec:conclusions}

This study presents a comprehensive dynamical model of the Milky Way (MW), utilizing an extensive dataset of 86,109 K-type giant stars. The combination of observational data from LAMOST DR8 and Gaia EDR3 provides a robust foundation for analyzing the kinematic and spatial distribution of these stars within the Galactic disk. 

Through careful data selection and the application of complex modeling techniques, we construct the best self-consistent model of the MW we can achieve. The main results of this paper can be summarized as follows.

(i) Our study employs an action-based modeling approach, incorporating both the circumgalactic medium (CGM) and a contracted dark matter halo into our models. 
By experimenting with multiple initial parameter sets, we explore a broader parameter space and identify the model parameters that best match our observational data.
  
(ii) Compared to the work of \cite{Wang2017}, we have not only updated the modeling methodology but have also utilized a new and more comprehensive observational dataset. The spatial distribution of the new K giants data is significantly more complete, allowing us to model the full three-dimensional velocity distribution of the K giants, thereby ensuring more precise constraints on the models.

(iii) The majority of our model parameters support a hotter and more extended thick disk (larger $\sigma_{\rm z0,\ thick}$ and $R_{\rm \sigma,\ thin}$), which aligns with the conclusions of \cite{Wang2017}. However, we also highlight that small thick disks ($R_{\sigma,\ \rm thick}$) cannot be entirely ruled out. Additionally, the thin disk is generally found to be thinner and cooler (smaller $\sigma_{\rm z0, thin}$). 

(iv) Our best-fitting model indicates that the velocity distribution of K giants near the Sun is consistent with the predictions of our dynamical model. However, there is still a discrepancy between the proper motion distributions of the model and data in regions far from the Galactic center. This indicates that the structure of the MW is more complex than we thought.
This might require the introduction of new structures to explain these discrepancies, such as additional disks or spiral arms, or an assumption of non-axisymmetry. In particular, the presence of the Galactic bar~\citep{2020MNRAS.495..895B, 2021MNRAS.505.2412C, 2021MNRAS.506.4687M}, and non-equilibrium features (e.g., phase spirals) \citep{Xu2020, Guo2022, McMillan2022, 2024ApJ...960..133G} could play significant roles.

Here, we have attempted to construct a self-consistent model of the MW, which can fit the observed kinematics in the regions near and far from the Sun. Although we have found a best-fitting model from exploring a large parameter space, a discrepancy between the model and the data still exists, which indicates the real MW is more complex than the model we adopt here.
Future work should expand upon these results by enlarging the dataset and refining the model to accommodate additional complexities, such as non-axisymmetric components and external perturbations.

\section*{Acknowledgments}
We thank the referees for the constructive and detailed comments for improving the paper. We are grateful to Chao Liu for the helpful discussion. This work is partly supported by the NSFC International (Regional) Cooperation and Exchange Project (No. 12361141814), by the National Key Research and Development Program of China (No. 2018YFA0404501 to Shude Mao, 2023YFB3002501 to Q. Wang), and by the National Science Foundation of China (Grant No. 11821303 to Shude Mao). We also acknowledge the science research grants from the China Manned Space Project with No. CMS-CSST-2021-A11. Q. Wang is also supported by the Strategic Priority Research Program of Chinese Academy of Sciences, Grant No. XDB0500203, SKA Program of China (Grant number 2020SKA0110401), National Natural Science Foundation of China (Grant numbers 11988101, 12033008), and K.C.Wong Education Foundation. Y. Li and X. Zhang acknowledge financial support from the National Natural Science Foundation of China
(Grant numbers 12473091, 12473001).

%

\vspace{5mm}
\facilities{LAMOST, Gaia}


\software{Astropy \citep{astropy2013, astropy2018}, AGAMA \citep{agama2018}}


\bibliography{arxiv}{}

\begin{thebibliography}{}
\expandafter\ifx\csname natexlab\endcsname\relax\def\natexlab#1{#1}\fi
\providecommand{\url}[1]{\href{#1}{#1}}
\providecommand{\dodoi}[1]{doi:~\href{http://doi.org/#1}{\nolinkurl{#1}}}
\providecommand{\doeprint}[1]{\href{http://ascl.net/#1}{\nolinkurl{http://ascl.net/#1}}}
\providecommand{\doarXiv}[1]{\href{https://arxiv.org/abs/#1}{\nolinkurl{https://arxiv.org/abs/#1}}}

\bibitem[{{Ablimit} {et~al.}(2020){Ablimit}, {Zhao}, {Flynn}, \& {Bird}}]{Ablimit2020}
{Ablimit}, I., {Zhao}, G., {Flynn}, C., \& {Bird}, S.~A. 2020, \apjl, 895, L12, \dodoi{10.3847/2041-8213/ab8d45}

\bibitem[{{Anguiano} {et~al.}(2018){Anguiano}, {Majewski}, {Allende Prieto}, {Meszaros}, {J{\"o}nsson}, {Garc{\'\i}a-Hern{\'a}ndez}, {Beaton}, {Stringfellow}, {Cunha}, \& {Smith}}]{anguiano2018}
{Anguiano}, B., {Majewski}, S.~R., {Allende Prieto}, C., {et~al.} 2018, \aap, 620, A76, \dodoi{10.1051/0004-6361/201833387}

\bibitem[{{Anguiano} {et~al.}(2020){Anguiano}, {Majewski}, {Hayes}, {Allende Prieto}, {Cheng}, {Bidin}, {Beaton}, {Beers}, \& {Minniti}}]{Anguiano2020}
{Anguiano}, B., {Majewski}, S.~R., {Hayes}, C.~R., {et~al.} 2020, \aj, 160, 43, \dodoi{10.3847/1538-3881/ab9813}

\bibitem[{{Astraatmadja} \& {Bailer-Jones}(2016)}]{2016ApJ...832..137A}
{Astraatmadja}, T.~L., \& {Bailer-Jones}, C. A.~L. 2016, \apj, 832, 137, \dodoi{10.3847/0004-637X/832/2/137}

\bibitem[{{Astropy Collaboration} {et~al.}(2013){Astropy Collaboration}, {Robitaille}, {Tollerud}, {Greenfield}, {Droettboom}, {Bray}, {Aldcroft}, {Davis}, {Ginsburg}, {Price-Whelan}, {Kerzendorf}, {Conley}, {Crighton}, {Barbary}, {Muna}, {Ferguson}, {Grollier}, {Parikh}, {Nair}, {Unther}, {Deil}, {Woillez}, {Conseil}, {Kramer}, {Turner}, {Singer}, {Fox}, {Weaver}, {Zabalza}, {Edwards}, {Azalee Bostroem}, {Burke}, {Casey}, {Crawford}, {Dencheva}, {Ely}, {Jenness}, {Labrie}, {Lim}, {Pierfederici}, {Pontzen}, {Ptak}, {Refsdal}, {Servillat}, \& {Streicher}}]{astropy2013}
{Astropy Collaboration}, {Robitaille}, T.~P., {Tollerud}, E.~J., {et~al.} 2013, \aap, 558, A33, \dodoi{10.1051/0004-6361/201322068}

\bibitem[{{Astropy Collaboration} {et~al.}(2018){Astropy Collaboration}, {Price-Whelan}, {Sip{\H{o}}cz}, {G{\"u}nther}, {Lim}, {Crawford}, {Conseil}, {Shupe}, {Craig}, {Dencheva}, {Ginsburg}, {VanderPlas}, {Bradley}, {P{\'e}rez-Su{\'a}rez}, {de Val-Borro}, {Aldcroft}, {Cruz}, {Robitaille}, {Tollerud}, {Ardelean}, {Babej}, {Bach}, {Bachetti}, {Bakanov}, {Bamford}, {Barentsen}, {Barmby}, {Baumbach}, {Berry}, {Biscani}, {Boquien}, {Bostroem}, {Bouma}, {Brammer}, {Bray}, {Breytenbach}, {Buddelmeijer}, {Burke}, {Calderone}, {Cano Rodr{\'\i}guez}, {Cara}, {Cardoso}, {Cheedella}, {Copin}, {Corrales}, {Crichton}, {D'Avella}, {Deil}, {Depagne}, {Dietrich}, {Donath}, {Droettboom}, {Earl}, {Erben}, {Fabbro}, {Ferreira}, {Finethy}, {Fox}, {Garrison}, {Gibbons}, {Goldstein}, {Gommers}, {Greco}, {Greenfield}, {Groener}, {Grollier}, {Hagen}, {Hirst}, {Homeier}, {Horton}, {Hosseinzadeh}, {Hu}, {Hunkeler}, {Ivezi{\'c}}, {Jain}, {Jenness}, {Kanarek}, {Kendrew}, {Kern}, {Kerzendorf}, {Khvalko}, {King}, {Kirkby}, {Kulkarni},
  {Kumar}, {Lee}, {Lenz}, {Littlefair}, {Ma}, {Macleod}, {Mastropietro}, {McCully}, {Montagnac}, {Morris}, {Mueller}, {Mumford}, {Muna}, {Murphy}, {Nelson}, {Nguyen}, {Ninan}, {N{\"o}the}, {Ogaz}, {Oh}, {Parejko}, {Parley}, {Pascual}, {Patil}, {Patil}, {Plunkett}, {Prochaska}, {Rastogi}, {Reddy Janga}, {Sabater}, {Sakurikar}, {Seifert}, {Sherbert}, {Sherwood-Taylor}, {Shih}, {Sick}, {Silbiger}, {Singanamalla}, {Singer}, {Sladen}, {Sooley}, {Sornarajah}, {Streicher}, {Teuben}, {Thomas}, {Tremblay}, {Turner}, {Terr{\'o}n}, {van Kerkwijk}, {de la Vega}, {Watkins}, {Weaver}, {Whitmore}, {Woillez}, {Zabalza}, \& {Astropy Contributors}}]{astropy2018}
{Astropy Collaboration}, {Price-Whelan}, A.~M., {Sip{\H{o}}cz}, B.~M., {et~al.} 2018, \aj, 156, 123, \dodoi{10.3847/1538-3881/aabc4f}

\bibitem[{{Bennett} \& {Bovy}(2019)}]{zsun2019}
{Bennett}, M., \& {Bovy}, J. 2019, \mnras, 482, 1417, \dodoi{10.1093/mnras/sty2813}

\bibitem[{{Binney}(2010)}]{binney2010}
{Binney}, J. 2010, \mnras, 401, 2318, \dodoi{10.1111/j.1365-2966.2009.15845.x}

\bibitem[{{Binney}(2012)}]{binney2012}
---. 2012, \mnras, 426, 1328, \dodoi{10.1111/j.1365-2966.2012.21692.x}

\bibitem[{{Binney}(2018)}]{Binney2018}
{Binney}, J. 2018, in Astrometry and Astrophysics in the Gaia Sky, ed. A.~{Recio-Blanco}, P.~{de Laverny}, A.~G.~A. {Brown}, \& T.~{Prusti}, Vol. 330, 111--118, \dodoi{10.1017/S1743921317007049}

\bibitem[{{Binney}(2020)}]{2020MNRAS.495..895B}
---. 2020, \mnras, 495, 895, \dodoi{10.1093/mnras/staa1103}

\bibitem[{{Binney} \& {Kumar}(1993)}]{1993MNRAS.261..584B}
{Binney}, J., \& {Kumar}, S. 1993, \mnras, 261, 584, \dodoi{10.1093/mnras/261.3.584}

\bibitem[{{Binney} \& {McMillan}(2011)}]{binney2011}
{Binney}, J., \& {McMillan}, P. 2011, \mnras, 413, 1889, \dodoi{10.1111/j.1365-2966.2011.18268.x}

\bibitem[{{Binney} \& {McMillan}(2016)}]{2016MNRAS.456.1982B}
{Binney}, J., \& {McMillan}, P.~J. 2016, \mnras, 456, 1982, \dodoi{10.1093/mnras/stv2734}

\bibitem[{{Binney} \& {Sanders}(2014)}]{Binney2014}
{Binney}, J., \& {Sanders}, J.~L. 2014, in Setting the scene for Gaia and LAMOST, ed. S.~{Feltzing}, G.~{Zhao}, N.~A. {Walton}, \& P.~{Whitelock}, Vol. 298, 117--129, \dodoi{10.1017/S1743921313006297}

\bibitem[{{Binney} \& {Tremaine}(2008)}]{binney2011galactic}
{Binney}, J., \& {Tremaine}, S. 2008, {Galactic Dynamics: Second Edition}

\bibitem[{{Binney} \& {Vasiliev}(2023)}]{binney2023}
{Binney}, J., \& {Vasiliev}, E. 2023, \mnras, 520, 1832, \dodoi{10.1093/mnras/stad094}

\bibitem[{{Binney} \& {Vasiliev}(2024)}]{binney2024}
---. 2024, \mnras, 527, 1915, \dodoi{10.1093/mnras/stad3312}

\bibitem[{{Binney} \& {Wong}(2017)}]{binney2017}
{Binney}, J., \& {Wong}, L.~K. 2017, \mnras, 467, 2446, \dodoi{10.1093/mnras/stx234}

\bibitem[{{Bird} {et~al.}(2022{\natexlab{a}}){Bird}, {Xue}, {Liu}, {Flynn}, {Shen}, {Wang}, {Yang}, {Zhai}, {Zhu}, {Zhao}, \& {Tian}}]{Bird2022}
{Bird}, S.~A., {Xue}, X.-X., {Liu}, C., {et~al.} 2022{\natexlab{a}}, \mnras, 516, 731, \dodoi{10.1093/mnras/stac2036}

\bibitem[{{Bird} {et~al.}(2022{\natexlab{b}}){Bird}, {Xue}, {Liu}, {Flynn}, {Shen}, {Wang}, {Yang}, {Zhai}, {Zhu}, {Zhao}, \& {Tian}}]{2022MNRAS.516..731B}
---. 2022{\natexlab{b}}, \mnras, 516, 731, \dodoi{10.1093/mnras/stac2036}

\bibitem[{{Bovy}(2015)}]{Bovy2015}
{Bovy}, J. 2015, \apjs, 216, 29, \dodoi{10.1088/0067-0049/216/2/29}

\bibitem[{{Carlin} {et~al.}(2015){Carlin}, {Liu}, {Newberg}, {Beers}, {Chen}, {Deng}, {Guhathakurta}, {Hou}, {Hou}, {L{\'e}pine}, {Li}, {Luo}, {Smith}, {Wu}, {Yang}, {Yanny}, {Zhang}, \& {Zheng}}]{carlin2015}
{Carlin}, J.~L., {Liu}, C., {Newberg}, H.~J., {et~al.} 2015, \aj, 150, 4, \dodoi{10.1088/0004-6256/150/1/4}

\bibitem[{{Cautun} {et~al.}(2020){Cautun}, {Ben{\'\i}tez-Llambay}, {Deason}, {Frenk}, {Fattahi}, {G{\'o}mez}, {Grand}, {Oman}, {Navarro}, \& {Simpson}}]{Cautun2020}
{Cautun}, M., {Ben{\'\i}tez-Llambay}, A., {Deason}, A.~J., {et~al.} 2020, \mnras, 494, 4291, \dodoi{10.1093/mnras/staa1017}

\bibitem[{{Chen} {et~al.}(2023){Chen}, {Li}, {Wang}, {Gong}, {Chen}, \& {Long}}]{Chen2023}
{Chen}, A., {Li}, Z., {Wang}, Y., {et~al.} 2023, \mnras, 525, 3075, \dodoi{10.1093/mnras/stad2296}

\bibitem[{{Chiba} \& {Sch{\"o}nrich}(2021)}]{2021MNRAS.505.2412C}
{Chiba}, R., \& {Sch{\"o}nrich}, R. 2021, \mnras, 505, 2412, \dodoi{10.1093/mnras/stab1094}

\bibitem[{{Das} {et~al.}(2023){Das}, {Ianjamasimanana}, {McGaugh}, {Schombert}, \& {Dwarakanath}}]{das2023}
{Das}, M., {Ianjamasimanana}, R., {McGaugh}, S.~S., {Schombert}, J., \& {Dwarakanath}, K.~S. 2023, \apjl, 946, L8, \dodoi{10.3847/2041-8213/acc10e}

\bibitem[{{Deng} {et~al.}(2012){Deng}, {Newberg}, {Liu}, {Carlin}, {Beers}, {Chen}, {Chen}, {Christlieb}, {Grillmair}, {Guhathakurta}, {Han}, {Hou}, {Lee}, {L{\'e}pine}, {Li}, {Liu}, {Pan}, {Sellwood}, {Wang}, {Wang}, {Yang}, {Yanny}, {Zhang}, {Zhang}, {Zheng}, \& {Zhu}}]{Deng2012}
{Deng}, L.-C., {Newberg}, H.~J., {Liu}, C., {et~al.} 2012, Research in Astronomy and Astrophysics, 12, 735, \dodoi{10.1088/1674-4527/12/7/003}

\bibitem[{{Ding} {et~al.}(2021){Ding}, {Xue}, {Yang}, {Zhao}, {Zhang}, \& {Zhu}}]{Ding2021}
{Ding}, P.-J., {Xue}, X.-X., {Yang}, C., {et~al.} 2021, \aj, 162, 112, \dodoi{10.3847/1538-3881/ac0892}

\bibitem[{{Dutton} {et~al.}(2016){Dutton}, {Macci{\`o}}, {Dekel}, {Wang}, {Stinson}, {Obreja}, {Di Cintio}, {Brook}, {Buck}, \& {Kang}}]{dutton2016}
{Dutton}, A.~A., {Macci{\`o}}, A.~V., {Dekel}, A., {et~al.} 2016, \mnras, 461, 2658, \dodoi{10.1093/mnras/stw1537}

\bibitem[{{Eilers} {et~al.}(2019){Eilers}, {Hogg}, {Rix}, \& {Ness}}]{eilers2019}
{Eilers}, A.-C., {Hogg}, D.~W., {Rix}, H.-W., \& {Ness}, M.~K. 2019, \apj, 871, 120, \dodoi{10.3847/1538-4357/aaf648}

\bibitem[{{Fritz} {et~al.}(2020){Fritz}, {Di Cintio}, {Battaglia}, {Brook}, \& {Taibi}}]{Fritz2020}
{Fritz}, T.~K., {Di Cintio}, A., {Battaglia}, G., {Brook}, C., \& {Taibi}, S. 2020, \mnras, 494, 5178, \dodoi{10.1093/mnras/staa1040}

\bibitem[{{Fu} {et~al.}(2022){Fu}, {Bragaglia}, {Liu}, {Zhang}, {Xu}, {Wang}, {Zhang}, {Zhong}, {Chang}, {Li}, {Chen}, {Chen}, {Wang}, {Gjergo}, {Wang}, {Yue}, \& {Zhang}}]{Fu2022}
{Fu}, X., {Bragaglia}, A., {Liu}, C., {et~al.} 2022, \aap, 668, A4, \dodoi{10.1051/0004-6361/202243590}

\bibitem[{{Gaia Collaboration} {et~al.}(2016){Gaia Collaboration}, {Prusti}, {de Bruijne}, {Brown}, {Vallenari}, {Babusiaux}, {Bailer-Jones}, {Bastian}, {Biermann}, {Evans}, {Eyer}, {Jansen}, {Jordi}, {Klioner}, {Lammers}, {Lindegren}, {Luri}, {Mignard}, {Milligan}, {Panem}, {Poinsignon}, {Pourbaix}, {Randich}, {Sarri}, {Sartoretti}, {Siddiqui}, {Soubiran}, {Valette}, {van Leeuwen}, {Walton}, {Aerts}, {Arenou}, {Cropper}, {Drimmel}, {H{\o}g}, {Katz}, {Lattanzi}, {O'Mullane}, {Grebel}, {Holland}, {Huc}, {Passot}, {Bramante}, {Cacciari}, {Casta{\~n}eda}, {Chaoul}, {Cheek}, {De Angeli}, {Fabricius}, {Guerra}, {Hern{\'a}ndez}, {Jean-Antoine-Piccolo}, {Masana}, {Messineo}, {Mowlavi}, {Nienartowicz}, {Ord{\'o}{\~n}ez-Blanco}, {Panuzzo}, {Portell}, {Richards}, {Riello}, {Seabroke}, {Tanga}, {Th{\'e}venin}, {Torra}, {Els}, {Gracia-Abril}, {Comoretto}, {Garcia-Reinaldos}, {Lock}, {Mercier}, {Altmann}, {Andrae}, {Astraatmadja}, {Bellas-Velidis}, {Benson}, {Berthier}, {Blomme}, {Busso}, {Carry}, {Cellino}, {Clementini},
  {Cowell}, {Creevey}, {Cuypers}, {Davidson}, {De Ridder}, {de Torres}, {Delchambre}, {Dell'Oro}, {Ducourant}, {Fr{\'e}mat}, {Garc{\'\i}a-Torres}, {Gosset}, {Halbwachs}, {Hambly}, {Harrison}, {Hauser}, {Hestroffer}, {Hodgkin}, {Huckle}, {Hutton}, {Jasniewicz}, {Jordan}, {Kontizas}, {Korn}, {Lanzafame}, {Manteiga}, {Moitinho}, {Muinonen}, {Osinde}, {Pancino}, {Pauwels}, {Petit}, {Recio-Blanco}, {Robin}, {Sarro}, {Siopis}, {Smith}, {Smith}, {Sozzetti}, {Thuillot}, {van Reeven}, {Viala}, {Abbas}, {Abreu Aramburu}, {Accart}, {Aguado}, {Allan}, {Allasia}, {Altavilla}, {{\'A}lvarez}, {Alves}, {Anderson}, {Andrei}, {Anglada Varela}, {Antiche}, {Antoja}, {Ant{\'o}n}, {Arcay}, {Atzei}, {Ayache}, {Bach}, {Baker}, {Balaguer-N{\'u}{\~n}ez}, {Barache}, {Barata}, {Barbier}, {Barblan}, {Baroni}, {Barrado y Navascu{\'e}s}, {Barros}, {Barstow}, {Becciani}, {Bellazzini}, {Bellei}, {Bello Garc{\'\i}a}, {Belokurov}, {Bendjoya}, {Berihuete}, {Bianchi}, {Bienaym{\'e}}, {Billebaud}, {Blagorodnova}, {Blanco-Cuaresma}, {Boch},
  {Bombrun}, {Borrachero}, {Bouquillon}, {Bourda}, {Bouy}, {Bragaglia}, {Breddels}, {Brouillet}, {Br{\"u}semeister}, {Bucciarelli}, {Budnik}, {Burgess}, {Burgon}, {Burlacu}, {Busonero}, {Buzzi}, {Caffau}, {Cambras}, {Campbell}, {Cancelliere}, {Cantat-Gaudin}, {Carlucci}, {Carrasco}, {Castellani}, {Charlot}, {Charnas}, {Charvet}, {Chassat}, {Chiavassa}, {Clotet}, {Cocozza}, {Collins}, {Collins}, {Costigan}, {Crifo}, {Cross}, {Crosta}, {Crowley}, {Dafonte}, {Damerdji}, {Dapergolas}, {David}, {David}, {De Cat}, {de Felice}, {de Laverny}, {De Luise}, {De March}, {de Martino}, {de Souza}, {Debosscher}, {del Pozo}, {Delbo}, {Delgado}, {Delgado}, {di Marco}, {Di Matteo}, {Diakite}, {Distefano}, {Dolding}, {Dos Anjos}, {Drazinos}, {Dur{\'a}n}, {Dzigan}, {Ecale}, {Edvardsson}, {Enke}, {Erdmann}, {Escolar}, {Espina}, {Evans}, {Eynard Bontemps}, {Fabre}, {Fabrizio}, {Faigler}, {Falc{\~a}o}, {Farr{\`a}s Casas}, {Faye}, {Federici}, {Fedorets}, {Fern{\'a}ndez-Hern{\'a}ndez}, {Fernique}, {Fienga}, {Figueras}, {Filippi},
  {Findeisen}, {Fonti}, {Fouesneau}, {Fraile}, {Fraser}, {Fuchs}, {Furnell}, {Gai}, {Galleti}, {Galluccio}, {Garabato}, {Garc{\'\i}a-Sedano}, {Gar{\'e}}, {Garofalo}, {Garralda}, {Gavras}, {Gerssen}, {Geyer}, {Gilmore}, {Girona}, {Giuffrida}, {Gomes}, {Gonz{\'a}lez-Marcos}, {Gonz{\'a}lez-N{\'u}{\~n}ez}, {Gonz{\'a}lez-Vidal}, {Granvik}, {Guerrier}, {Guillout}, {Guiraud}, {G{\'u}rpide}, {Guti{\'e}rrez-S{\'a}nchez}, {Guy}, {Haigron}, {Hatzidimitriou}, {Haywood}, {Heiter}, {Helmi}, {Hobbs}, {Hofmann}, {Holl}, {Holland}, {Hunt}, {Hypki}, {Icardi}, {Irwin}, {Jevardat de Fombelle}, {Jofr{\'e}}, {Jonker}, {Jorissen}, {Julbe}, {Karampelas}, {Kochoska}, {Kohley}, {Kolenberg}, {Kontizas}, {Koposov}, {Kordopatis}, {Koubsky}, {Kowalczyk}, {Krone-Martins}, {Kudryashova}, {Kull}, {Bachchan}, {Lacoste-Seris}, {Lanza}, {Lavigne}, {Le Poncin-Lafitte}, {Lebreton}, {Lebzelter}, {Leccia}, {Leclerc}, {Lecoeur-Taibi}, {Lemaitre}, {Lenhardt}, {Leroux}, {Liao}, {Licata}, {Lindstr{\o}m}, {Lister}, {Livanou}, {Lobel}, {L{\"o}ffler},
  {L{\'o}pez}, {Lopez-Lozano}, {Lorenz}, {Loureiro}, {MacDonald}, {Magalh{\~a}es Fernandes}, {Managau}, {Mann}, {Mantelet}, {Marchal}, {Marchant}, {Marconi}, {Marie}, {Marinoni}, {Marrese}, {Marschalk{\'o}}, {Marshall}, {Mart{\'\i}n-Fleitas}, {Martino}, {Mary}, {Matijevi{\v{c}}}, {Mazeh}, {McMillan}, {Messina}, {Mestre}, {Michalik}, {Millar}, {Miranda}, {Molina}, {Molinaro}, {Molinaro}, {Moln{\'a}r}, {Moniez}, {Montegriffo}, {Monteiro}, {Mor}, {Mora}, {Morbidelli}, {Morel}, {Morgenthaler}, {Morley}, {Morris}, {Mulone}, {Muraveva}, {Musella}, {Narbonne}, {Nelemans}, {Nicastro}, {Noval}, {Ord{\'e}novic}, {Ordieres-Mer{\'e}}, {Osborne}, {Pagani}, {Pagano}, {Pailler}, {Palacin}, {Palaversa}, {Parsons}, {Paulsen}, {Pecoraro}, {Pedrosa}, {Pentik{\"a}inen}, {Pereira}, {Pichon}, {Piersimoni}, {Pineau}, {Plachy}, {Plum}, {Poujoulet}, {Pr{\v{s}}a}, {Pulone}, {Ragaini}, {Rago}, {Rambaux}, {Ramos-Lerate}, {Ranalli}, {Rauw}, {Read}, {Regibo}, {Renk}, {Reyl{\'e}}, {Ribeiro}, {Rimoldini}, {Ripepi}, {Riva}, {Rixon},
  {Roelens}, {Romero-G{\'o}mez}, {Rowell}, {Royer}, {Rudolph}, {Ruiz-Dern}, {Sadowski}, {Sagrist{\`a} Sell{\'e}s}, {Sahlmann}, {Salgado}, {Salguero}, {Sarasso}, {Savietto}, {Schnorhk}, {Schultheis}, {Sciacca}, {Segol}, {Segovia}, {Segransan}, {Serpell}, {Shih}, {Smareglia}, {Smart}, {Smith}, {Solano}, {Solitro}, {Sordo}, {Soria Nieto}, {Souchay}, {Spagna}, {Spoto}, {Stampa}, {Steele}, {Steidelm{\"u}ller}, {Stephenson}, {Stoev}, {Suess}, {S{\"u}veges}, {Surdej}, {Szabados}, {Szegedi-Elek}, {Tapiador}, {Taris}, {Tauran}, {Taylor}, {Teixeira}, {Terrett}, {Tingley}, {Trager}, {Turon}, {Ulla}, {Utrilla}, {Valentini}, {van Elteren}, {Van Hemelryck}, {van Leeuwen}, {Varadi}, {Vecchiato}, {Veljanoski}, {Via}, {Vicente}, {Vogt}, {Voss}, {Votruba}, {Voutsinas}, {Walmsley}, {Weiler}, {Weingrill}, {Werner}, {Wevers}, {Whitehead}, {Wyrzykowski}, {Yoldas}, {{\v{Z}}erjal}, {Zucker}, {Zurbach}, {Zwitter}, {Alecu}, {Allen}, {Allende Prieto}, {Amorim}, {Anglada-Escud{\'e}}, {Arsenijevic}, {Azaz}, {Balm}, {Beck}, {Bernstein},
  {Bigot}, {Bijaoui}, {Blasco}, {Bonfigli}, {Bono}, {Boudreault}, {Bressan}, {Brown}, {Brunet}, {Bunclark}, {Buonanno}, {Butkevich}, {Carret}, {Carrion}, {Chemin}, {Ch{\'e}reau}, {Corcione}, {Darmigny}, {de Boer}, {de Teodoro}, {de Zeeuw}, {Delle Luche}, {Domingues}, {Dubath}, {Fodor}, {Fr{\'e}zouls}, {Fries}, {Fustes}, {Fyfe}, {Gallardo}, {Gallegos}, {Gardiol}, {Gebran}, {Gomboc}, {G{\'o}mez}, {Grux}, {Gueguen}, {Heyrovsky}, {Hoar}, {Iannicola}, {Isasi Parache}, {Janotto}, {Joliet}, {Jonckheere}, {Keil}, {Kim}, {Klagyivik}, {Klar}, {Knude}, {Kochukhov}, {Kolka}, {Kos}, {Kutka}, {Lainey}, {LeBouquin}, {Liu}, {Loreggia}, {Makarov}, {Marseille}, {Martayan}, {Martinez-Rubi}, {Massart}, {Meynadier}, {Mignot}, {Munari}, {Nguyen}, {Nordlander}, {Ocvirk}, {O'Flaherty}, {Olias Sanz}, {Ortiz}, {Osorio}, {Oszkiewicz}, {Ouzounis}, {Palmer}, {Park}, {Pasquato}, {Peltzer}, {Peralta}, {P{\'e}turaud}, {Pieniluoma}, {Pigozzi}, {Poels}, {Prat}, {Prod'homme}, {Raison}, {Rebordao}, {Risquez}, {Rocca-Volmerange}, {Rosen},
  {Ruiz-Fuertes}, {Russo}, {Sembay}, {Serraller Vizcaino}, {Short}, {Siebert}, {Silva}, {Sinachopoulos}, {Slezak}, {Soffel}, {Sosnowska}, {Strai{\v{z}}ys}, {ter Linden}, {Terrell}, {Theil}, {Tiede}, {Troisi}, {Tsalmantza}, {Tur}, {Vaccari}, {Vachier}, {Valles}, {Van Hamme}, {Veltz}, {Virtanen}, {Wallut}, {Wichmann}, {Wilkinson}, {Ziaeepour}, \& {Zschocke}}]{2016A&A...595A...1G}
{Gaia Collaboration}, {Prusti}, T., {de Bruijne}, J.~H.~J., {et~al.} 2016, \aap, 595, A1, \dodoi{10.1051/0004-6361/201629272}

\bibitem[{{Gaia Collaboration} {et~al.}(2018){Gaia Collaboration}, {Brown}, {Vallenari}, {Prusti}, {de Bruijne}, {Babusiaux}, {Bailer-Jones}, {Biermann}, {Evans}, {Eyer}, {Jansen}, {Jordi}, {Klioner}, {Lammers}, {Lindegren}, {Luri}, {Mignard}, {Panem}, {Pourbaix}, {Randich}, {Sartoretti}, {Siddiqui}, {Soubiran}, {van Leeuwen}, {Walton}, {Arenou}, {Bastian}, {Cropper}, {Drimmel}, {Katz}, {Lattanzi}, {Bakker}, {Cacciari}, {Casta{\~n}eda}, {Chaoul}, {Cheek}, {De Angeli}, {Fabricius}, {Guerra}, {Holl}, {Masana}, {Messineo}, {Mowlavi}, {Nienartowicz}, {Panuzzo}, {Portell}, {Riello}, {Seabroke}, {Tanga}, {Th{\'e}venin}, {Gracia-Abril}, {Comoretto}, {Garcia-Reinaldos}, {Teyssier}, {Altmann}, {Andrae}, {Audard}, {Bellas-Velidis}, {Benson}, {Berthier}, {Blomme}, {Burgess}, {Busso}, {Carry}, {Cellino}, {Clementini}, {Clotet}, {Creevey}, {Davidson}, {De Ridder}, {Delchambre}, {Dell'Oro}, {Ducourant}, {Fern{\'a}ndez-Hern{\'a}ndez}, {Fouesneau}, {Fr{\'e}mat}, {Galluccio}, {Garc{\'\i}a-Torres},
  {Gonz{\'a}lez-N{\'u}{\~n}ez}, {Gonz{\'a}lez-Vidal}, {Gosset}, {Guy}, {Halbwachs}, {Hambly}, {Harrison}, {Hern{\'a}ndez}, {Hestroffer}, {Hodgkin}, {Hutton}, {Jasniewicz}, {Jean-Antoine-Piccolo}, {Jordan}, {Korn}, {Krone-Martins}, {Lanzafame}, {Lebzelter}, {L{\"o}ffler}, {Manteiga}, {Marrese}, {Mart{\'\i}n-Fleitas}, {Moitinho}, {Mora}, {Muinonen}, {Osinde}, {Pancino}, {Pauwels}, {Petit}, {Recio-Blanco}, {Richards}, {Rimoldini}, {Robin}, {Sarro}, {Siopis}, {Smith}, {Sozzetti}, {S{\"u}veges}, {Torra}, {van Reeven}, {Abbas}, {Abreu Aramburu}, {Accart}, {Aerts}, {Altavilla}, {{\'A}lvarez}, {Alvarez}, {Alves}, {Anderson}, {Andrei}, {Anglada Varela}, {Antiche}, {Antoja}, {Arcay}, {Astraatmadja}, {Bach}, {Baker}, {Balaguer-N{\'u}{\~n}ez}, {Balm}, {Barache}, {Barata}, {Barbato}, {Barblan}, {Barklem}, {Barrado}, {Barros}, {Barstow}, {Bartholom{\'e} Mu{\~n}oz}, {Bassilana}, {Becciani}, {Bellazzini}, {Berihuete}, {Bertone}, {Bianchi}, {Bienaym{\'e}}, {Blanco-Cuaresma}, {Boch}, {Boeche}, {Bombrun}, {Borrachero},
  {Bossini}, {Bouquillon}, {Bourda}, {Bragaglia}, {Bramante}, {Breddels}, {Bressan}, {Brouillet}, {Br{\"u}semeister}, {Brugaletta}, {Bucciarelli}, {Burlacu}, {Busonero}, {Butkevich}, {Buzzi}, {Caffau}, {Cancelliere}, {Cannizzaro}, {Cantat-Gaudin}, {Carballo}, {Carlucci}, {Carrasco}, {Casamiquela}, {Castellani}, {Castro-Ginard}, {Charlot}, {Chemin}, {Chiavassa}, {Cocozza}, {Costigan}, {Cowell}, {Crifo}, {Crosta}, {Crowley}, {Cuypers}, {Dafonte}, {Damerdji}, {Dapergolas}, {David}, {David}, {de Laverny}, {De Luise}, {De March}, {de Martino}, {de Souza}, {de Torres}, {Debosscher}, {del Pozo}, {Delbo}, {Delgado}, {Delgado}, {Di Matteo}, {Diakite}, {Diener}, {Distefano}, {Dolding}, {Drazinos}, {Dur{\'a}n}, {Edvardsson}, {Enke}, {Eriksson}, {Esquej}, {Eynard Bontemps}, {Fabre}, {Fabrizio}, {Faigler}, {Falc{\~a}o}, {Farr{\`a}s Casas}, {Federici}, {Fedorets}, {Fernique}, {Figueras}, {Filippi}, {Findeisen}, {Fonti}, {Fraile}, {Fraser}, {Fr{\'e}zouls}, {Gai}, {Galleti}, {Garabato}, {Garc{\'\i}a-Sedano}, {Garofalo},
  {Garralda}, {Gavel}, {Gavras}, {Gerssen}, {Geyer}, {Giacobbe}, {Gilmore}, {Girona}, {Giuffrida}, {Glass}, {Gomes}, {Granvik}, {Gueguen}, {Guerrier}, {Guiraud}, {Guti{\'e}rrez-S{\'a}nchez}, {Haigron}, {Hatzidimitriou}, {Hauser}, {Haywood}, {Heiter}, {Helmi}, {Heu}, {Hilger}, {Hobbs}, {Hofmann}, {Holland}, {Huckle}, {Hypki}, {Icardi}, {Jan{\ss}en}, {Jevardat de Fombelle}, {Jonker}, {Juh{\'a}sz}, {Julbe}, {Karampelas}, {Kewley}, {Klar}, {Kochoska}, {Kohley}, {Kolenberg}, {Kontizas}, {Kontizas}, {Koposov}, {Kordopatis}, {Kostrzewa-Rutkowska}, {Koubsky}, {Lambert}, {Lanza}, {Lasne}, {Lavigne}, {Le Fustec}, {Le Poncin-Lafitte}, {Lebreton}, {Leccia}, {Leclerc}, {Lecoeur-Taibi}, {Lenhardt}, {Leroux}, {Liao}, {Licata}, {Lindstr{\o}m}, {Lister}, {Livanou}, {Lobel}, {L{\'o}pez}, {Managau}, {Mann}, {Mantelet}, {Marchal}, {Marchant}, {Marconi}, {Marinoni}, {Marschalk{\'o}}, {Marshall}, {Martino}, {Marton}, {Mary}, {Massari}, {Matijevi{\v{c}}}, {Mazeh}, {McMillan}, {Messina}, {Michalik}, {Millar}, {Molina}, {Molinaro},
  {Moln{\'a}r}, {Montegriffo}, {Mor}, {Morbidelli}, {Morel}, {Morris}, {Mulone}, {Muraveva}, {Musella}, {Nelemans}, {Nicastro}, {Noval}, {O'Mullane}, {Ord{\'e}novic}, {Ord{\'o}{\~n}ez-Blanco}, {Osborne}, {Pagani}, {Pagano}, {Pailler}, {Palacin}, {Palaversa}, {Panahi}, {Pawlak}, {Piersimoni}, {Pineau}, {Plachy}, {Plum}, {Poggio}, {Poujoulet}, {Pr{\v{s}}a}, {Pulone}, {Racero}, {Ragaini}, {Rambaux}, {Ramos-Lerate}, {Regibo}, {Reyl{\'e}}, {Riclet}, {Ripepi}, {Riva}, {Rivard}, {Rixon}, {Roegiers}, {Roelens}, {Romero-G{\'o}mez}, {Rowell}, {Royer}, {Ruiz-Dern}, {Sadowski}, {Sagrist{\`a} Sell{\'e}s}, {Sahlmann}, {Salgado}, {Salguero}, {Sanna}, {Santana-Ros}, {Sarasso}, {Savietto}, {Schultheis}, {Sciacca}, {Segol}, {Segovia}, {S{\'e}gransan}, {Shih}, {Siltala}, {Silva}, {Smart}, {Smith}, {Solano}, {Solitro}, {Sordo}, {Soria Nieto}, {Souchay}, {Spagna}, {Spoto}, {Stampa}, {Steele}, {Steidelm{\"u}ller}, {Stephenson}, {Stoev}, {Suess}, {Surdej}, {Szabados}, {Szegedi-Elek}, {Tapiador}, {Taris}, {Tauran}, {Taylor},
  {Teixeira}, {Terrett}, {Teyssandier}, {Thuillot}, {Titarenko}, {Torra Clotet}, {Turon}, {Ulla}, {Utrilla}, {Uzzi}, {Vaillant}, {Valentini}, {Valette}, {van Elteren}, {Van Hemelryck}, {van Leeuwen}, {Vaschetto}, {Vecchiato}, {Veljanoski}, {Viala}, {Vicente}, {Vogt}, {von Essen}, {Voss}, {Votruba}, {Voutsinas}, {Walmsley}, {Weiler}, {Wertz}, {Wevers}, {Wyrzykowski}, {Yoldas}, {{\v{Z}}erjal}, {Ziaeepour}, {Zorec}, {Zschocke}, {Zucker}, {Zurbach}, \& {Zwitter}}]{gaia2018a}
{Gaia Collaboration}, {Brown}, A.~G.~A., {Vallenari}, A., {et~al.} 2018, \aap, 616, A1, \dodoi{10.1051/0004-6361/201833051}

\bibitem[{{Gaia Collaboration} {et~al.}(2021){Gaia Collaboration}, {Brown}, {Vallenari}, {Prusti}, {de Bruijne}, {Babusiaux}, {Biermann}, {Creevey}, {Evans}, {Eyer}, {Hutton}, {Jansen}, {Jordi}, {Klioner}, {Lammers}, {Lindegren}, {Luri}, {Mignard}, {Panem}, {Pourbaix}, {Randich}, {Sartoretti}, {Soubiran}, {Walton}, {Arenou}, {Bailer-Jones}, {Bastian}, {Cropper}, {Drimmel}, {Katz}, {Lattanzi}, {van Leeuwen}, {Bakker}, {Cacciari}, {Casta{\~n}eda}, {De Angeli}, {Ducourant}, {Fabricius}, {Fouesneau}, {Fr{\'e}mat}, {Guerra}, {Guerrier}, {Guiraud}, {Jean-Antoine Piccolo}, {Masana}, {Messineo}, {Mowlavi}, {Nicolas}, {Nienartowicz}, {Pailler}, {Panuzzo}, {Riclet}, {Roux}, {Seabroke}, {Sordo}, {Tanga}, {Th{\'e}venin}, {Gracia-Abril}, {Portell}, {Teyssier}, {Altmann}, {Andrae}, {Bellas-Velidis}, {Benson}, {Berthier}, {Blomme}, {Brugaletta}, {Burgess}, {Busso}, {Carry}, {Cellino}, {Cheek}, {Clementini}, {Damerdji}, {Davidson}, {Delchambre}, {Dell'Oro}, {Fern{\'a}ndez-Hern{\'a}ndez}, {Galluccio}, {Garc{\'\i}a-Lario},
  {Garcia-Reinaldos}, {Gonz{\'a}lez-N{\'u}{\~n}ez}, {Gosset}, {Haigron}, {Halbwachs}, {Hambly}, {Harrison}, {Hatzidimitriou}, {Heiter}, {Hern{\'a}ndez}, {Hestroffer}, {Hodgkin}, {Holl}, {Jan{\ss}en}, {Jevardat de Fombelle}, {Jordan}, {Krone-Martins}, {Lanzafame}, {L{\"o}ffler}, {Lorca}, {Manteiga}, {Marchal}, {Marrese}, {Moitinho}, {Mora}, {Muinonen}, {Osborne}, {Pancino}, {Pauwels}, {Petit}, {Recio-Blanco}, {Richards}, {Riello}, {Rimoldini}, {Robin}, {Roegiers}, {Rybizki}, {Sarro}, {Siopis}, {Smith}, {Sozzetti}, {Ulla}, {Utrilla}, {van Leeuwen}, {van Reeven}, {Abbas}, {Abreu Aramburu}, {Accart}, {Aerts}, {Aguado}, {Ajaj}, {Altavilla}, {{\'A}lvarez}, {{\'A}lvarez Cid-Fuentes}, {Alves}, {Anderson}, {Anglada Varela}, {Antoja}, {Audard}, {Baines}, {Baker}, {Balaguer-N{\'u}{\~n}ez}, {Balbinot}, {Balog}, {Barache}, {Barbato}, {Barros}, {Barstow}, {Bartolom{\'e}}, {Bassilana}, {Bauchet}, {Baudesson-Stella}, {Becciani}, {Bellazzini}, {Bernet}, {Bertone}, {Bianchi}, {Blanco-Cuaresma}, {Boch}, {Bombrun}, {Bossini},
  {Bouquillon}, {Bragaglia}, {Bramante}, {Breedt}, {Bressan}, {Brouillet}, {Bucciarelli}, {Burlacu}, {Busonero}, {Butkevich}, {Buzzi}, {Caffau}, {Cancelliere}, {C{\'a}novas}, {Cantat-Gaudin}, {Carballo}, {Carlucci}, {Carnerero}, {Carrasco}, {Casamiquela}, {Castellani}, {Castro-Ginard}, {Castro Sampol}, {Chaoul}, {Charlot}, {Chemin}, {Chiavassa}, {Cioni}, {Comoretto}, {Cooper}, {Cornez}, {Cowell}, {Crifo}, {Crosta}, {Crowley}, {Dafonte}, {Dapergolas}, {David}, {David}, {de Laverny}, {De Luise}, {De March}, {De Ridder}, {de Souza}, {de Teodoro}, {de Torres}, {del Peloso}, {del Pozo}, {Delbo}, {Delgado}, {Delgado}, {Delisle}, {Di Matteo}, {Diakite}, {Diener}, {Distefano}, {Dolding}, {Eappachen}, {Edvardsson}, {Enke}, {Esquej}, {Fabre}, {Fabrizio}, {Faigler}, {Fedorets}, {Fernique}, {Fienga}, {Figueras}, {Fouron}, {Fragkoudi}, {Fraile}, {Franke}, {Gai}, {Garabato}, {Garcia-Gutierrez}, {Garc{\'\i}a-Torres}, {Garofalo}, {Gavras}, {Gerlach}, {Geyer}, {Giacobbe}, {Gilmore}, {Girona}, {Giuffrida}, {Gomel}, {Gomez},
  {Gonzalez-Santamaria}, {Gonz{\'a}lez-Vidal}, {Granvik}, {Guti{\'e}rrez-S{\'a}nchez}, {Guy}, {Hauser}, {Haywood}, {Helmi}, {Hidalgo}, {Hilger}, {H{\l}adczuk}, {Hobbs}, {Holland}, {Huckle}, {Jasniewicz}, {Jonker}, {Juaristi Campillo}, {Julbe}, {Karbevska}, {Kervella}, {Khanna}, {Kochoska}, {Kontizas}, {Kordopatis}, {Korn}, {Kostrzewa-Rutkowska}, {Kruszy{\'n}ska}, {Lambert}, {Lanza}, {Lasne}, {Le Campion}, {Le Fustec}, {Lebreton}, {Lebzelter}, {Leccia}, {Leclerc}, {Lecoeur-Taibi}, {Liao}, {Licata}, {Lindstr{\o}m}, {Lister}, {Livanou}, {Lobel}, {Madrero Pardo}, {Managau}, {Mann}, {Marchant}, {Marconi}, {Marcos Santos}, {Marinoni}, {Marocco}, {Marshall}, {Martin Polo}, {Mart{\'\i}n-Fleitas}, {Masip}, {Massari}, {Mastrobuono-Battisti}, {Mazeh}, {McMillan}, {Messina}, {Michalik}, {Millar}, {Mints}, {Molina}, {Molinaro}, {Moln{\'a}r}, {Montegriffo}, {Mor}, {Morbidelli}, {Morel}, {Morris}, {Mulone}, {Munoz}, {Muraveva}, {Murphy}, {Musella}, {Noval}, {Ord{\'e}novic}, {Orr{\`u}}, {Osinde}, {Pagani}, {Pagano},
  {Palaversa}, {Palicio}, {Panahi}, {Pawlak}, {Pe{\~n}alosa Esteller}, {Penttil{\"a}}, {Piersimoni}, {Pineau}, {Plachy}, {Plum}, {Poggio}, {Poretti}, {Poujoulet}, {Pr{\v{s}}a}, {Pulone}, {Racero}, {Ragaini}, {Rainer}, {Raiteri}, {Rambaux}, {Ramos}, {Ramos-Lerate}, {Re Fiorentin}, {Regibo}, {Reyl{\'e}}, {Ripepi}, {Riva}, {Rixon}, {Robichon}, {Robin}, {Roelens}, {Rohrbasser}, {Romero-G{\'o}mez}, {Rowell}, {Royer}, {Rybicki}, {Sadowski}, {Sagrist{\`a} Sell{\'e}s}, {Sahlmann}, {Salgado}, {Salguero}, {Samaras}, {Sanchez Gimenez}, {Sanna}, {Santove{\~n}a}, {Sarasso}, {Schultheis}, {Sciacca}, {Segol}, {Segovia}, {S{\'e}gransan}, {Semeux}, {Shahaf}, {Siddiqui}, {Siebert}, {Siltala}, {Slezak}, {Smart}, {Solano}, {Solitro}, {Souami}, {Souchay}, {Spagna}, {Spoto}, {Steele}, {Steidelm{\"u}ller}, {Stephenson}, {S{\"u}veges}, {Szabados}, {Szegedi-Elek}, {Taris}, {Tauran}, {Taylor}, {Teixeira}, {Thuillot}, {Tonello}, {Torra}, {Torra}, {Turon}, {Unger}, {Vaillant}, {van Dillen}, {Vanel}, {Vecchiato}, {Viala}, {Vicente},
  {Voutsinas}, {Weiler}, {Wevers}, {Wyrzykowski}, {Yoldas}, {Yvard}, {Zhao}, {Zorec}, {Zucker}, {Zurbach}, \& {Zwitter}}]{GaiaEDR32020}
---. 2021, \aap, 649, A1, \dodoi{10.1051/0004-6361/202039657}

\bibitem[{{Gravity Collaboration} {et~al.}(2019){Gravity Collaboration}, {Abuter}, {Amorim}, {Baub{\"o}ck}, {Berger}, {Bonnet}, {Brandner}, {Cl{\'e}net}, {Coud{\'e} Du Foresto}, {de Zeeuw}, {Dexter}, {Duvert}, {Eckart}, {Eisenhauer}, {F{\"o}rster Schreiber}, {Garcia}, {Gao}, {Gendron}, {Genzel}, {Gerhard}, {Gillessen}, {Habibi}, {Haubois}, {Henning}, {Hippler}, {Horrobin}, {Jim{\'e}nez-Rosales}, {Jocou}, {Kervella}, {Lacour}, {Lapeyr{\`e}re}, {Le Bouquin}, {L{\'e}na}, {Ott}, {Paumard}, {Perraut}, {Perrin}, {Pfuhl}, {Rabien}, {Rodriguez Coira}, {Rousset}, {Scheithauer}, {Sternberg}, {Straub}, {Straubmeier}, {Sturm}, {Tacconi}, {Vincent}, {von Fellenberg}, {Waisberg}, {Widmann}, {Wieprecht}, {Wiezorrek}, {Woillez}, \& {Yazici}}]{GravityCollaboration2019}
{Gravity Collaboration}, {Abuter}, R., {Amorim}, A., {et~al.} 2019, \aap, 625, L10, \dodoi{10.1051/0004-6361/201935656}

\bibitem[{{Guo} {et~al.}(2024){Guo}, {Li}, {Shen}, {Mao}, \& {Liu}}]{2024ApJ...960..133G}
{Guo}, R., {Li}, Z.-Y., {Shen}, J., {Mao}, S., \& {Liu}, C. 2024, \apj, 960, 133, \dodoi{10.3847/1538-4357/ad037b}

\bibitem[{{Guo} {et~al.}(2020){Guo}, {Liu}, {Mao}, {Xue}, {Long}, \& {Zhang}}]{2020MNRAS.495.4828G}
{Guo}, R., {Liu}, C., {Mao}, S., {et~al.} 2020, \mnras, 495, 4828, \dodoi{10.1093/mnras/staa1483}

\bibitem[{{Guo} {et~al.}(2022){Guo}, {Shen}, {Li}, {Liu}, \& {Mao}}]{Guo2022}
{Guo}, R., {Shen}, J., {Li}, Z.-Y., {Liu}, C., \& {Mao}, S. 2022, \apj, 936, 103, \dodoi{10.3847/1538-4357/ac86cd}

\bibitem[{{Hao} {et~al.}(2021){Hao}, {Xu}, {Hou}, {Bian}, {Li}, {Wu}, {He}, {Li}, \& {Liu}}]{Hao2021}
{Hao}, C.~J., {Xu}, Y., {Hou}, L.~G., {et~al.} 2021, \aap, 652, A102, \dodoi{10.1051/0004-6361/202140608}

\bibitem[{{Hattori} {et~al.}(2021){Hattori}, {Valluri}, \& {Vasiliev}}]{hattori2021}
{Hattori}, K., {Valluri}, M., \& {Vasiliev}, E. 2021, \mnras, 508, 5468, \dodoi{10.1093/mnras/stab2898}

\bibitem[{{Hellwing} {et~al.}(2016){Hellwing}, {Frenk}, {Cautun}, {Bose}, {Helly}, {Jenkins}, {Sawala}, \& {Cytowski}}]{hellwing2016}
{Hellwing}, W.~A., {Frenk}, C.~S., {Cautun}, M., {et~al.} 2016, \mnras, 457, 3492, \dodoi{10.1093/mnras/stw214}

\bibitem[{{Hunt} \& {Kawata}(2013)}]{Hunt2013}
{Hunt}, J. A.~S., \& {Kawata}, D. 2013, \mnras, 430, 1928, \dodoi{10.1093/mnras/stt021}

\bibitem[{{Jeans}(1915)}]{Jeans1915}
{Jeans}, J.~H. 1915, \mnras, 76, 70, \dodoi{10.1093/mnras/76.2.70}

\bibitem[{{Kaasalainen} \& {Binney}(1994)}]{1994MNRAS.268.1033K}
{Kaasalainen}, M., \& {Binney}, J. 1994, \mnras, 268, 1033, \dodoi{10.1093/mnras/268.4.1033}

\bibitem[{{Kafle} {et~al.}(2014){Kafle}, {Sharma}, {Lewis}, \& {Bland-Hawthorn}}]{Kafle2014}
{Kafle}, P.~R., {Sharma}, S., {Lewis}, G.~F., \& {Bland-Hawthorn}, J. 2014, \apj, 794, 59, \dodoi{10.1088/0004-637X/794/1/59}

\bibitem[{{Klypin} {et~al.}(2016){Klypin}, {Yepes}, {Gottl{\"o}ber}, {Prada}, \& {He{\ss}}}]{klypin2016}
{Klypin}, A., {Yepes}, G., {Gottl{\"o}ber}, S., {Prada}, F., \& {He{\ss}}, S. 2016, \mnras, 457, 4340, \dodoi{10.1093/mnras/stw248}

\bibitem[{{Lim} {et~al.}(2025){Lim}, {Putney}, {Buckley}, \& {Shih}}]{2025JCAP...01..021L}
{Lim}, S.~H., {Putney}, E., {Buckley}, M.~R., \& {Shih}, D. 2025, \jcap, 2025, 021, \dodoi{10.1088/1475-7516/2025/01/021}

\bibitem[{{Lin} \& {Li}(2019)}]{lin2019}
{Lin}, H.-N., \& {Li}, X. 2019, \mnras, 487, 5679, \dodoi{10.1093/mnras/stz1698}

\bibitem[{{Lindegren} {et~al.}(2021){Lindegren}, {Bastian}, {Biermann}, {Bombrun}, {de Torres}, {Gerlach}, {Geyer}, {Hern{\'a}ndez}, {Hilger}, {Hobbs}, {Klioner}, {Lammers}, {McMillan}, {Ramos-Lerate}, {Steidelm{\"u}ller}, {Stephenson}, \& {van Leeuwen}}]{lingdian2021}
{Lindegren}, L., {Bastian}, U., {Biermann}, M., {et~al.} 2021, \aap, 649, A4, \dodoi{10.1051/0004-6361/202039653}

\bibitem[{{Liu} {et~al.}(2014){Liu}, {Deng}, {Carlin}, {Smith}, {Li}, {Newberg}, {Gao}, {Yang}, {Xue}, {Xu}, {Zhang}, {Xin}, {Wu}, \& {Jin}}]{Liu2014}
{Liu}, C., {Deng}, L.-C., {Carlin}, J.~L., {et~al.} 2014, \apj, 790, 110, \dodoi{10.1088/0004-637X/790/2/110}

\bibitem[{{Long} {et~al.}(2013){Long}, {Mao}, {Shen}, \& {Wang}}]{Long2013}
{Long}, R.~J., {Mao}, S., {Shen}, J., \& {Wang}, Y. 2013, \mnras, 428, 3478, \dodoi{10.1093/mnras/sts285}

\bibitem[{{Lovell} {et~al.}(2018){Lovell}, {Pillepich}, {Genel}, {Nelson}, {Springel}, {Pakmor}, {Marinacci}, {Weinberger}, {Torrey}, {Vogelsberger}, {Alabi}, \& {Hernquist}}]{lovell2018}
{Lovell}, M.~R., {Pillepich}, A., {Genel}, S., {et~al.} 2018, \mnras, 481, 1950, \dodoi{10.1093/mnras/sty2339}

\bibitem[{{Luri} {et~al.}(2018){Luri}, {Brown}, {Sarro}, {Arenou}, {Bailer-Jones}, {Castro-Ginard}, {de Bruijne}, {Prusti}, {Babusiaux}, \& {Delgado}}]{Luri2018}
{Luri}, X., {Brown}, A.~G.~A., {Sarro}, L.~M., {et~al.} 2018, \aap, 616, A9, \dodoi{10.1051/0004-6361/201832964}

\bibitem[{{Mackereth} {et~al.}(2019){Mackereth}, {Bovy}, {Leung}, {Schiavon}, {Trick}, {Chaplin}, {Cunha}, {Feuillet}, {Majewski}, {Martig}, {Miglio}, {Nidever}, {Pinsonneault}, {Aguirre}, {Sobeck}, {Tayar}, \& {Zasowski}}]{Mackereth2019}
{Mackereth}, J.~T., {Bovy}, J., {Leung}, H.~W., {et~al.} 2019, \mnras, 489, 176, \dodoi{10.1093/mnras/stz1521}

\bibitem[{{McGill} \& {Binney}(1990)}]{1990MNRAS.244..634M}
{McGill}, C., \& {Binney}, J. 1990, \mnras, 244, 634

\bibitem[{{McMillan}(2017)}]{McMillan2017}
{McMillan}, P.~J. 2017, \mnras, 465, 76, \dodoi{10.1093/mnras/stw2759}

\bibitem[{{McMillan} \& {Binney}(2008)}]{2008MNRAS.390..429M}
{McMillan}, P.~J., \& {Binney}, J.~J. 2008, \mnras, 390, 429, \dodoi{10.1111/j.1365-2966.2008.13767.x}

\bibitem[{{McMillan} {et~al.}(2022){McMillan}, {Petersson}, {Tepper-Garcia}, {Bland-Hawthorn}, {Antoja}, {Chemin}, {Figueras}, {Khanna}, {Kordopatis}, {Ramos}, {Romero-G{\'o}mez}, \& {Seabroke}}]{McMillan2022}
{McMillan}, P.~J., {Petersson}, J., {Tepper-Garcia}, T., {et~al.} 2022, \mnras, 516, 4988, \dodoi{10.1093/mnras/stac2571}

\bibitem[{{Monari} {et~al.}(2018){Monari}, {Famaey}, {Carrillo}, {Piffl}, {Steinmetz}, {Wyse}, {Anders}, {Chiappini}, \& {Jan{\ss}en}}]{monai2018}
{Monari}, G., {Famaey}, B., {Carrillo}, I., {et~al.} 2018, \aap, 616, L9, \dodoi{10.1051/0004-6361/201833748}

\bibitem[{{Moreno} {et~al.}(2021){Moreno}, {Fern{\'a}ndez-Trincado}, {Schuster}, {P{\'e}rez-Villegas}, \& {Chaves-Velasquez}}]{2021MNRAS.506.4687M}
{Moreno}, E., {Fern{\'a}ndez-Trincado}, J.~G., {Schuster}, W.~J., {P{\'e}rez-Villegas}, A., \& {Chaves-Velasquez}, L. 2021, \mnras, 506, 4687, \dodoi{10.1093/mnras/stab1908}

\bibitem[{{Navarro} {et~al.}(1997){Navarro}, {Frenk}, \& {White}}]{navarro1997}
{Navarro}, J.~F., {Frenk}, C.~S., \& {White}, S. D.~M. 1997, \apj, 490, 493, \dodoi{10.1086/304888}

\bibitem[{{Pang} {et~al.}(2022){Pang}, {Tang}, {Li}, {Yu}, {Wang}, {Li}, {Li}, {Wang}, {Wang}, {Zhang}, {Pasquato}, \& {Kouwenhoven}}]{Pang2022}
{Pang}, X., {Tang}, S.-Y., {Li}, Y., {et~al.} 2022, \apj, 931, 156, \dodoi{10.3847/1538-4357/ac674e}

\bibitem[{{Piffl} {et~al.}(2014){Piffl}, {Binney}, {McMillan}, {Steinmetz}, {Helmi}, {Wyse}, {Bienaym{\'e}}, {Bland-Hawthorn}, {Freeman}, {Gibson}, {Gilmore}, {Grebel}, {Kordopatis}, {Navarro}, {Parker}, {Reid}, {Seabroke}, {Siebert}, {Watson}, \& {Zwitter}}]{piffl2014}
{Piffl}, T., {Binney}, J., {McMillan}, P.~J., {et~al.} 2014, \mnras, 445, 3133, \dodoi{10.1093/mnras/stu1948}

\bibitem[{{Planck Collaboration} {et~al.}(2014){Planck Collaboration}, {Ade}, {Aghanim}, {Alves}, {Armitage-Caplan}, {Arnaud}, {Ashdown}, {Atrio-Barandela}, {Aumont}, {Aussel}, {Baccigalupi}, {Banday}, {Barreiro}, {Barrena}, {Bartelmann}, {Bartlett}, {Bartolo}, {Basak}, {Battaner}, {Battye}, {Benabed}, {Beno{\^\i}t}, {Benoit-L{\'e}vy}, {Bernard}, {Bersanelli}, {Bertincourt}, {Bethermin}, {Bielewicz}, {Bikmaev}, {Blanchard}, {Bobin}, {Bock}, {B{\"o}hringer}, {Bonaldi}, {Bonavera}, {Bond}, {Borrill}, {Bouchet}, {Boulanger}, {Bourdin}, {Bowyer}, {Bridges}, {Brown}, {Bucher}, {Burenin}, {Burigana}, {Butler}, {Calabrese}, {Cappellini}, {Cardoso}, {Carr}, {Carvalho}, {Casale}, {Castex}, {Catalano}, {Challinor}, {Chamballu}, {Chary}, {Chen}, {Chiang}, {Chiang}, {Chon}, {Christensen}, {Churazov}, {Church}, {Clemens}, {Clements}, {Colombi}, {Colombo}, {Combet}, {Comis}, {Couchot}, {Coulais}, {Crill}, {Cruz}, {Curto}, {Cuttaia}, {Da Silva}, {Dahle}, {Danese}, {Davies}, {Davis}, {de Bernardis}, {de Rosa}, {de Zotti},
  {D{\'e}chelette}, {Delabrouille}, {Delouis}, {D{\'e}mocl{\`e}s}, {D{\'e}sert}, {Dick}, {Dickinson}, {Diego}, {Dolag}, {Dole}, {Donzelli}, {Dor{\'e}}, {Douspis}, {Ducout}, {Dunkley}, {Dupac}, {Efstathiou}, {Elsner}, {En{\ss}lin}, {Eriksen}, {Fabre}, {Falgarone}, {Falvella}, {Fantaye}, {Fergusson}, {Filliard}, {Finelli}, {Flores-Cacho}, {Foley}, {Forni}, {Fosalba}, {Frailis}, {Fraisse}, {Franceschi}, {Freschi}, {Fromenteau}, {Frommert}, {Gaier}, {Galeotta}, {Gallegos}, {Galli}, {Gandolfo}, {Ganga}, {Gauthier}, {G{\'e}nova-Santos}, {Ghosh}, {Giard}, {Giardino}, {Gilfanov}, {Girard}, {Giraud-H{\'e}raud}, {Gjerl{\o}w}, {Gonz{\'a}lez-Nuevo}, {G{\'o}rski}, {Gratton}, {Gregorio}, {Gruppuso}, {Gudmundsson}, {Haissinski}, {Hamann}, {Hansen}, {Hansen}, {Hanson}, {Harrison}, {Heavens}, {Helou}, {Hempel}, {Henrot-Versill{\'e}}, {Hern{\'a}ndez-Monteagudo}, {Herranz}, {Hildebrandt}, {Hivon}, {Ho}, {Hobson}, {Holmes}, {Hornstrup}, {Hou}, {Hovest}, {Huey}, {Huffenberger}, {Hurier}, {Ili{\'c}}, {Jaffe}, {Jaffe}, {Jasche},
  {Jewell}, {Jones}, {Juvela}, {Kalberla}, {Kangaslahti}, {Keih{\"a}nen}, {Kerp}, {Keskitalo}, {Khamitov}, {Kiiveri}, {Kim}, {Kisner}, {Kneissl}, {Knoche}, {Knox}, {Kunz}, {Kurki-Suonio}, {Lacasa}, {Lagache}, {L{\"a}hteenm{\"a}ki}, {Lamarre}, {Langer}, {Lasenby}, {Lattanzi}, {Laureijs}, {Lavabre}, {Lawrence}, {Le Jeune}, {Leach}, {Leahy}, {Leonardi}, {Le{\'o}n-Tavares}, {Leroy}, {Lesgourgues}, {Lewis}, {Li}, {Liddle}, {Liguori}, {Lilje}, {Linden-V{\o}rnle}, {Lindholm}, {L{\'o}pez-Caniego}, {Lowe}, {Lubin}, {Mac{\'\i}as-P{\'e}rez}, {MacTavish}, {Maffei}, {Maggio}, {Maino}, {Mandolesi}, {Mangilli}, {Marcos-Caballero}, {Marinucci}, {Maris}, {Marleau}, {Marshall}, {Martin}, {Mart{\'\i}nez-Gonz{\'a}lez}, {Masi}, {Massardi}, {Matarrese}, {Matsumura}, {Matthai}, {Maurin}, {Mazzotta}, {McDonald}, {McEwen}, {McGehee}, {Mei}, {Meinhold}, {Melchiorri}, {Melin}, {Mendes}, {Menegoni}, {Mennella}, {Migliaccio}, {Mikkelsen}, {Millea}, {Miniscalco}, {Mitra}, {Miville-Desch{\^e}nes}, {Molinari}, {Moneti}, {Montier},
  {Morgante}, {Morisset}, {Mortlock}, {Moss}, {Munshi}, {Murphy}, {Naselsky}, {Nati}, {Natoli}, {Negrello}, {Nesvadba}, {Netterfield}, {N{\o}rgaard-Nielsen}, {North}, {Noviello}, {Novikov}, {Novikov}, {O'Dwyer}, {Orieux}, {Osborne}, {O'Sullivan}, {Oxborrow}, {Paci}, {Pagano}, {Pajot}, {Paladini}, {Pandolfi}, {Paoletti}, {Partridge}, {Pasian}, {Patanchon}, {Paykari}, {Pearson}, {Pearson}, {Peel}, {Peiris}, {Perdereau}, {Perotto}, {Perrotta}, {Pettorino}, {Piacentini}, {Piat}, {Pierpaoli}, {Pietrobon}, {Plaszczynski}, {Platania}, {Pogosyan}, {Pointecouteau}, {Polenta}, {Ponthieu}, {Popa}, {Poutanen}, {Pratt}, {Pr{\'e}zeau}, {Prunet}, {Puget}, {Pullen}, {Rachen}, {Racine}, {Rahlin}, {R{\"a}th}, {Reach}, {Rebolo}, {Reinecke}, {Remazeilles}, {Renault}, {Renzi}, {Riazuelo}, {Ricciardi}, {Riller}, {Ringeval}, {Ristorcelli}, {Robbers}, {Rocha}, {Roman}, {Rosset}, {Rossetti}, {Roudier}, {Rowan-Robinson}, {Rubi{\~n}o-Mart{\'\i}n}, {Ruiz-Granados}, {Rusholme}, {Salerno}, {Sandri}, {Sanselme}, {Santos}, {Savelainen},
  {Savini}, {Schaefer}, {Schiavon}, {Scott}, {Seiffert}, {Serra}, {Shellard}, {Smith}, {Smoot}, {Souradeep}, {Spencer}, {Starck}, {Stolyarov}, {Stompor}, {Sudiwala}, {Sunyaev}, {Sureau}, {Sutter}, {Sutton}, {Suur-Uski}, {Sygnet}, {Tauber}, {Tavagnacco}, {Taylor}, {Terenzi}, {Texier}, {Toffolatti}, {Tomasi}, {Torre}, {Tristram}, {Tucci}, {Tuovinen}, {T{\"u}rler}, {Tuttlebee}, {Umana}, {Valenziano}, {Valiviita}, {Van Tent}, {Varis}, {Vibert}, {Viel}, {Vielva}, {Villa}, {Vittorio}, {Wade}, {Wandelt}, {Watson}, {Watson}, {Wehus}, {Welikala}, {Weller}, {White}, {White}, {Wilkinson}, {Winkel}, {Xia}, {Yvon}, {Zacchei}, {Zibin}, \& {Zonca}}]{planck2014}
{Planck Collaboration}, {Ade}, P.~A.~R., {Aghanim}, N., {et~al.} 2014, \aap, 571, A1, \dodoi{10.1051/0004-6361/201321529}

\bibitem[{{Planck Collaboration} {et~al.}(2020){Planck Collaboration}, {Aghanim}, {Akrami}, {Ashdown}, {Aumont}, {Baccigalupi}, {Ballardini}, {Banday}, {Barreiro}, {Bartolo}, {Basak}, {Battye}, {Benabed}, {Bernard}, {Bersanelli}, {Bielewicz}, {Bock}, {Bond}, {Borrill}, {Bouchet}, {Boulanger}, {Bucher}, {Burigana}, {Butler}, {Calabrese}, {Cardoso}, {Carron}, {Challinor}, {Chiang}, {Chluba}, {Colombo}, {Combet}, {Contreras}, {Crill}, {Cuttaia}, {de Bernardis}, {de Zotti}, {Delabrouille}, {Delouis}, {Di Valentino}, {Diego}, {Dor{\'e}}, {Douspis}, {Ducout}, {Dupac}, {Dusini}, {Efstathiou}, {Elsner}, {En{\ss}lin}, {Eriksen}, {Fantaye}, {Farhang}, {Fergusson}, {Fernandez-Cobos}, {Finelli}, {Forastieri}, {Frailis}, {Fraisse}, {Franceschi}, {Frolov}, {Galeotta}, {Galli}, {Ganga}, {G{\'e}nova-Santos}, {Gerbino}, {Ghosh}, {Gonz{\'a}lez-Nuevo}, {G{\'o}rski}, {Gratton}, {Gruppuso}, {Gudmundsson}, {Hamann}, {Handley}, {Hansen}, {Herranz}, {Hildebrandt}, {Hivon}, {Huang}, {Jaffe}, {Jones}, {Karakci}, {Keih{\"a}nen},
  {Keskitalo}, {Kiiveri}, {Kim}, {Kisner}, {Knox}, {Krachmalnicoff}, {Kunz}, {Kurki-Suonio}, {Lagache}, {Lamarre}, {Lasenby}, {Lattanzi}, {Lawrence}, {Le Jeune}, {Lemos}, {Lesgourgues}, {Levrier}, {Lewis}, {Liguori}, {Lilje}, {Lilley}, {Lindholm}, {L{\'o}pez-Caniego}, {Lubin}, {Ma}, {Mac{\'\i}as-P{\'e}rez}, {Maggio}, {Maino}, {Mandolesi}, {Mangilli}, {Marcos-Caballero}, {Maris}, {Martin}, {Martinelli}, {Mart{\'\i}nez-Gonz{\'a}lez}, {Matarrese}, {Mauri}, {McEwen}, {Meinhold}, {Melchiorri}, {Mennella}, {Migliaccio}, {Millea}, {Mitra}, {Miville-Desch{\^e}nes}, {Molinari}, {Montier}, {Morgante}, {Moss}, {Natoli}, {N{\o}rgaard-Nielsen}, {Pagano}, {Paoletti}, {Partridge}, {Patanchon}, {Peiris}, {Perrotta}, {Pettorino}, {Piacentini}, {Polastri}, {Polenta}, {Puget}, {Rachen}, {Reinecke}, {Remazeilles}, {Renzi}, {Rocha}, {Rosset}, {Roudier}, {Rubi{\~n}o-Mart{\'\i}n}, {Ruiz-Granados}, {Salvati}, {Sandri}, {Savelainen}, {Scott}, {Shellard}, {Sirignano}, {Sirri}, {Spencer}, {Sunyaev}, {Suur-Uski}, {Tauber}, {Tavagnacco},
  {Tenti}, {Toffolatti}, {Tomasi}, {Trombetti}, {Valenziano}, {Valiviita}, {Van Tent}, {Vibert}, {Vielva}, {Villa}, {Vittorio}, {Wandelt}, {Wehus}, {White}, {White}, {Zacchei}, \& {Zonca}}]{planck2020}
{Planck Collaboration}, {Aghanim}, N., {Akrami}, Y., {et~al.} 2020, \aap, 641, A6, \dodoi{10.1051/0004-6361/201833910}

\bibitem[{{Poggio} {et~al.}(2021){Poggio}, {Drimmel}, {Cantat-Gaudin}, {Ramos}, {Ripepi}, {Zari}, {Andrae}, {Blomme}, {Chemin}, {Clementini}, {Figueras}, {Fouesneau}, {Fr{\'e}mat}, {Lobel}, {Marshall}, {Muraveva}, \& {Romero-G{\'o}mez}}]{Poggio2021}
{Poggio}, E., {Drimmel}, R., {Cantat-Gaudin}, T., {et~al.} 2021, \aap, 651, A104, \dodoi{10.1051/0004-6361/202140687}

\bibitem[{{Portail} {et~al.}(2017){Portail}, {Wegg}, {Gerhard}, \& {Ness}}]{2017MNRAS.470.1233P}
{Portail}, M., {Wegg}, C., {Gerhard}, O., \& {Ness}, M. 2017, \mnras, 470, 1233, \dodoi{10.1093/mnras/stx1293}

\bibitem[{{Posti} \& {Helmi}(2019)}]{posti2019}
{Posti}, L., \& {Helmi}, A. 2019, \aap, 621, A56, \dodoi{10.1051/0004-6361/201833355}

\bibitem[{{Price-Whelan}(2017)}]{Price-Whelan2017}
{Price-Whelan}, A.~M. 2017, The Journal of Open Source Software, 2, 388, \dodoi{10.21105/joss.00388}

\bibitem[{{Robin} {et~al.}(2022){Robin}, {Bienaym{\'e}}, {Salomon}, {Reyl{\'e}}, {Lagarde}, {Figueras}, {Mor}, {Fern{\'a}ndez-Trincado}, \& {Montillaud}}]{Robin2022}
{Robin}, A.~C., {Bienaym{\'e}}, O., {Salomon}, J.~B., {et~al.} 2022, \aap, 667, A98, \dodoi{10.1051/0004-6361/202243686}

\bibitem[{{Sanders} \& {Binney}(2016)}]{2016MNRAS.457.2107S}
{Sanders}, J.~L., \& {Binney}, J. 2016, \mnras, 457, 2107, \dodoi{10.1093/mnras/stw106}

\bibitem[{{Sanders} \& {Das}(2018)}]{2018MNRAS.481.4093S}
{Sanders}, J.~L., \& {Das}, P. 2018, \mnras, 481, 4093, \dodoi{10.1093/mnras/sty2490}

\bibitem[{{Schaller} {et~al.}(2015){Schaller}, {Frenk}, {Bower}, {Theuns}, {Jenkins}, {Schaye}, {Crain}, {Furlong}, {Dalla Vecchia}, \& {McCarthy}}]{schaller2015}
{Schaller}, M., {Frenk}, C.~S., {Bower}, R.~G., {et~al.} 2015, \mnras, 451, 1247, \dodoi{10.1093/mnras/stv1067}

\bibitem[{{Sch{\"o}nrich} {et~al.}(2010){Sch{\"o}nrich}, {Binney}, \& {Dehnen}}]{schonrich2010}
{Sch{\"o}nrich}, R., {Binney}, J., \& {Dehnen}, W. 2010, \mnras, 403, 1829, \dodoi{10.1111/j.1365-2966.2010.16253.x}

\bibitem[{{Schwarzschild}(1979)}]{1979ApJ...232..236S}
{Schwarzschild}, M. 1979, \apj, 232, 236, \dodoi{10.1086/157282}

\bibitem[{{Schwarzschild}(1993)}]{1993ApJ...409..563S}
---. 1993, \apj, 409, 563, \dodoi{10.1086/172687}

\bibitem[{{Sun} {et~al.}(2023){Sun}, {Wang}, {Liu}, {Long}, {Chen}, \& {Gao}}]{Sun2023}
{Sun}, G., {Wang}, Y., {Liu}, C., {et~al.} 2023, Research in Astronomy and Astrophysics, 23, 015013, \dodoi{10.1088/1674-4527/ac9e91}

\bibitem[{{Sun} {et~al.}(2024){Sun}, {Huang}, {Shen}, {Wang}, {Zhang}, {Tian}, {Liu}, \& {Jiang}}]{Sun2024}
{Sun}, W., {Huang}, Y., {Shen}, H., {et~al.} 2024, \apj, 961, 141, \dodoi{10.3847/1538-4357/ad06ad}

\bibitem[{{Syer} \& {Tremaine}(1996)}]{1996MNRAS.282..223S}
{Syer}, D., \& {Tremaine}, S. 1996, \mnras, 282, 223, \dodoi{10.1093/mnras/282.1.223}

\bibitem[{{Tarricq} {et~al.}(2021){Tarricq}, {Soubiran}, {Casamiquela}, {Cantat-Gaudin}, {Chemin}, {Anders}, {Antoja}, {Romero-G{\'o}mez}, {Figueras}, {Jordi}, {Bragaglia}, {Balaguer-N{\'u}{\~n}ez}, {Carrera}, {Castro-Ginard}, {Moitinho}, {Ramos}, \& {Bossini}}]{Tarricq2021}
{Tarricq}, Y., {Soubiran}, C., {Casamiquela}, L., {et~al.} 2021, \aap, 647, A19, \dodoi{10.1051/0004-6361/202039388}

\bibitem[{{Tian} {et~al.}(2015){Tian}, {Liu}, {Carlin}, {Zhao}, {Chen}, {Wu}, {Li}, {Hou}, \& {Zhang}}]{Tian2015}
{Tian}, H.-J., {Liu}, C., {Carlin}, J.~L., {et~al.} 2015, \apj, 809, 145, \dodoi{10.1088/0004-637X/809/2/145}

\bibitem[{{Valluri} \& {Merritt}(1999)}]{1999ASPC..182..178V}
{Valluri}, M., \& {Merritt}, D. 1999, in Astronomical Society of the Pacific Conference Series, Vol. 182, Galaxy Dynamics - A Rutgers Symposium, ed. D.~R. {Merritt}, M.~{Valluri}, \& J.~A. {Sellwood}, 178, \dodoi{10.48550/arXiv.astro-ph/9906176}

\bibitem[{{Vasiliev}(2018)}]{agama2018}
{Vasiliev}, E. 2018, arXiv e-prints, arXiv:1802.08255, \dodoi{10.48550/arXiv.1802.08255}

\bibitem[{{Vasiliev}(2019{\natexlab{a}})}]{agama2019}
---. 2019{\natexlab{a}}, \mnras, 482, 1525, \dodoi{10.1093/mnras/sty2672}

\bibitem[{{Vasiliev}(2019{\natexlab{b}})}]{Vasiliev2019}
---. 2019{\natexlab{b}}, \mnras, 484, 2832, \dodoi{10.1093/mnras/stz171}

\bibitem[{{Vieira} {et~al.}(2022){Vieira}, {Carraro}, {Korchagin}, {Lutsenko}, {Girard}, \& {van Altena}}]{vieira2022}
{Vieira}, K., {Carraro}, G., {Korchagin}, V., {et~al.} 2022, \apj, 932, 28, \dodoi{10.3847/1538-4357/ac6b9b}

\bibitem[{{Wang} {et~al.}(2023){Wang}, {Chrob{\'a}kov{\'a}}, {L{\'o}pez-Corredoira}, \& {Sylos Labini}}]{Wang2023}
{Wang}, H.-F., {Chrob{\'a}kov{\'a}}, {\v{Z}}., {L{\'o}pez-Corredoira}, M., \& {Sylos Labini}, F. 2023, \apj, 942, 12, \dodoi{10.3847/1538-4357/aca27c}

\bibitem[{{Wang} {et~al.}(2017){Wang}, {Wang}, {Liu}, {Mao}, \& {Long}}]{Wang2017}
{Wang}, Q., {Wang}, Y., {Liu}, C., {Mao}, S., \& {Long}, R.~J. 2017, \mnras, 470, 2949, \dodoi{10.1093/mnras/stx1382}

\bibitem[{{Wang} {et~al.}(2020){Wang}, {Han}, {Cautun}, {Li}, \& {Ishigaki}}]{Wang2020}
{Wang}, W., {Han}, J., {Cautun}, M., {Li}, Z., \& {Ishigaki}, M.~N. 2020, Science China Physics, Mechanics, and Astronomy, 63, 109801, \dodoi{10.1007/s11433-019-1541-6}

\bibitem[{{Wang} {et~al.}(2013){Wang}, {Mao}, {Long}, \& {Shen}}]{Wang2013}
{Wang}, Y., {Mao}, S., {Long}, R.~J., \& {Shen}, J. 2013, \mnras, 435, 3437, \dodoi{10.1093/mnras/stt1537}

\bibitem[{{Wang} {et~al.}(2012){Wang}, {Zhao}, {Mao}, \& {Rich}}]{Wang2012}
{Wang}, Y., {Zhao}, H., {Mao}, S., \& {Rich}, R.~M. 2012, \mnras, 427, 1429, \dodoi{10.1111/j.1365-2966.2012.22063.x}

\bibitem[{{Watkins} {et~al.}(2010){Watkins}, {Evans}, \& {An}}]{Watkins2010}
{Watkins}, L.~L., {Evans}, N.~W., \& {An}, J.~H. 2010, \mnras, 406, 264, \dodoi{10.1111/j.1365-2966.2010.16708.x}

\bibitem[{{Webb} {et~al.}(2023){Webb}, {Hunt}, \& {Bovy}}]{Webb2023}
{Webb}, J.~J., {Hunt}, J. A.~S., \& {Bovy}, J. 2023, \mnras, 521, 3898, \dodoi{10.1093/mnras/stad762}

\bibitem[{{Wenger} {et~al.}(2000){Wenger}, {Ochsenbein}, {Egret}, {Dubois}, {Bonnarel}, {Borde}, {Genova}, {Jasniewicz}, {Lalo{\"e}}, {Lesteven}, \& {Monier}}]{Wenger2000}
{Wenger}, M., {Ochsenbein}, F., {Egret}, D., {et~al.} 2000, \aaps, 143, 9, \dodoi{10.1051/aas:2000332}

\bibitem[{{Xiang} {et~al.}(2018){Xiang}, {Shi}, {Liu}, {Yuan}, {Chen}, {Huang}, {Wang}, {Wu}, {Tian}, {Huo}, {Zhang}, \& {Zhang}}]{2018ApJS..237...33X}
{Xiang}, M., {Shi}, J., {Liu}, X., {et~al.} 2018, \apjs, 237, 33, \dodoi{10.3847/1538-4365/aad237}

\bibitem[{{Xu} {et~al.}(2020){Xu}, {Liu}, {Tian}, {Newberg}, {Laporte}, {Zhang}, {Wang}, {Fu}, {Li}, \& {Deng}}]{Xu2020}
{Xu}, Y., {Liu}, C., {Tian}, H., {et~al.} 2020, \apj, 905, 6, \dodoi{10.3847/1538-4357/abc2cb}

\bibitem[{{Xue} {et~al.}(2008){Xue}, {Rix}, {Zhao}, {Re Fiorentin}, {Naab}, {Steinmetz}, {van den Bosch}, {Beers}, {Lee}, {Bell}, {Rockosi}, {Yanny}, {Newberg}, {Wilhelm}, {Kang}, {Smith}, \& {Schneider}}]{Xue2008}
{Xue}, X.~X., {Rix}, H.~W., {Zhao}, G., {et~al.} 2008, \apj, 684, 1143, \dodoi{10.1086/589500}

\bibitem[{{Zhao}(1996)}]{Zhao1996}
{Zhao}, H. 1996, \mnras, 283, 149, \dodoi{10.1093/mnras/283.1.149}

\bibitem[{{Zhu} {et~al.}(2014){Zhu}, {Long}, {Mao}, {Peng}, {Liu}, {Caldwell}, {Li}, {Blakeslee}, {C{\^o}t{\'e}}, {Cuillandre}, {Durrell}, {Emsellem}, {Ferrarese}, {Gwyn}, {Jord{\'a}n}, {Lan{\c{c}}on}, {Mei}, {Mu{\~n}oz}, \& {Puzia}}]{Zhu2014}
{Zhu}, L., {Long}, R.~J., {Mao}, S., {et~al.} 2014, \apj, 792, 59, \dodoi{10.1088/0004-637X/792/1/59}

\end{thebibliography}
\bibliographystyle{aasjournal}

\appendix
\numberwithin{figure}{section}
\section{velocity distribution}
\label{sec:appendix}
In this appendix, we show the rest velocity distributions for the regions we consider.
\begin{figure*} [!htb]
\includegraphics[width=\textwidth]{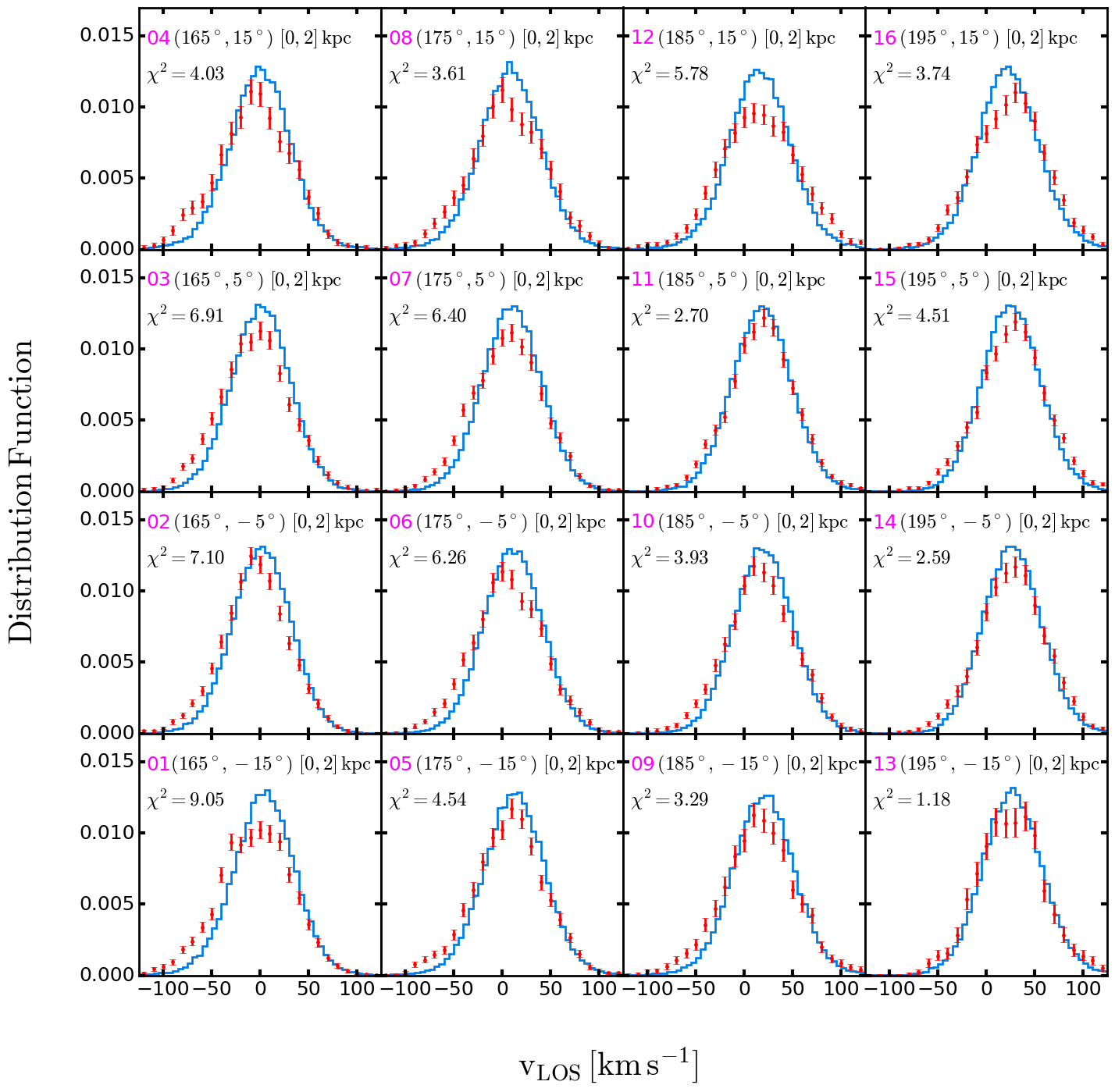}
\caption{The distribution function of the line-of-sight velocity ($v_{\rm LOS}$) for sky regions 01-16, as defined in Table~\ref{tab:sky area}. Each sky region covers an area of $10^\circ \times 10^\circ$. The reduced $\chi^2$ is indicated in each panel.}
\label{fig:rv01-16}
\end{figure*}

\begin{figure*}[!htb]
\includegraphics[width=\textwidth]{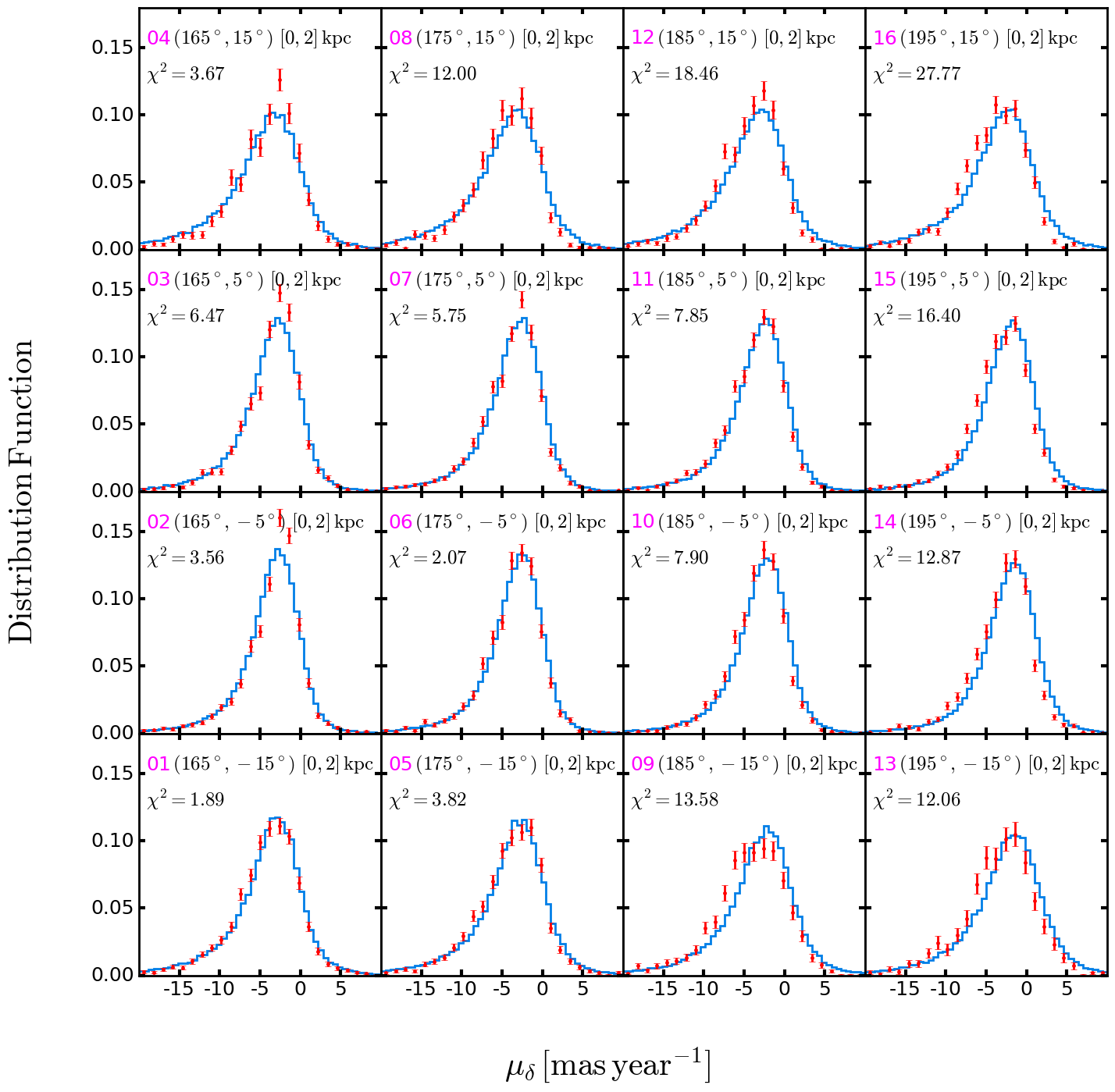}
\caption{The distribution function of proper motion in the declination direction ($\mu_\delta$) for sky regions 01-16, as defined in Table~\ref{tab:sky area}. Each sky region covers an area of $10^\circ \times 10^\circ$.}
\label{fig:dec01-16}
\end{figure*}

\begin{figure}[!htb]
\includegraphics[width=\columnwidth]{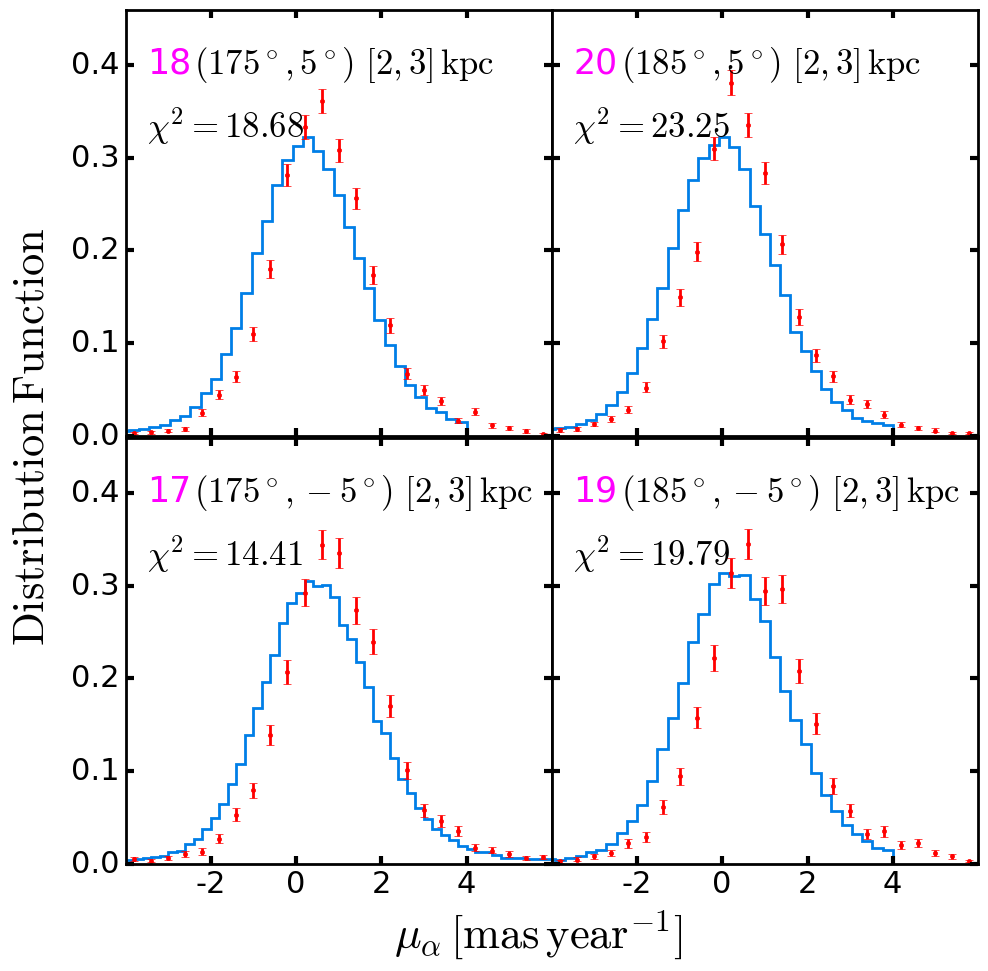}
\caption{The distribution function of proper motion along the right ascension ($\mu_\alpha$) for sky regions 17-20, as defined in Table~\ref{tab:sky area}. Each sky region covers an area of $10^\circ \times 10^\circ$.}
\label{fig:ra17-20}
\end{figure}

\begin{figure}[!htb]
\includegraphics[width=\columnwidth]{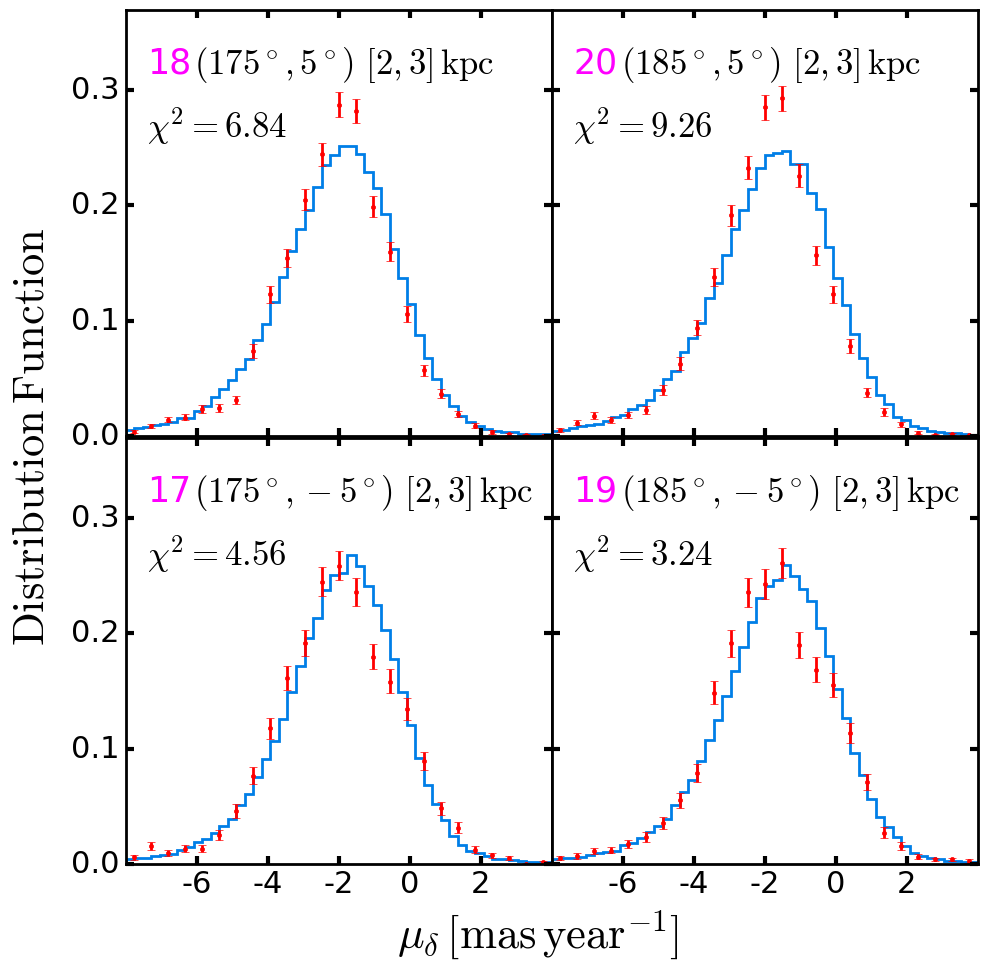}
\caption{The distribution function of proper motion in the declination direction ($\mu_\delta$) for sky regions 17-20, as defined in Table~\ref{tab:sky area}. Each sky region covers an area of $10^\circ \times 10^\circ$.}
\label{fig:dec17-20}
\end{figure}

\begin{figure}[!htb]
\includegraphics[width=\columnwidth]{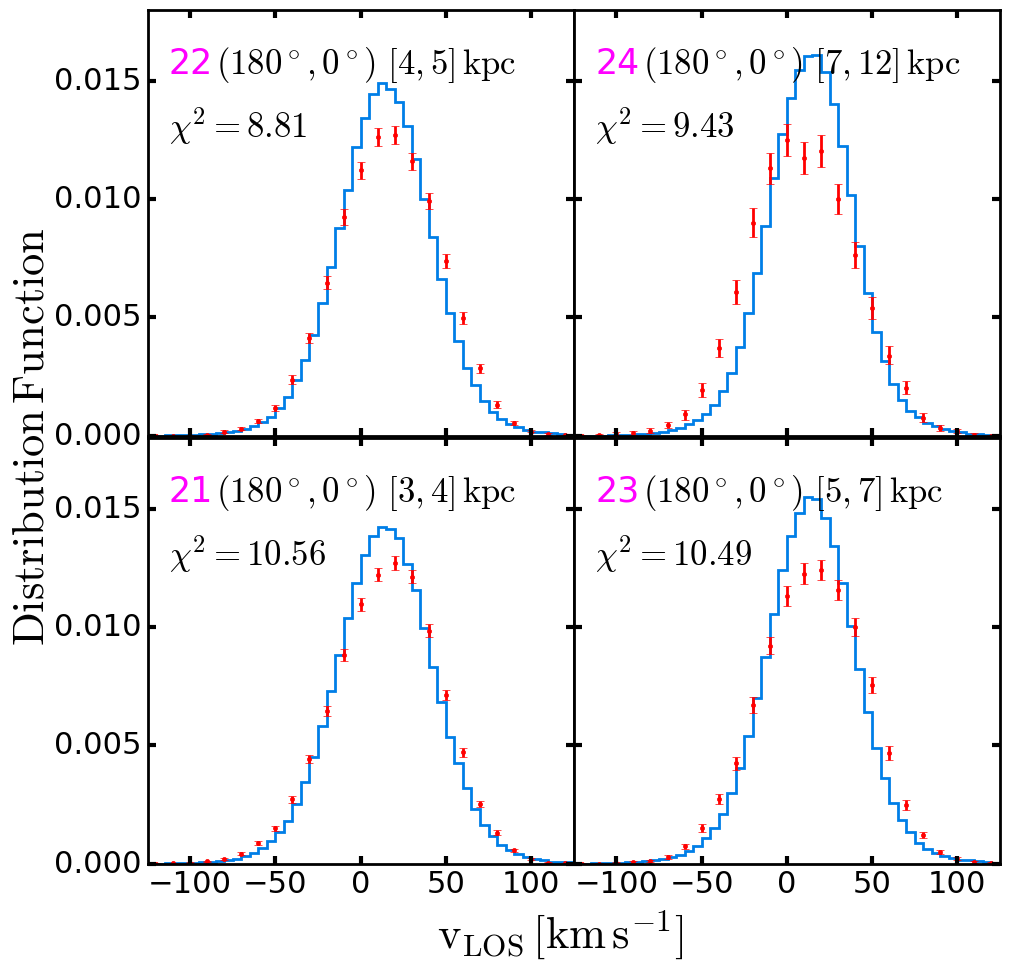}
\caption{The distribution function of the line-of-sight velocity ($v_{\rm LOS}$) for sky regions 21-24, as defined in Table~\ref{tab:sky area}. Each sky region covers an area of $20^\circ \times 20^\circ$.}
\label{fig:rv21-24}
\end{figure}

\begin{figure}[!htb]
\includegraphics[width=\columnwidth]{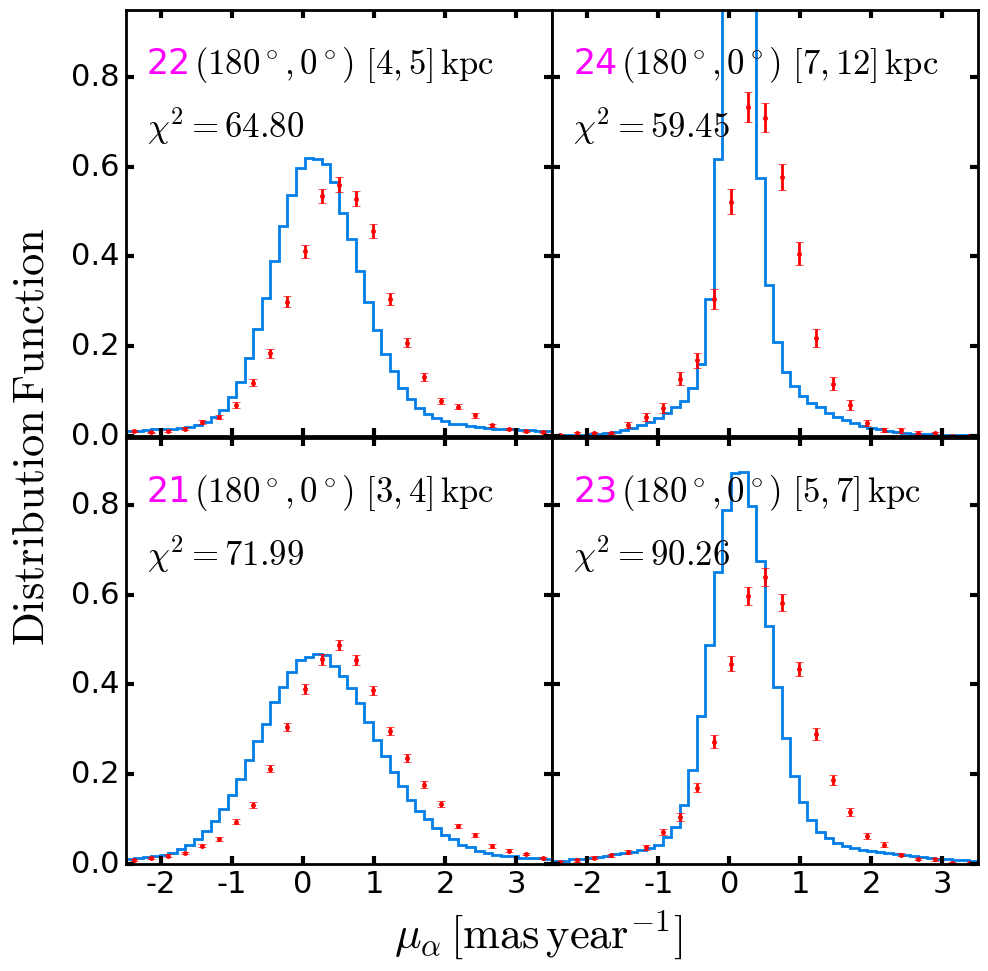}
\caption{The distribution function of proper motion along the right ascension ($\mu_\alpha$) for sky regions 21-24, as defined in Table~\ref{tab:sky area}. Each sky region covers an area of $20^\circ \times 20^\circ$.}
\label{fig:ra21-24}
\end{figure}

\begin{figure}[!htb]
\includegraphics[width=\columnwidth]{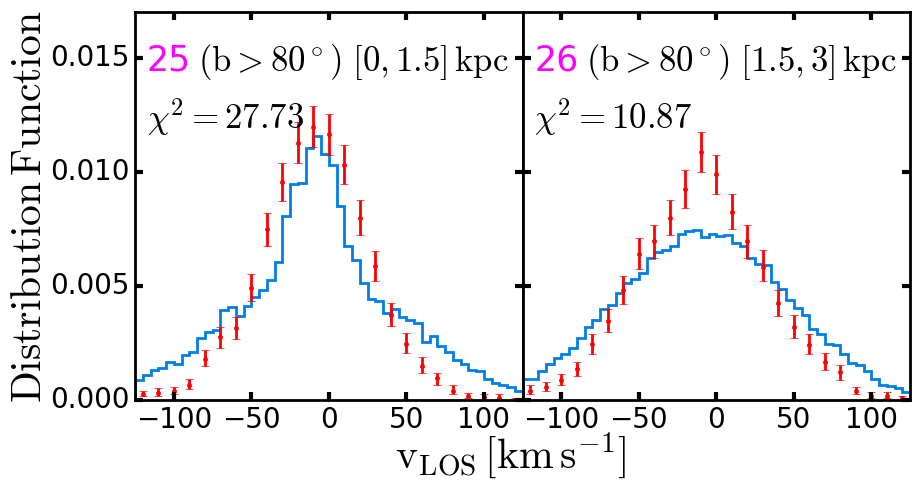}
\caption{The distribution function of the line-of-sight velocity ($v_{\rm LOS}$)  for sky regions 25 and 26, as defined in Table~\ref{tab:sky area}. Each sky region covers a circle with a diameter of $10^\circ$.}
\label{fig:rv30-32}
\end{figure}

\begin{figure}[!htb]
\includegraphics[width=\columnwidth]{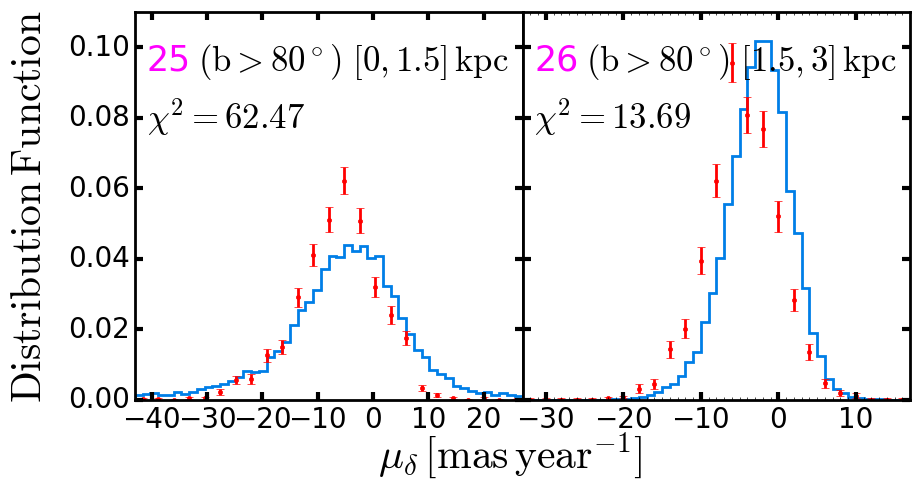}
\caption{The distribution function of proper motion in the declination direction ($\mu_\delta$) for sky regions 25 and 26, as defined in Table~\ref{tab:sky area}. Each sky region covers a circle with a diameter of $10^\circ$.}
\label{fig:dec30-32}
\end{figure}

\begin{figure}[!htb]
\includegraphics[width=\columnwidth]{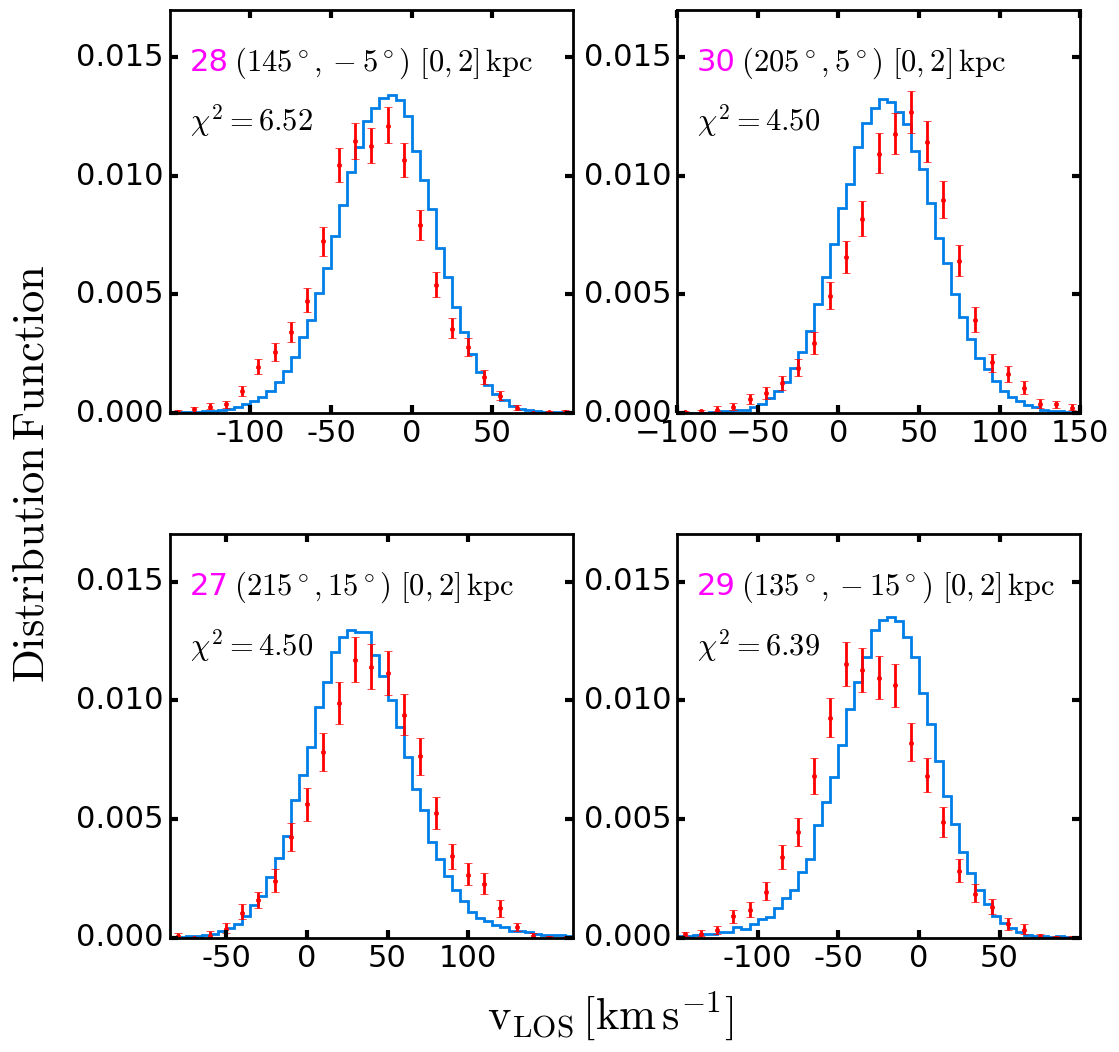}
\caption{The distribution function of the line-of-sight velocity ($v_{\rm LOS}$) for sky regions 27-30, as defined in Table~\ref{tab:sky area}. Each sky region covers an area of $10^\circ \times 10^\circ$.}
\label{fig:rv40-43}
\end{figure}

\begin{figure}[!htb]
\includegraphics[width=\columnwidth]{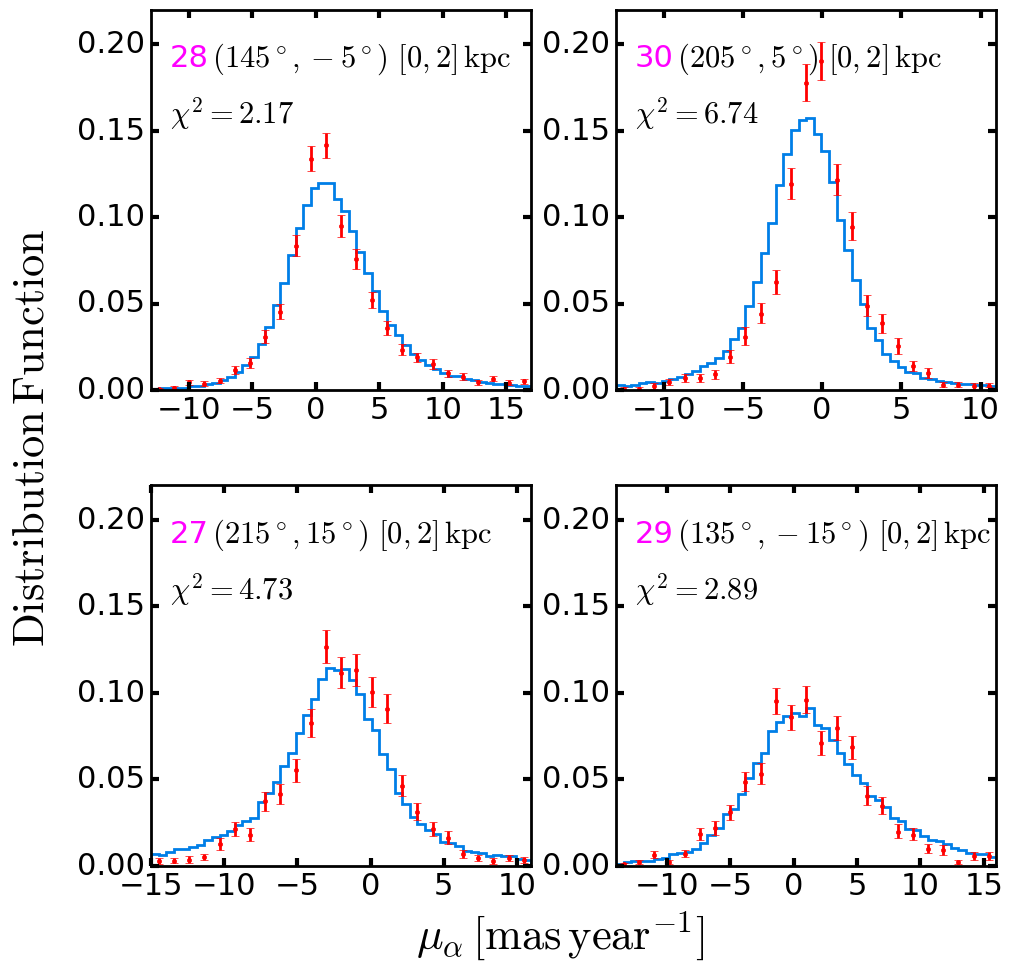}
\caption{The distribution function of proper motion along the right ascension ($\mu_\alpha$) for sky regions 27-30, as defined in Table~\ref{tab:sky area}. Each sky region covers an area of $10^\circ \times 10^\circ$.}
\label{fig:ra40-43}
\end{figure}



\end{document}